\documentclass[10pt,twocolumn,letterpaper]{article}

\usepackage[pagenumbers]{cvpr}              

\usepackage{tocloft}

\advance\cftsecnumwidth 0.5em\relax
\advance\cftsubsecindent 0.5em\relax
\advance\cftsubsecnumwidth 0.5em\relax

\usepackage[dvipsnames]{xcolor}
\usepackage[most]{tcolorbox}

\usepackage{multirow}
\usepackage{listings}
\newcommand{\ignore}[1]{}
\usepackage{lipsum} 
\usepackage{float} 

\definecolor{codegreen}{rgb}{0,0.6,0}
\definecolor{codegray}{rgb}{0.5,0.5,0.5}
\definecolor{codepurple}{rgb}{0.58,0,0.82}
\definecolor{backcolour}{rgb}{0.95,0.95,0.92}

\lstdefinestyle{mystyle}{
    backgroundcolor=\color{backcolour},
    commentstyle=\color{codegreen},
    keywordstyle=\color{codepurple},
    numberstyle=\tiny\color{codegray},
    stringstyle=\color{codepurple},
    basicstyle=\ttfamily\footnotesize,
    breakatwhitespace=false,
    breaklines=true,
    captionpos=b,
    keepspaces=true,
    numbersep=5pt,
    frame = shadowbox,
    showspaces=false,
    showstringspaces=false,
    showtabs=false,
    tabsize=2
}

\newcommand{\boxedthm}[1]{
\begin{tcolorbox}[colback=gray!20,
                  colframe=black,
                  width=\linewidth,
                  arc=1mm, auto outer arc,
                  boxrule=1pt,
                  boxsep=-1mm,
                 ]
  #1
\end{tcolorbox}
}

\newcommand{\boxedeg}[1]{
\begin{tcolorbox}[colback=gray!00,
                  colframe=black,
                  width=\linewidth,
                  arc=1mm, auto outer arc,
                  boxrule=1pt,
                  boxsep=-1mm,
                 ]
  #1
\end{tcolorbox}
}

\definecolor{cvprblue}{rgb}{0.21,0.49,0.74}
\usepackage[pagebackref,breaklinks,colorlinks,citecolor=cvprblue]{hyperref}

\newcommand{\excludefromtoc}[1]{%
    \addtocontents{toc}{\protect\setcounter{tocdepth}{-10}}%
    #1
    \addtocontents{toc}{\protect\setcounter{tocdepth}{3}}%
}

\def\paperID{14975} 
\def\confName{CVPR}
\def\confYear{2024}

\title{Resolution Limit of Single-Photon LiDAR}

\author{Stanley H. Chan$^{\dag 1}$
\and
Hashan K. Weerasooriya$^{1}$
\and
Weijian Zhang$^{1}$
\and
Pamela Abshire$^{2}$
\and
Istvan Gyongy$^{3}$
\and
Robert K. Henderson$^{3}$
\and
$^1$ Purdue University, $^2$ Univ. Maryland College Park, $^3$ University of Edinburgh
}

\begin{document}
\maketitle
\begin{abstract}
Single-photon Light Detection and Ranging (LiDAR) systems are often equipped with an array of detectors for improved spatial resolution and sensing speed. However, given a fixed amount of flux produced by the laser transmitter across the scene, the per-pixel Signal-to-Noise Ratio (SNR) will decrease when more pixels are packed in a unit space. This presents a fundamental trade-off between the spatial resolution of the sensor array and the SNR received at each pixel. Theoretical characterization of this fundamental limit is explored. By deriving the photon arrival statistics and introducing a series of new approximation techniques, the Mean Squared Error (MSE) of the maximum-likelihood estimator of the time delay is derived. The theoretical predictions align well with simulations and real data.
\end{abstract} 
\excludefromtoc{
\section{Introduction}
\label{sec:intro}

Single-photon LiDAR has a wide range of applications in navigation and object identification \cite{Rapp_2020_SPM, Morimoto_2020_SPAD, Li_2023_SPAD, McCarthy_2013_km_1560_depth, Qian_2024_SP_LiDAR_SOTA_for_autonomous, Mora-Martin_2021_dynamic_object_detection}. By actively illuminating the scene with a laser pulse of a known shape, we measure the time delays of single photons upon their return, which correspond to the distance of the object \cite{Boretti_2024_SP_Lidar_System,   Kirmani_2014_FirstPhoton, Shin_2015_3D}. The advancement of photo detectors has significantly improved the resolution of today's LiDAR ~\cite{wayne_2022_500x500_SPAD, Richardson_2009_32x32, Dutton_2016_QVGA, Henderson_2019_192x128, Ulku_2019_512x512, Villa_2021_Sensors, Hadfield_2023_spad_sensor_review}. Moreover, algorithms have shown how to reconstruct both the scene reflectivity and 3D structure~\cite{Shin_2015_3D, Lee_2023_CASPI, Poisson_2022_Compressed, Lindell_2018_SIGGRAPH, Altmann_2016_LiDAR, Tachella_2019_SP_PnP_reconstriction, Wei_2023_wideband_lidar, Xin_2019_fermat_NLS_reconstruction, Choi_2018_QIS_reconstruction_DL, Heide_2018_sub_ps_SP_LiDAR_reconstruction, Sizhuo_2020_QBP}.

\begin{figure}[t!]
\centering
\includegraphics[width=\linewidth]{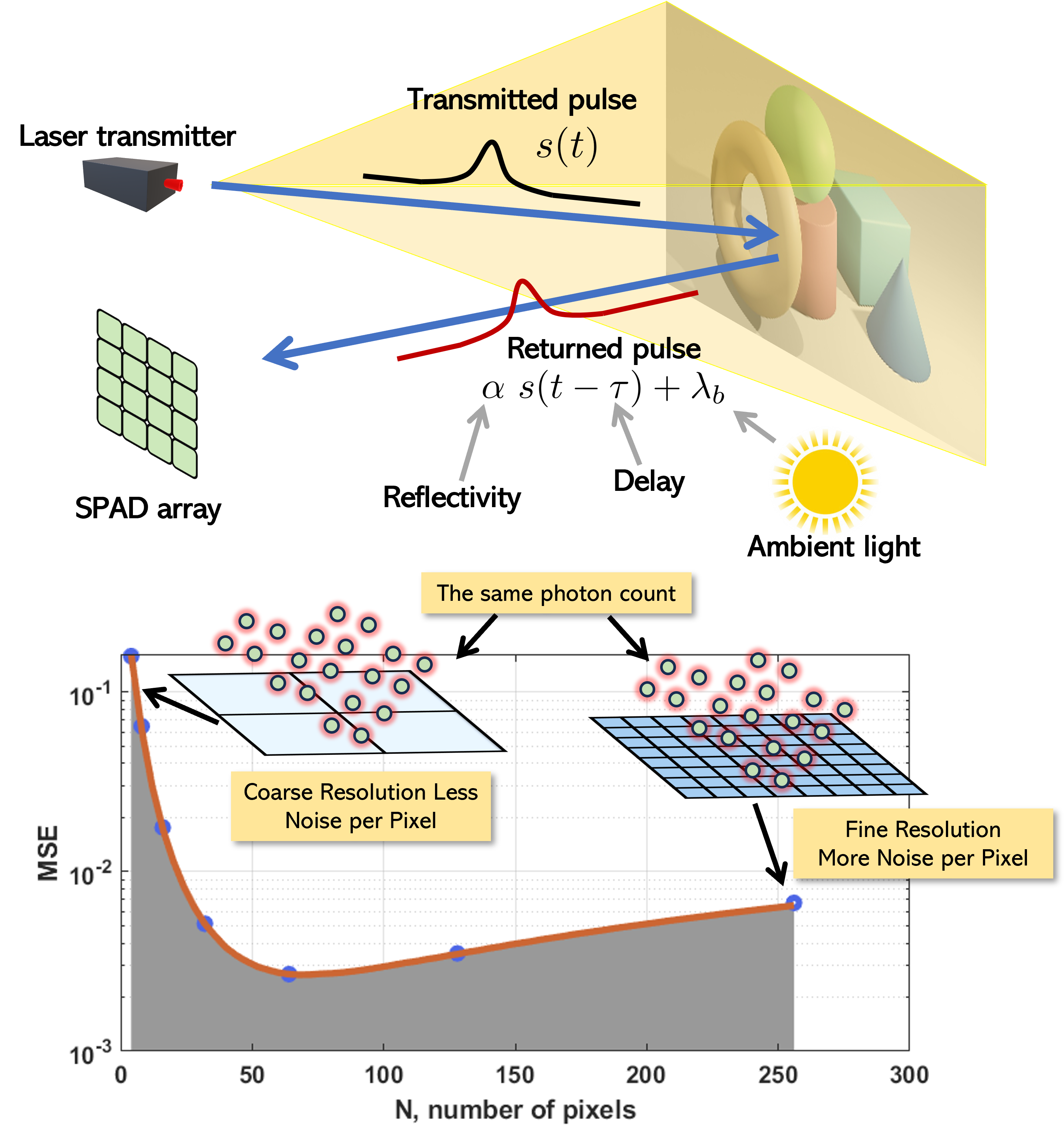}
\caption{As we pack more pixels in a unit space, we gain the spatial resolution with a reduction in the SNR. The goal of this paper is to understand the trade-off between the two factors.}
\label{fig: main}
\end{figure}

As an imaging device, a photodetector used in LiDAR faces similar problems as any other CCD or CMOS pixels. Packing more pixels into a unit space decreases the SNR because the amount of photon flux seen by each pixel diminishes \cite{Fossum_2016_QIS_review}. This fundamental limit is linked to the stochastic nature of the underlying Poisson arrival process of the photons \cite{Snyder_1991_book, Fossum_2023_CMOS_dev}. Unless noise mitigation schemes are employed~\cite{Altmann_2016_LiDAR, Lindell_2018_SIGGRAPH, Rapp_2017_Unmixing, Gupta_2019_Flooded}, there is a trade-off between the number of pixels one can pack in a unit space and the SNR we will observe at each pixel. The situation can be visualized in \cref{fig: main}, where we highlight the phenomenon that if we use many small pixels, the spatial resolution is good but the per pixel noise caused by the random fluctuation of photons will be high. The bias and variance trade-off will then lead to a performance curve that tells us how the accuracy of the depth estimate will behave as we vary the spatial resolution.

The goal of this paper is to rigorously derive the above phenomenon. In particular, we want to answer the following question:

\boxedthm{
Can we theoretically derive, ideally in closed-form, the mean squared error of the LiDAR depth estimate as a function of the number of pixels per unit space?
}

The theoretical analysis presented in this paper is unique from several perspectives:
\begin{itemize}
\item \textbf{Beyond Single Pixel}. The majority of the computer vision papers in single-photon LiDAR are algorithmic. Few papers have theoretical derivations, but they all focus on a single pixel~\cite{Altmann_2016_LiDAR, Lindell_2018_SIGGRAPH, Rapp_2017_Unmixing, Gupta_2019_Flooded}, of which the foundation can be traced back to the original work of Bar-David (1969) ~\cite{Bar-David_1969}. Our paper departs from these results by generalizing the mean square estimation to an array of pixels.
\item \textbf{New Proof Techniques}. A brute force derivation of the mean squared error is notoriously difficult. We overcome the hurdles by introducing a series of new theoretical approximation techniques in terms of modeling depth, approximating pixels, and utilizing convolutions.
\item \textbf{Closed-form Results}. Under appropriate assumptions about the scene and sensors, our result has a simple interpretable \emph{closed-form} expression that provides an excellent match with the practical scenarios in both real-world and simulated experiments.
\end{itemize}
}


\excludefromtoc{
\section{Background: Photon Arrival Statistics}
\label{sec:Photon Arrival Statistics}

In this section, we discuss the mathematical preliminaries. Our result is based on Bar-David~\cite{Bar-David_1969} which precedes many of the more recently published work \cite{Rapp_2017_Unmixing, Gupta_2019_Asynchronous, Snyder_1991_book}. For notation simplicity, our models are derived in 1D. Moreover, to make the main text concise, proofs of theorems are presented in the supplementary material.

\subsection{Pulse Model}
Let $c = 3\times 10^8$ [m/s] be the speed of light, and let $d(x)$ be the distance [m] of the object at coordinate $x \in  \R$. Hence, the total time [s] for the pulse to travel forward and then back is $\tau(x) = \frac{2d(x)}{c}$. We assume that $\tau(x)$ is a continuous-space function with a continuous amplitude.

The laser pulse is defined as a symmetric time-invarying function $s(t)$. Given a delay $\tau$, the shifted pulse is $s(t-\tau)$.

\boxedeg{
\begin{example}
If the pulse is Gaussian, then
\begin{equation}
s(t-\tau) =
\underset{=\calN(t \,|\, \tau, \sigma^2_t)}{\underbrace{\frac{1}{\sqrt{2\pi\sigma^2_t
}}\exp\left\{-\frac{(t-\tau)^2}{2\sigma^2_t}\right\}}},
\end{equation}
where $\sigma_t$ denotes the standard deviation.
\end{example}
}

For simplicity, we ignore the boundary conditions by assuming that the observation interval $(-T,T)$ is significantly larger than the width of the pulse, i.e., $\sigma_t \ll T$. Moreover, We assume that the delay $\tau$ lies well inside the observation interval, and the pulse is normalized so that
\begin{equation}
\int_{-T}^T s(t-\tau) \; dt = 1.
\label{eq: integrate s}
\end{equation}

As the pulse reaches an object and is reflected back to the receiver, the received pulse takes the form of
\begin{equation}
\lambda(t) = \alpha\cdot s(t - \tau) + \lambda_b,
\label{eq: lambda(t)}
\end{equation}
In this equation, $\alpha$ denotes the reflectivity of the object. For simplicity, we assume that $\alpha$ is a constant. The constant $\lambda_b \in \R$ denotes the background flux due to ambient light. The energy $Q$ carried by $\lambda(t)$ is measured by
\begin{equation}
Q\bydef \int_{-T}^T \lambda(t) \, dt = \alpha + 2T\lambda_b,
\label{eq: energy per pulse}
\end{equation}
Which can be obtained by inserting \cref{eq: lambda(t)} into the integrand shown in \cref{eq: energy per pulse}, and then using \cref{eq: integrate s} to evaluate the integral.

\subsection{Time of Arrival}
Given $\lambda(t)$, we assume that $M$ number of time stamps are generated over $[-T,T]$. Denote these time stamps as $\vt_M = \{t_j\}_{j=1}^M$, where $-T \le t_1 < t_2 < \ldots < t_M \le T$. The joint distribution of $\vt_M$ and $M$ is as follows.

\boxedthm{
\begin{theorem}[\cite{Bar-David_1969} Joint distribution of $M$ time stamps]
\label{thm: joint time stamp PDF}
Let $\vt_M = \{t_j\}_{j=1}^M$ such that $-T \le t_1 < t_2 < \ldots < t_M \le T$. For $M \ge 1$,
\begin{equation}
p(\vt_M, M) = e^{-Q} \prod_{j=1}^M \lambda(t_j).
\label{eq: p(t_M)}
\end{equation}
\end{theorem}
}

The number $M$ is a random variable. The probability mass function of $M$ can be computed by marginalizing the joint distribution.

\boxedthm{
\begin{corollary}[Probability of $M$ occurrence]
\label{cor: M occurrence}
For any $M \ge 1$, the probability that there are $M$ occurrences is
\begin{equation}
p(M) = \frac{e^{-Q} Q^M}{M!}.
\end{equation}
\end{corollary}
}

If $M=0$, then there is no occurrence in $[-T,T]$. In this case, the probability is defined as
\begin{equation}
p(\vt_0, 0) = e^{-Q}.
\end{equation}

\boxedeg{
\begin{example}{}
Suppose $s(t)$ is a Gaussian pulse and assume that $\lambda_b = 0$ and $\alpha = 1$. Then,
\begin{align*}
p(\vt_M,M) = \frac{e^{-Q} }{(\sqrt{2\pi\sigma^2_t})^M} \exp\left\{-\sum_{j=1}^M \frac{(t_j-\tau)^2}{2\sigma^2_t}\right\}.
\end{align*}
\end{example}
}

The conditional probability of seeing $\vt_M$ given $M$ can be obtained by taking the ratio of the joint distribution $p(\vt_M,M)$ and $p(M)$, yielding the following result.
\begin{equation}
p(\vt_M \,|\, M) = \frac{p(\vt_M, M)}{p(M)} = Q^{-M}M! \prod_{j=1}^M \lambda(t_j).
\end{equation}
The other conditional probability of seeing $M$ given $\vt_M$ is 1. Putting these together, we can show that
\begin{equation}
p( \vt_M, M ) = p(M \,|\, \vt_M) p(\vt_M) = p(\vt_M).
\end{equation}
We can show that the integration of $p(\vt_M,M)$ over the entire sample space is 1:

\boxedthm{
\begin{corollary}[Probability over the sample space]
\label{cor: sample space}
\begin{equation}
\sum_{M=0}^{\infty} \int_{\Omega_M} p(\vt_M, M) \; d \vt_M = 1,
\end{equation}
where $\Omega_M = \{\vt_M \,|\, -T \le t_1 < t_2 < \ldots t_M \le T\}$.
\end{corollary}
}

\subsection{Sampling from {\boldmath{$p(t_M)$}}}
When the pulse is Gaussian, Monte Carlo simulations of the time stamps can be performed in a two-step process:

\begin{itemize}
\item Step 1: Determine the number of samples $M$. This can be done by recognizing that the total energy of the pulse is $Q = \alpha + 2T \lambda_b$. The total number of samples $M$ is a Poisson random variable such that $M \sim \text{Poisson}(Q)$. However, since the two summands of $Q$ are independent, Raikov theorem states that $M$ can be decomposed into a sum of two independent Poisson random variables. Thus, the number of samples is determined based on
\begin{equation}
M_s \sim \text{Poisson}(\alpha), \qquad M_b \sim \text{Poisson}(2T\lambda_b).
\end{equation}
We let $M = M_s + M_b$.
\item Step 2: Draw $M_s$ samples from $\calN(t \,|\, \tau, \sigma^2)$ and $M_b$ samples from a uniform distribution of a PDF:
\begin{align*}
t_j  \,|\, M_s &\sim \calN(t \,|\, \tau, \sigma^2), \qquad j = 1,\ldots,M_s,\\
t_i  \,|\, M_b
 &\sim \text{Uniform}(-T, T), \qquad i = 1,\ldots,M_b.
\end{align*}
The overall set of samples is $\vt_M = \{t_j\}_{j=1}^{M_s} \cup \{t_i\}_{i=1}^{M_b}$.
\end{itemize}
As we can see, the distribution of the samples is nothing but the shape of the pulse. This is consistent with the literature where we draw a random number representing the height of each histogram bin. In our sampling procedure, we draw the time stamps \emph{without} quantizing them into bins. For pulses of an arbitrary shape, we can perform an inverse CDF technique outlined in the supplementary material.

\subsection{Assumptions For Theoretical Analysis}
The goal of this paper is to derive \emph{closed-form} results. As such, a series of assumptions are required to minimize the notational burden. Our assumptions are summarized below:
\begin{itemize}
\item We do not assume any dark current. In the supplementary material, we have a discussion about the dark current effects.
\item We assume that $\alpha$ is a constant. To relax this assumption, we can replace $\alpha$ with $\alpha(x)$ in the proof. However, the final equation will involve an integration over $\alpha(x)$. 
\item Dead-time and Pile-up \cite{Coates_1968, Pediredla_2018, Gupta_2019_Asynchronous, Ryan_2022_adaptive_gating, Gupta_2019_Flooded}. We assume there is no dead-time and hence no pile-up. The empirical analysis in the supplementary material, however, includes a case study that involves pile-up effect.
\item Self-excitation process. Prior work such as \cite{Rapp_2017_Unmixing} and \cite{Gupta_2019_Asynchronous} use self-excitation process (a variant of the Markov chain) to model the photon arrivals \cite{Snyder_1991_book}. While this is accurate, deriving closed-form expressions is infeasible. Since we do not assume any dead-time, we follow Bar-David's inhomogeneous Poisson process \cite{Bar-David_1969} instead.
\item Single-bounce and no multiple path. This is a standard assumption in LiDAR theory.
\end{itemize}
}

\excludefromtoc{
\section{MSE Analysis}
\subsection{Single-Pixel MSE}
\label{subsec: single pixel mse}
To quantify the performance of a LiDAR pixel, we recall that the decision process involves estimating the delay $\tau$ given the measurements $\vt_M$. Therefore, we need to specify the estimation procedure. Based on the estimates, we can then discuss the performance by evaluating the variance of the estimate.

\noindent \textbf{Maximum-Likelihood Estimation (MLE)}. When no knowledge about $\tau$ is known a priori, we use MLE \cite{Kay_2013_Estimation, Shin_2015_3D}. MLE has been thoroughly exploited in single-depth estimation problems \cite{Altmann_2016_LiDAR}. Given the measured time stamps $\vt_M = [t_1,\ldots,t_M]$, we consider the log-likelihood
\begin{align*}
\widehat{\tau}
&= \argmax{\tau} \;\; \calL(\tau) \bydef \sum_{j=1}^M \log\left[ \alpha s(t_j-\tau) + \lambda_b\right],
\end{align*}
 Since the variable $\tau$ in the ML estimation is the time shift, the optimization can be solved by running a matched filter. Given the shape $s(t)$, we shift the pulse left and right until we see the best match with the data. \cref{fig: matched filter} shows a pictorial illustration.

\begin{figure}[h]
	\centering
	\includegraphics[width=0.95\linewidth]{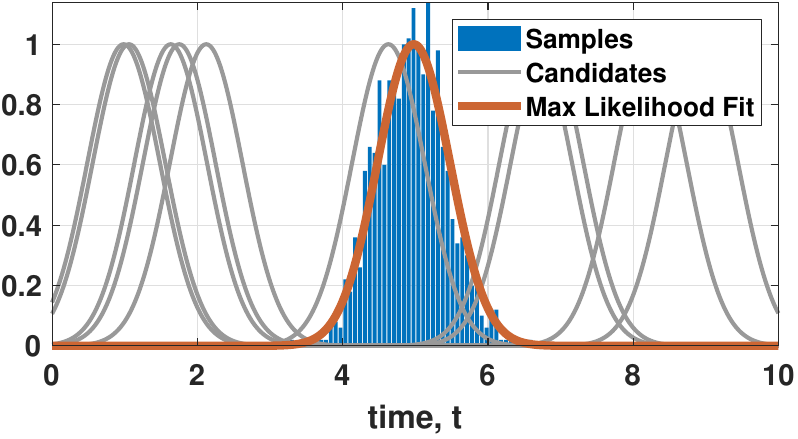}
	\caption{Matched filter: Given a known pulse shape, we shift the pulse until it matches with the measured samples.}
	\label{fig: matched filter}
\end{figure}

\noindent If $\widehat{\tau}$ is the ML estimate, it is necessary that
\begin{equation}
\left.\frac{d \calL}{d\tau}\right \vert_{\tau = \widehat{\tau}} = \left.\sum_{j=1}^M \frac{\alpha \dot{s}(t_j-\tau)}{\alpha s(t_j-\tau) + \lambda_b}\right \vert_{\tau = \widehat{\tau}} = 0.
\end{equation}
This result is used in the implementation. Details can be found in the supplementary material.

\noindent \textbf{MSE calculation}.When $\tau_0$ denotes the true time of arrival, the Taylor expansion of $\dot{\calL}(\tau) = d\calL/d\tau$ will give us
\begin{align*}
\dot{\calL}(\tau) = \dot{\calL}(\tau_0) + (\tau-\tau_0)\ddot{\calL}(\tau_0) + \ldots
\end{align*}
Substituting $\tau = \widehat{\tau}$, and using the fact that $\dot{\calL}(\widehat{\tau}) = 0$ because $\widehat{\tau}$ is the maximizer, we can show that
\begin{align*}
0 = \dot{\calL}(\widehat{\tau}) = \dot{\calL}(\tau_0) + (\widehat{\tau}-\tau_0)\ddot{\calL}(\tau_0) + \ldots
\end{align*}
Therefore, the error is $\widehat{\tau} -\tau_0 \approx -\frac{\dot{\calL}(\tau_0)}{\ddot{\calL}(\tau_0)}$. By using this result, the variance of the estimate $\widehat{\tau}$ can be shown as follows.

\boxedthm{
\begin{theorem}{}
\label{thm: MSE single pixel}
~\cite{Bar-David_1969, Abshire_1987_JOSA} Let $\lambda(t) = \alpha s(t-\tau_0) + \lambda_b$. Then
\begin{align}
\E[(\widehat{\tau}-\tau_0)^2] = \left[ \int_{-T}^T \frac{(\alpha \dot{s}(t))^2}{\alpha s(t) + \lambda_b} \; dt \right]^{-1},
\end{align}
where $\dot{s}(t)$ is the derivative of $s$ with respect to $t$.
\end{theorem}
}

\boxedeg{
\begin{example}{}
\label{example: Eg 3}
In the special case where $s(t) = \calN(t \,|\, \tau_0, \sigma^2_t)$, and assume that $\lambda_b = 0$, we have
\begin{align*}
&\E[(\widehat{\tau}-\tau_0)^2]
= \left[\int_{-T}^T \frac{(\alpha \dot{s}(t))^2}{\alpha s(t)+\lambda_b}  \; dt \right]^{-1}\\
&= \left[\bigintsss_{-T}^T \frac{\left(-\frac{t}{\sigma^2_t} \cdot \frac{\alpha }{\sqrt{2\pi \sigma_t^2}}e^{-\frac{t^2}{2\sigma^2_t}}\right)^2}{\frac{\alpha }{\sqrt{2\pi \sigma^2_t}}e^{-\frac{t^2}{2\sigma^2_t}}} \; dt\right]^{-1}\\
&= \left[\int_{-T}^T \frac{t^2}{\sigma^4_t} \frac{\alpha }{\sqrt{2\pi \sigma_t^2}}e^{-\frac{t^2}{2\sigma^2_t}} \; dt \right]^{-1}
 \approx \left(\frac{\alpha}{\sigma^2_t}\right)^{-1} = \frac{\sigma^2_t}{\alpha}.
\end{align*}
The last integration is the second moment of a zero-mean Gaussian, which will give us $\sigma^2_t$.
\end{example}
}

We remark that the per-pixel error calculated in \cref{thm: MSE single pixel} reaches the equality of the Cramer-Rao lower bound previously reported in~\cite{Shin_2013_ICIP, Erkmen_2009_ISIT, Scholes_2023_Fundamental}. Thus, no other unbiased estimator is better than what is reported here.

\subsection{Space-Time Model}
\label{subsec: space time model}
\noindent \textbf{Continuous {\boldmath$\lambda(x,t)$}}. Our resolution-noise trade-off analysis requires a model of an \emph{array} of pixels. To this end, we need to generalize from a single time delay $\tau$ to a function of time of arrivals $\tau(x)$ where $x$ is the spatial coordinate. Thus, at every location $x$, and given the pulse shape $s(t)$, the ideal return pulse is
\begin{equation}
\lambda(x,t) = \alpha \cdot s(t - \tau(x)) + \lambda_b.
\end{equation}
\cref{fig: lambda} shows a typical $\lambda(x,t)$ where the time delay $\tau(x)$ is translated to a space-time signal with a Gaussian pulse at every $x$. The discretization of $\lambda(x,t)$ will play a key role in our analysis.

\begin{figure}[h!]
\centering
\includegraphics[width=\linewidth]{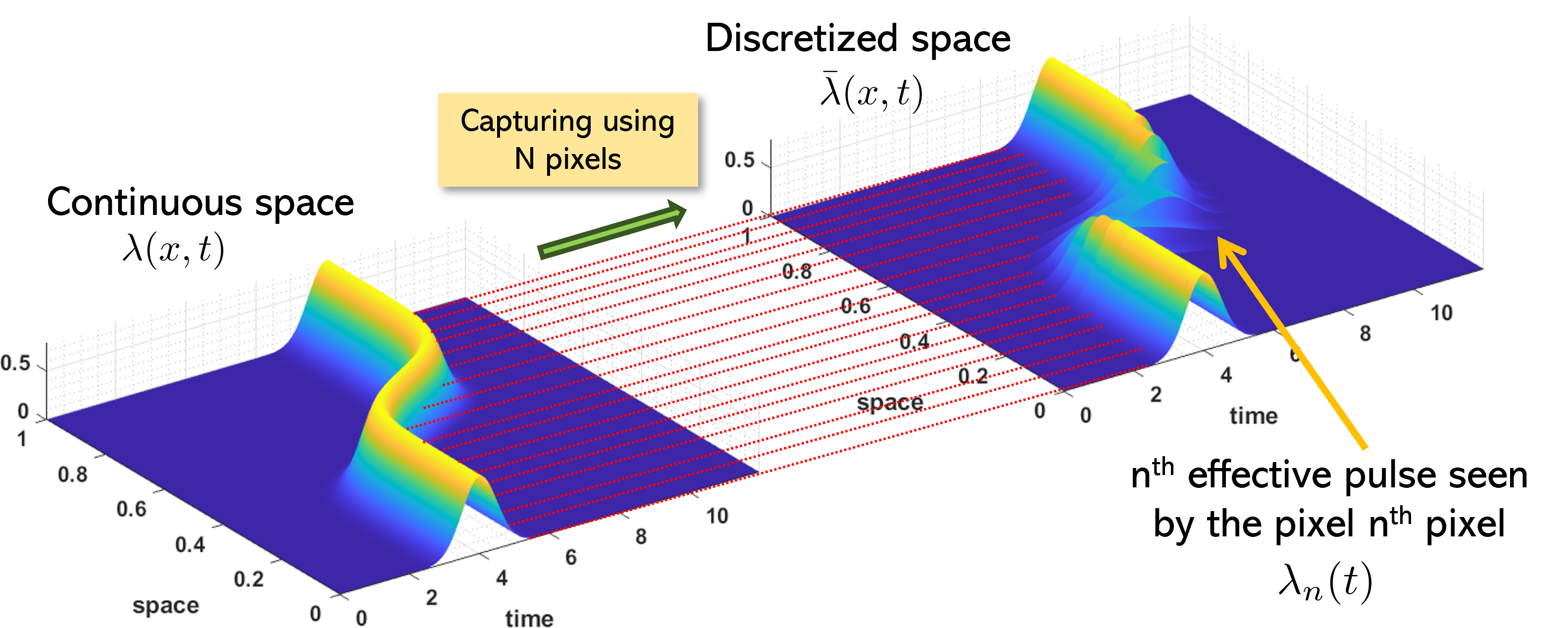}
\caption{The space-time signal $\lambda(x,t)$ in the unit length $0 \le x \le 1$ and time span $[0,T]$, and its corresponding ``effective'' returned pulse $\overline{\lambda}(x,t)$ where each individual returned pulse is $\lambda_n(t)$.}
\label{fig: lambda}
\end{figure}

\noindent \textbf{Observing {\boldmath$\lambda(x,t)$} through $N$ pixels}. Suppose that we allocate $N$ pixels in the unit space to measure the returned time of arrivals. These times of arrivals are generated according to the joint distribution specified in \cref{eq: lambda(t)}. However, since at the $n^\text{th}$ pixel the function $\lambda(x,t)$ occupies the interval $\tfrac{n}{N} \le x \le \tfrac{n+1}{N}$, we can define the \emph{effective} return pulse $\lambda_n(t)$ by absorbing the coordinate $x$ through integration. Specifically, we define $\lambda_n(t)$ as
\begin{align}
\lambda_n(t) = \int_{\tfrac{n}{N}}^{\tfrac{n+1}{N}} \lambda(x, t) dx. \quad n = 0,\ldots,N-1 \label{eq: lambda_n}
\end{align}

The resulting space-time approximation of $\lambda(x,t)$ is thus a piecewise function
\begin{equation}
\overline{\lambda}(x,t) = N \sum_{n=0}^{N-1} \lambda_n(t) \varphi(Nx-n),
\end{equation}
where $\varphi(x)$ is a boxcar function defined as
\begin{equation}
\varphi(x) =
\begin{cases}
1, &\quad 0 \le x \le 1,\\
0, &\quad \text{otherwise}.
\end{cases}
\label{eq: boxcar}
\end{equation}
The definition here is consistent with how the spatial-oversampled quanta image sensor was defined \cite{Yang_2012_BitsfromPhotons, Elgendy_2018_Threshold}.

\noindent \textbf{Remark 1: Can we approximate {\boldmath$\tau(x)$} instead?} By looking at \cref{eq: lambda_n}, it is tempted to think that we can approximate $\tau(x)$ via a piecewise \emph{constant} function
\begin{equation}
\tau(x) \approx \overline{\tau}(x) = \sum_{n=0}^{N-1} \overline{\tau}_n \varphi(Nx-n),
\label{eq: overline{tau}(x)}
\end{equation}
where
\begin{equation}
\overline{\tau}_n = N \int_{\tfrac{n}{N}}^{\tfrac{n+1}{N}} \tau(x) dx.
\label{eq: overline{tau}_n}
\end{equation}
This will give us a plausible candidate for $\lambda_n(t)$:
\begin{equation}
\lambda_n(t) = \alpha s(t-\overline{\tau}_n) + \lambda_b , \quad n = 0,\ldots,N-1.
\label{eq: lambda tilde wrong}
\end{equation}
However, the problem with this approximation is that physically it is invalid. As light propagates, the energy carried by the wave follows the ``scattering'' process via the superposition of the electromagnetic field \cite{Chimitt_2024_TSP}. When energy is distributed from the source, we need to integrate $\lambda(x,t)$ and not $\tau(x)$.

\subsection{New Approximation Techniques}
\label{subsec: approximation for lambda}
While \cref{eq: lambda_n} is a physically valid way to perform spatial discretization, it does not have a simple analytic expression. For example, when $\alpha = 1$ and $\lambda_b = 0$, if we plug a Gaussian pulse $s(t) = \calN(t \,|\, 0, \sigma_{t}^2)$  into \cref{eq: lambda_n}, we will need to evaluate the integral
\begin{align*}
\lambda_n(t) = \int_{\tfrac{n}{N}}^{\tfrac{n+1}{N}} \frac{1}{\sqrt{2\pi\sigma_t^2}} \exp\left\{-\frac{(t-\tau(x))^2}{2\sigma_t^2}\right\}dx.
\end{align*}
Since $\tau(x)$ is a function of $x$, it is impossible to arrive at a closed-form expression.

Our plan of deriving the theoretical bound involves several steps. At the core of our proof technique is the approximation of the boxcar function using a Gaussian kernel, as illustrated in \cref{fig: convolution in space time}. If we assume that the pulse is also a Gaussian, then a convolution of two Gaussians will remain a Gaussian. This will substantially improve the tractability of our equations.

\begin{figure}[h]
\centering
\includegraphics[width=\linewidth]{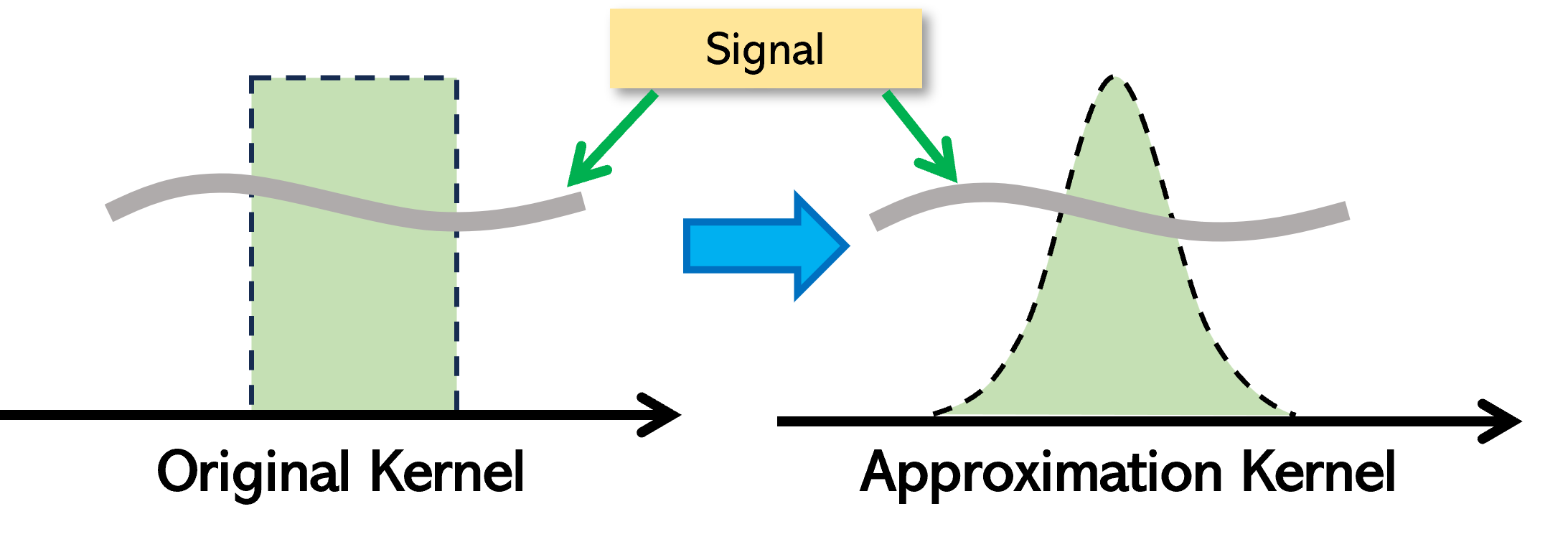}
\caption{Our core proof involves an approximation of the boxcar kernel by a Gaussian. Doing so will allow us to replace the integration with a convolution.}
\vspace{-2ex}
\label{fig: convolution in space time}
\end{figure}

\noindent \textbf{Approximation 1: Linearize \boldmath{$\tau(x)$}}. We approximate the time of arrival function $\tau(x)$ by a piecewise \emph{linear} function. Suppose that there are $N$ pixels in $[0,1]$. We define the mid point $x_n$ of each interval $[\tfrac{n}{N}, \tfrac{n+1}{N}]$ as
\begin{equation*}
x_n \bydef \frac{\tfrac{n}{N} + \tfrac{n+1}{N}}{2}= \frac{2n+1}{2N}.
\end{equation*}
Expanding $\tau(x)$ around $x_n$ will give us
\begin{equation}
\tau(x) \approx \underset{\bydef \tau_n}{\underbrace{\tau(x_n)}} + \underset{\bydef c_n}{\underbrace{\tau'(x_n)}}(x-x_n), \qquad \tfrac{n}{N} \le x \le \tfrac{n+1}{N}.
\label{eq: tau(x) linear approx}
\end{equation}
 Thus, for the entire $0 \le x \le 1$, $\tau(x)$ is approximated by
\begin{equation}
\tau(x) \approx \sum_{n=0}^{N-1} [\tau_n + c_n(x-x_n)] \varphi(Nx-n),
\end{equation}
where $\varphi$ is the boxcar function defined in \cref{eq: boxcar}.

\noindent \textbf{Approximation 2: Replace boxcar by Gaussian}. The second approximation is to give up the boxcar function $\varphi(x)$ because it does not allow us to derive a closed-form expression of $\lambda_n(t)$. We replace it with a Gaussian $\phi(x)$:
\begin{equation}
\phi(x) = \frac{1}{\sqrt{2\pi\sigma_x^2}} \exp\left\{-\frac{x^2}{2\sigma_x^2}\right\} = \calN(x \,|\, 0, \sigma_x^2).
\end{equation}
However, if we want to approximate a boxcar function (with width $W$) by a Gaussian (with a standard deviation $\sigma_x$), what should be the relationship between $W$ and $\sigma_x$ so that the approximation is optimized? The answer is $\sigma_x = W/\sqrt{12}$.

\boxedthm{
\begin{lemma}{}
\label{lemma: KL divergence}
Let $\varphi(x)$ be a boxcar function over the interval $[-\tfrac{W}{2},\tfrac{W}{2}]$ and $\phi(x) = \calN(x \,|\, 0, \sigma_x^2)$ be a Gaussian function. The optimal $\sigma_x$ that offers the best match between $\varphi(x)$ and $\phi(x)$ is
\begin{equation}
\sigma_x = \frac{W}{\sqrt{12}}.
\end{equation}
\end{lemma}
}

If there are $N$ pixels in $[0,1]$, then the width of each pixel is $1/N$. This means that $\varphi(Nx-n)$ has a width of $1/N$. Therefore, the standard deviation of the shifted Gaussian $\phi(Nx-n)$ is $\sigma_x = 1/(\sqrt{12}N)$.

\noindent\textbf{Approximation 3: Replace projection by convolution}. One of the difficulties in \cref{eq: lambda_n} is the integration over the spatial interval. With the introduction of the Gaussian kernel, we replace the projection step by a spatially invariant convolution:
\begin{align*}
\widetilde{\lambda}(x,t)
&= \phi(x) \circledast \lambda(x,t) \qquad \text{[previously it was $\varphi(x)$]}\\
&= \calN(x \,|\, 0, \sigma_x^2) \circledast [\alpha \calN(t \,|\, \tau(x), \sigma_t^2) + \lambda_b]\\
&= \alpha \Big(\calN(x \,|\, 0, \sigma_x^2) \circledast \calN(t \,|\, \tau(x), \sigma_t^2)\Big) + \lambda_b.
\end{align*}
The resulting $\lambda_n(t)$ can then be determined as the value of $\widetilde{\lambda}(x,t)$ at the mid point $x_n$ of each pixel interval, i.e.,
\begin{equation}
\lambda_n(t) = \widetilde{\lambda}(x_n,t).
\end{equation}
The following theorem summarizes the result of this series of approximations:
\boxedthm{
\begin{theorem}{}
	\label{thm: lambda_n}
	Under Approximations 1-3, the effective return pulse received by the $n^\text{th}$ pixel is
	\begin{equation}
		\lambda_n(t)
		= \alpha \cdot \frac{1}{\sqrt{2\pi \sigma_n^2}} \exp\left\{-\frac{(t-\tau_n)^2}{2\sigma_n^2}\right\} + \lambda_b,
		\label{eq: lambda n correct}
	\end{equation}
	where $\tau_n = \tau(x_n)$, and $\sigma_n^2 = c_n^2\sigma_x^2+\sigma_t^2$.
\end{theorem}
}
The biggest difference between \cref{eq: lambda n correct} and \cref{eq: lambda tilde wrong} is the standard deviation of the Gaussian. In \cref{eq: lambda tilde wrong}, the pulse width is always $\sigma_t$. Thus, the shape of the Gaussian is never changed no matter which pixel we consider. This problem is fixed in \cref{eq: lambda n correct} where the standard deviation now depends on three things: (i) the temporal pulse width $\sigma_t$, (ii) the width of the pixel $\sigma_x$, (iii) the first derivative $c_n$ of the time of arrival function $\tau(x)$.

\subsection{Derivation of the MSE}
\label{subsec: derivation of mse}

\noindent \textbf{Bias-Variance Decomposition}. We are now in the position to derive the overall MSE. The MSE is measured between the true function $\tau(x)$ and the reconstructed function $\widehat{\tau}(x)$:
\begin{equation}
\text{MSE}(\widehat{\tau}, \tau) \bydef \E\left[ \int_0^1 ( \widehat{\tau}(x) - \tau(x) )^2 \, dx\right].
\label{eq: MSE def}
\end{equation}
In this equation, the reconstructed function $\widehat{\tau}(x)$ is a piecewise constant function defined by
\begin{align*}
	\widehat{\tau}(x) \bydef \sum_{n=0}^{N-1} \widehat{\tau}_n \varphi(Nx - n),
\end{align*}
where $\widehat{\tau}_n$ is the ML estimate of the time of arrival at the $n^{\text{th}}$ pixel, and $\varphi(x)$ is the boxcar function.

As will be shown in the supplementary material, the MSE defined \cref{eq: MSE def} can be decomposed into bias and variance:
\begin{align*}
\text{MSE}(\widehat{\tau}, \tau) &=
\underset{\text{bias}}{\underbrace{\|\tau-\overline{\tau}\|_{L_2}^2}} +
\underset{\text{variance}}{\underbrace{\E\left[ \|\widehat{\tau}-\overline{\tau}\|_{L_2}^2 \right]}}.
\end{align*}
The bias measures how much resolution will drop when we use piecewise constant function $\overline{\tau}$ to approximate the continuous $\tau$. The variance measures the noise fluctuation caused by the random ML estimate $\widehat{\tau}$.

\noindent \textbf{Main Theoretical Result}. The main result is stated in the theorem below.

\boxedthm{
\begin{theorem}[Overall MSE]
\label{thm: overall MSE}
The MSE is
	\begin{equation}
		\text{MSE}(\widehat{\tau},\tau) =
		\underset{\text{bias}}{\underbrace{\frac{c^2}{12N^2}}} +
		\underset{\text{variance}}{\underbrace{\frac{N}{\alpha_0}\left(c^2\sigma_x^2 + \sigma_t^2\right)}}.
	\end{equation}
where $c^2 = (1/N)\sum_{j=1}^N c_n^2$, and $\alpha_0$ is the total flux of the scene.
\end{theorem}
}
When deriving this main result, we assume that the pulse is Gaussian and the floor noise $\lambda_b$ is zero. We will relax these assumptions in the supplementary material to consider more realistic situations.

\noindent \textbf{Significance of \cref{thm: overall MSE}}. The main result is the first \emph{closed-form expression} about the noise-resolution trade-off that we are aware of. As we will demonstrate in the experiment section, this simple formula matches well with the Monte Carlo simulation, albeit with minor numerical precision errors.

The closed-form expression in \cref{thm: overall MSE} offers many important insights about the behavior of the problem.
\begin{itemize}
\item $\alpha_0$: Since $\alpha_0$ is the total flux of the scene, a large $\alpha_0$ will generate more time stamps which will in turn improve the variance. $\alpha_0$ has no impact on the bias.
\item $\sigma_t$: The pulse width determines the uncertainty of the time of arrivals, which affects the variance. $\sigma_t$ does not affect the bias because the bias is independent of $t$.
\item $c$: The slope of $\tau(x)$ specifies ``how difficult'' the scene is. In the easiest case where the scene is flat so that $c_n = 0$, the bias term drops to zero. If the slope is large, both bias and variance will suffer.
\item $\sigma_x$: The parameter $\sigma_x$ is a modeling constant. $\sigma_x$ can be considered as a proxy to any diffraction limit caused by the optical system. A large point spread function of the optics will result in a large $\sigma_x$.
\end{itemize}
}

\excludefromtoc{
\section{Experiments}

\subsection{Simulated 1D Experiment}
We consider multiple 1D ground truth time of arrival functions $\tau(x)$ outlined in the supplementary material Sec. {\color{red}9}.\ignore {\cref{sec: Appendix Experiment Config}.}
The configurations can be found in Tab {\color{red}1}\ignore {\cref{table: parameters}}, also in the supplementary material.

\noindent \textbf{Simulation}. During simulation, we construct a space-time function $\lambda(x,t)$ with a very fine-grained spatial grid. At each $x$ in the grid, there is a pulse function $s(t-\tau(x))$. We integrate $\lambda(x,t)$ for $\tfrac{n}{N} \le x \le \tfrac{n+1}{N}$ for each interval $n$ to obtain the effective pulse $\lambda_n(t)$. $M$ random time stamps are drawn from the inverse CDF of $\lambda_n(t)$, where $M$ is a Poisson random variable with a rate $\alpha_0/N$. The $M$ time stamps (per each $n$) will give us an estimate $\widehat{\tau}_n$, which is then used to construct the reconstructed delay profile $\widehat{\tau}(x) = \sum_{n=0}^{N-1} \widehat{\tau}_n \varphi(Nx-n)$. We numerically compute the MSE for this $\widehat{\tau}(x)$.

\noindent \textbf{Theory}. The theoretical prediction follows the equation $\text{MSE}(\widehat{\tau},\tau)$ described in Theorem~\ref{thm: overall MSE}. This is a one-line formula.

\noindent \textbf{Result}. The result of our experiment is reported in \cref{fig: MSE}. As evident from the figure, the theoretical prediction matches very well with the simulation. The optimal number of pixels for this particular problem is around $N = 64$.

\begin{figure}[t!]
\centering
\includegraphics[width=0.98\linewidth]{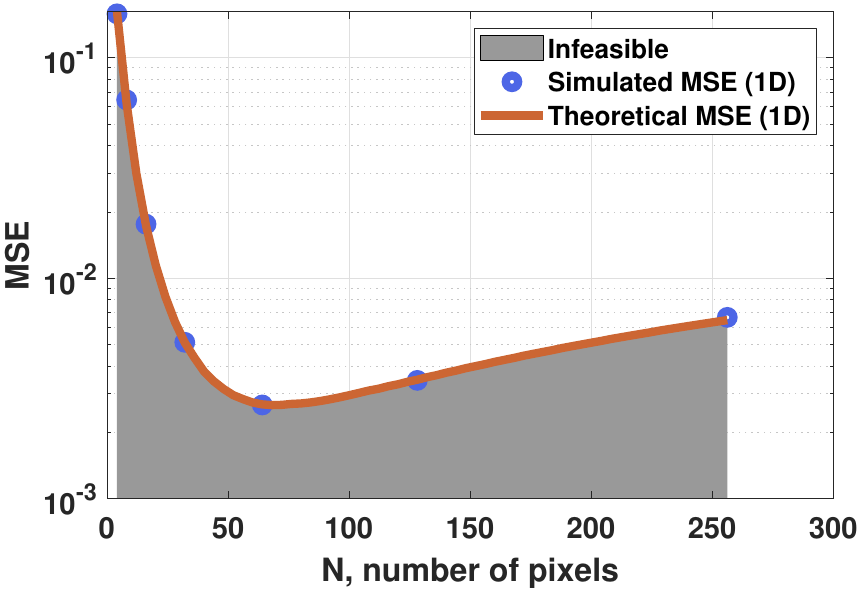}
\caption{1D simulation. Comparing simulation and the theoretically predicted MSE. Note the excellent match between the theory and the simulation.}
\label{fig: MSE}
\end{figure}

\begin{figure}[b!]
\centering
\includegraphics[width=0.98\linewidth]{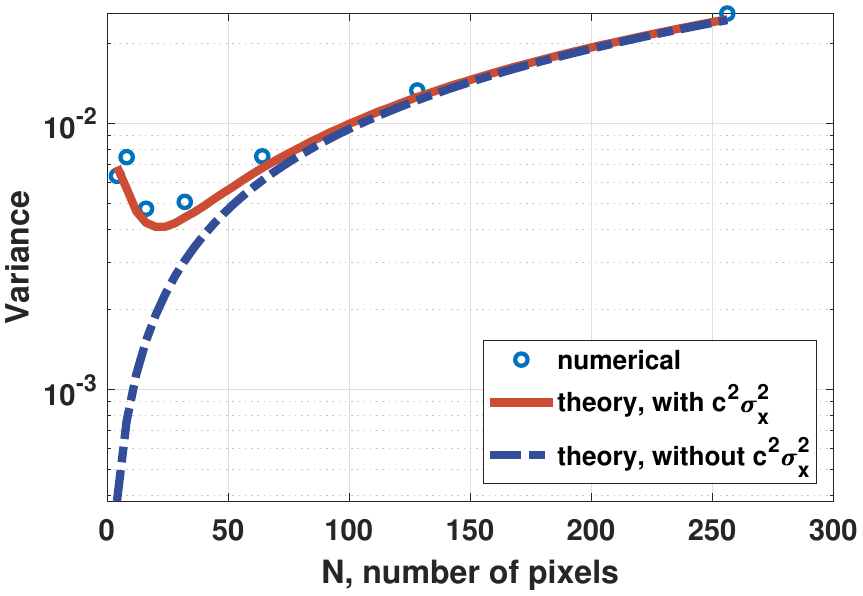}
\caption{We compare two theoretical bounds: One with $c^2\sigma_x^2$ included (which is our full model), and one with $c^2\sigma_x^2$ missing (which is the simplified model). Note the excellent match between the theoretical prediction and the simulation result.}
\label{fig: ablation variance}
\end{figure}

\begin{figure}[!t]
\centering
\includegraphics[width=0.98\linewidth]{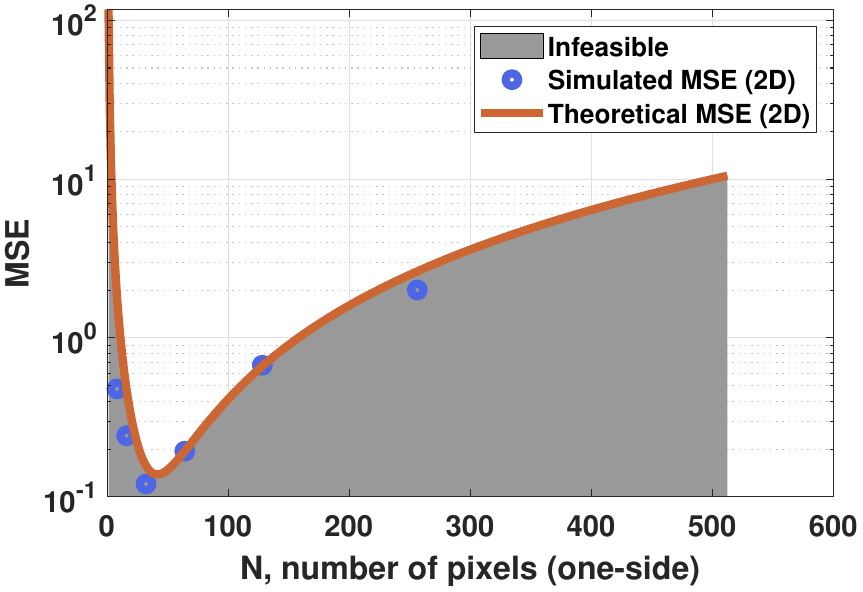}
\caption{2D simulation. Comparing simulation and the theoretically predicted MSE. Note the excellent match between the theory and the simulation.}
\label{fig: 2D MSE}
\end{figure}

\subsection{Analysis of Variance}
Our second experiment concerns about the validity of the approximations. Suppose we take a ``lazy'' route by using the ``cheap'' approximation outlined in \cref{eq: lambda tilde wrong}. Then, under the condition that $s(t)$ is Gaussian and $\lambda_b = 0$, \cref{thm: MSE single pixel} will give us (via \cref{example: Eg 3})
\begin{equation}
\text{MSE}(\tau,\widehat{\tau}) = \underset{\text{bias}}{\underbrace{\frac{c^2}{12N^2}}} + \underset{\text{variance}}{\underbrace{\frac{N}{\alpha_0}\sigma_t^2}} .
\end{equation}
Compared with Theorem~\ref{thm: overall MSE}, the term $c^2\sigma_x^2$ is omitted. For the particular example shown in \cref{fig: MSE}, we show in \cref{fig: ablation variance} a side-by-side comparison when the term $c^2\sigma_x^2$ is included and not included. It is clear from the figure that only the one with $c^2\sigma_x^2$ included can match with the simulation.

\subsection{Simulated 2D Experiment}
\label{sec: simulated 2D}
\noindent\textbf{Simulation}. For 2D experiments, we use a ground truth depth map to generate the true time of arrival signal $\tau(\vx)$. Then, following a similar procedure outlined for the 1D case, we generate time stamps according to the required spatial resolution. For simplicity, we assume that the pulses are Gaussian, and that there is no noise floor. A piecewise constant 2D signal is reconstructed and the MSE is calculated.

\noindent \textbf{Theory}. The derivation of the theoretical MSE is outlined in the supplementary material Sec {\color{red}14}.\ignore{\cref{sec: theory 2D}} Summarizing it here, the MSE is (with $N$ being the number of pixels in one direction)

\begin{equation}
\text{MSE}(\tau,\widehat{\tau}) = \frac{\|\vc\|^2}{12N^2} + \frac{N^2}{\alpha_0}\left(\|\vc\|^2 \sigma_x^2 + \sigma_t^2 \right),
\label{eq: 2D mse}
\end{equation}
where $\|\vc\|^2 = \int_{[0,1]^2} \|\nabla \tau(\vx)\|^2 d\vx$ is the average gradient of the 2D time of arrival function.

\begin{figure*}[h]
\centering
\begin{tabular}{ccccc}
\hspace{-2ex}\includegraphics[width=0.125\linewidth]{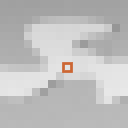}&
\hspace{-2ex}\includegraphics[width=0.125\linewidth]{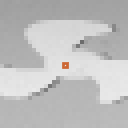}&
\hspace{-2ex}\includegraphics[width=0.125\linewidth]{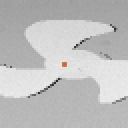}&
\hspace{-2ex}\includegraphics[width=0.125\linewidth]{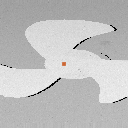}&
\multirow{2}{*}[2cm]{\includegraphics[width=0.4\linewidth]{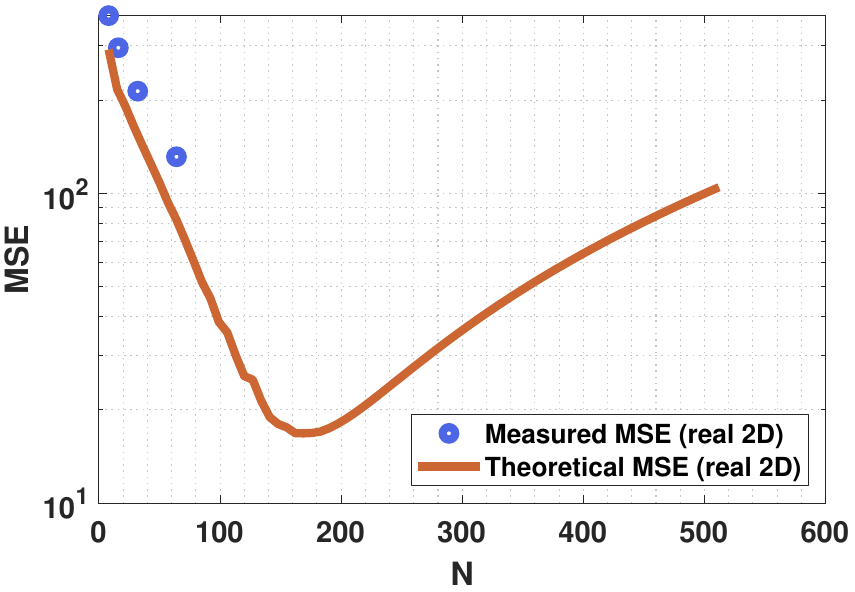}}\\
\hspace{-2ex}\includegraphics[width=0.125\linewidth]{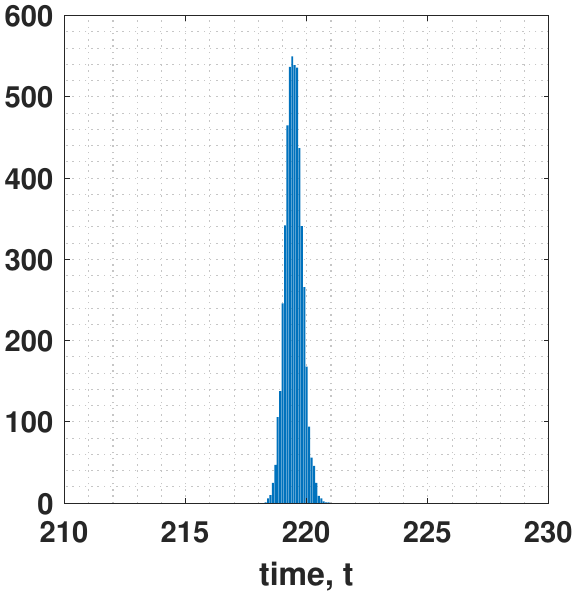}&
\hspace{-2ex}\includegraphics[width=0.125\linewidth]{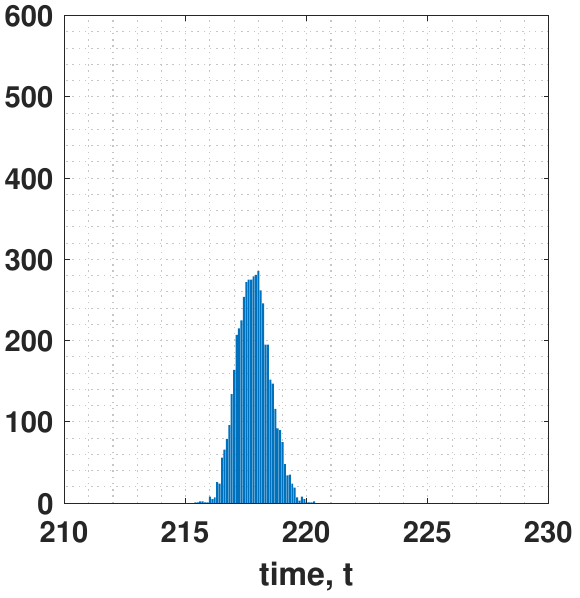}&
\hspace{-2ex}\includegraphics[width=0.125\linewidth]{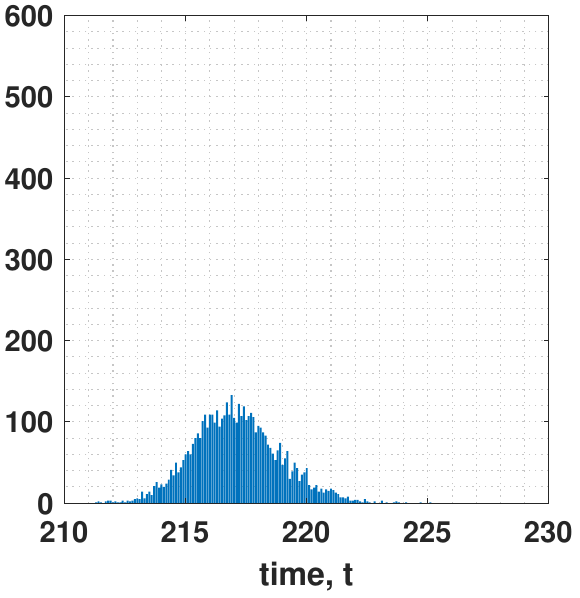}&
\hspace{-2ex}\includegraphics[width=0.125\linewidth]{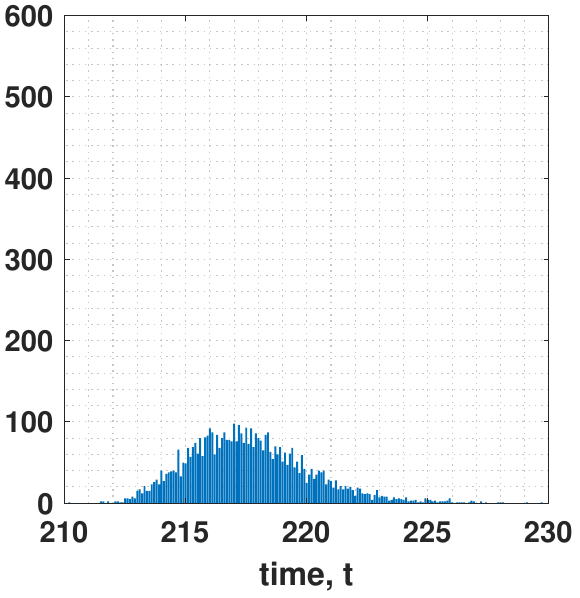}&\\
\end{tabular}
\caption{Real 2D experiment using a $192 \times 128$ SPAD reported in \cite{Henderson_2019_192x128}. [Left-Top] ML estimate of the time of arrivals at different resolutions. As we reduce the spatial resolution of the SPAD, the noise per pixel reduces whereas the resolution becomes worse. [Left-Bottom] The distribution of ML estimate at the orange location. As we use a larger pixel, the variance of the estimated time of arrival reduces. [Right] The MSE curve compares the estimate and the pseudo ground truth, and the corresponding theoretical predictions.}
\label{fig: 2D}
\end{figure*}

\begin{figure}[t]
\centering
\includegraphics[width=0.91\linewidth]{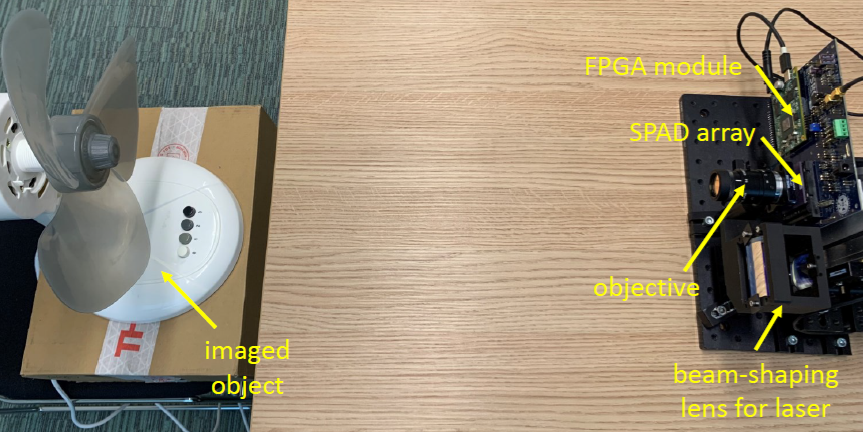}
\caption{Experimental setup to capture the real SPAD data. The sensor we used here is $192\times 128$ SPAD developed by Henderson et al. \cite{Henderson_2019_192x128}.}
\label{fig: optical setup}
\end{figure}

\noindent \textbf{Result}. The result is outlined in \cref{fig: 2D MSE}. As we can see, the theoretical MSE again provides an excellent match with the simulated MSE.

\subsection{Real 2D Experiment}
In this experiment we analyze the real SPAD data collected by a sensor reported in \cite{Henderson_2019_192x128}. The indoor scene consists of a static fan with a flat background, which is flood-illuminated using a picosecond pulsed laser source (Picoquant LDH series 670 nm laser diode with ~1nJ pulse energy, operated with 25 MHz repetition rate). An f/4 objective was used in front of the SPAD, and binary time stamp frames (with a maximum of a single time stamp per pixel) were captured with an exposure time of 1 ms per frame. A total of 10,000 time stamps with a timing resolution of 35ps were thereby collected for each pixel. Pre-processing is performed to remove outliers. More descriptions of how this is done can be found in the supplementary material. The outcomes of the real 2D experiment are depicted in \cref{fig: 2D}, whereas the schematic diagram of the experimental setup is shown in \cref{fig: optical setup}.

The top row of \cref{fig: 2D} shows the estimated depth map at four different resolutions. The estimation is done using the ML estimation. The bottom row of \cref{fig: 2D} shows the distribution of the ML estimates. This distribution is obtained through a bootstrap procedure where we sample with replacement $M = 3$ time stamps to estimate the time, and we bootstrap for $5,000$ times. The shrinking variance confirms that as we use fewer pixels, the estimation quality improves. The right hand side of \cref{fig: 2D} shows the theoretically predicted MSE and the measured MSE. The measured MSE is obtained by first constructing a pseudo ground truth from all the 10,000 frames (with pre-processing). We draw $M = 3$ samples from each pixel, sampled with replacement repeatedly 100 times, to compute the MSE.

The result in \cref{fig: 2D} does not show a valley. This is because the optimal $N$, by taking derivative of \cref{eq: 2D mse}, is $N = \left(\frac{\sqrt{\alpha_0}\|\vc\|}{\sqrt{12} \sigma_t}\right)^{1/2}$. Therefore, if the pulse width is short so $\sigma_t$ is small, it is possible that the optimal $N$ is larger than the physical resolution of the SPAD. In this case, maximizing the resolution is the best option.

}

\excludefromtoc{

\section{Conclusion}
A closed-form expression of the resolution limit for a SPAD sensor array is presented. It is found that the MSE decreases when the total amount of flux is high, the scene is smooth, and the pulse width is small. The MSE demonstrates a U-shape as a function of the number of pixels $N$ in a unit space. When the optimal $N$ is beyond the physical resolution of the sensor, no binning would be required. Extension of the theory to pile-up effects and non-Gaussian pulse is achievable with numerical integration. Advanced post-processing can possibly outperform the predicted bound which is based on ML estimation.

\subsection*{Acknowledgement}
We thank the anonymous reviewers and the area chair for the meticulous review of this paper and, for offering insightful and constructive feedback. We are particularly thankful to Reviewer \texttt{SxN3} for trusting us on this revision. We thank Abdullah Al Shabili for the initial investigation of the LiDAR statistics. We thank Jonathan Leach's lab at Heriot-Watt University, and German Mora-Martin of the University of Edinburgh for collecting and sharing the data with us.

The work is supported, in part, by the DARPA / SRC CogniSense JUMP 2.0 Center, NSF IIS-2133032, and NSF ECCS-2030570. 

All plots presented in this paper can be reproduced using the code provided at \url{https://github.itap.purdue.edu/StanleyChanGroup/}.

}

\newpage

{
    \small
    \bibliographystyle{ieeenat_fullname}
    \bibliography{main}
}


\clearpage

\maketitlesupplementary

\setcounter{lemma}{1}
\setcounter{section}{5}
\setcounter{figure}{10}
\setcounter{corollary}{2}
\setcounter{example}{3}
\setcounter{theorem}{4}
\setcounter{equation}{30}

\tableofcontents

\section{Proof of Theorems}
In this section, we present the proofs of the theorems. To clarify the contributions of this paper, we add comments to each proof to highlight whether this is based on existing theory or it is a new proof.

\subsection{Axioms}
We first state the three axioms for inhomogeneous Poisson processes \cite{Snyder_1991_book}.

\boxedthm{
\begin{definition}{Axioms for inhomogeneous Poisson process}
\begin{itemize}
\item The probability of one occurrence in an infinitesimal interval $\Delta t$ is given by
\begin{equation}
\Pb[1, \Delta t] = \lambda(t) \Delta t, \qquad \Delta t \rightarrow 0.
\end{equation}
\item The probability of more than one occurrence in $\Delta t$ is zero for $\Delta t \rightarrow 0$. Therefore, $\Pb[0, \Delta t]$ is the complement of $\Pb[1, \Delta t]$, which will give us
\begin{equation}
\Pb[0, \Delta t] = 1-\lambda(t) \Delta t, \qquad \Delta t \rightarrow 0.
\end{equation}
\item The number of occurrences in any interval is independent of those in all other disjoint intervals.
\end{itemize}
\end{definition}
}

\subsection{Proof of Theorem {\color{red} 1}}

$\maltese$ \emph{Remark: This proof is adopted from Bar-David \cite{Bar-David_1969} with new elaborations provided for each step.}

Consider a set of time stamps $-T \le t_1 < t_2 < \ldots < t_M \le T$. For each timestamp $t_j$, we consider an infinitesimal width $\pm \Delta t_j/2$. The probability that one and only one photon falls in each of the intervals and none outside is
\begin{align*}
&p(\vt_M, M) \Delta t_1 \Delta t_2 \ldots \Delta t_M\\
&\qquad = \Pb[\text{observing one and only one in each interval}] \\
&\qquad = \Pb[0, (-T,t_1-\tfrac{1}{2}\Delta t_1)]   \\
&\qquad\qquad\qquad  \times \Pb[1, (t_1-\tfrac{1}{2}\Delta t_1, t_1+\tfrac{1}{2}\Delta t_1 ) ] \\
&\qquad\qquad\qquad  \times \Pb[0, (t_1+\tfrac{1}{2}\Delta t_1, t_2-\tfrac{1}{2}\Delta t_2 ) ] \\
&\qquad\qquad\qquad\qquad\qquad  \ldots \\
&\qquad\qquad\qquad  \times \Pb[1, (t_M-\tfrac{1}{2}\Delta t_M, t_M+\tfrac{1}{2}\Delta t_M ) ] \\
&\qquad\qquad\qquad  \times \Pb[0, (t_M+\tfrac{1}{2}\Delta t_M, T ) ]\\
&\qquad = \Pb[0, (-T,t_1-\tfrac{1}{2}\Delta t_1)]   \\
&\qquad\qquad\qquad  \times \lambda(t_1) \Delta t_1  \times \Pb[0, (t_1+\tfrac{1}{2}\Delta t_1, t_2-\tfrac{1}{2}\Delta t_2 ) ] \\
&\qquad\qquad\qquad\qquad\qquad  \ldots \\
&\qquad\qquad\qquad  \times \lambda(t_M) \Delta t_M  \times \Pb[0, (t_M+\tfrac{1}{2}\Delta t_M, T ) ]
\end{align*}
As $\Delta t_j \rightarrow 0$, the intervals merge to become $(t_1+\tfrac{1}{2}\Delta t_1, t_2-\tfrac{1}{2}\Delta t_2 ) \rightarrow (t_1,t_2)$. Therefore,
\begin{align*}
&\Pb[0, (-T,t_1-\tfrac{1}{2}\Delta t_1)] \times \ldots \times \Pb[0, (t_M+\tfrac{1}{2}\Delta t_M, T ) ]\\
&= \Pb[0, (-T,t_1)] \times \Pb[0, (t_1,t_2)] \ldots \times \Pb[0, (t_M,T) ]\\
&= \Pb[0, (-T,T)] \\
&= \exp\left[-\int_{-T}^T \lambda(t) \; dt \right] = e^{-Q}.
\end{align*}
Regrouping the terms, we can show that
\begin{equation*}
p(\vt_M, M) = p(\vt_M) = e^{-Q}\prod_{j=1}^M \lambda(t_j).
\end{equation*}

\subsection{Proof of Theorem {\color{red} 2}}

$\maltese$  \emph{Remark: This proof is adopted from Bar-David \cite{Bar-David_1969}.}

Let's assume that $F(\tau)$ is a twice differentiable function. Then, let $F_1 = \dot{F}(\tau_0)$, and decompose $F_1$ as $F_1 = \mu_1 + F_1'$, where $F_1'$ is a zero-mean random variable. Similarly, decompose $F_2 = \ddot{F}(\tau_0)$ as $F_2 = \mu_2 + F_2'$. Then,
\begin{align*}
\epsilon = - \frac{\dot{F}(\tau_0)}{\ddot{F}(\tau_0)} = -\frac{F_1}{F_2} = -\frac{\mu_1 + F_1'}{\mu_2 + F_2'}.
\end{align*}

Using Lemma \ref{lemma: mu1, var1}, we know that $\mu_1 = 0$. Moreover, let's assume that $F_2'\ll\mu_2$. We can, hence, simplify the above as
\begin{align*}
\epsilon^2 = \left(\frac{F_1'}{\mu_2 + F_2'}\right)^2
&= \left(\frac{F_1'}{\mu_2} \left[\frac{1}{1+\frac{F_2'}{\mu_2}}\right] \right)^2\\
&\approx \left(\frac{F_1'}{\mu_2} \left[1-\frac{F_2'}{\mu_2}\right] \right)^2.
\end{align*}
Taking the expectation will give
\begin{align*}
\E[\epsilon^2]
&\approx \frac{\E[F_1^{'2}]}{\mu_2^2}
-\underset{\rightarrow 0}{\underbrace{2\frac{\E[(F_1^{'2}F_2^{'}]}{\mu_2^3}}}
+ \underset{\rightarrow 0}{\underbrace{\frac{\E[F_1^{'2} F_2^{'2}]}{\mu_2^4}}}.
\end{align*}
Using Lemma~\ref{lemma: mu1, var1} and Lemma~\ref{lemma: mu2, var2}, and assuming that $\tau_0 = 0$ without loss of generality, it follows that
\begin{align*}
\E[\epsilon^2] = \frac{\E[F_1^{'2}]}{\mu_2^2}
&= \frac{ \int_{-T}^T \frac{(\alpha \dot{s}(t))^2}{\alpha s(t) + \lambda_b} \; dt}{\left( \int_{-T}^T \frac{(\alpha \dot{s}(t))^2}{\alpha s(t) + \lambda_b} \; dt\right)^2}\\
&= \left[ \int_{-T}^T \frac{(\alpha \dot{s}(t))^2}{\alpha s(t) + \lambda_b} \; dt\right]^{-1}.
\end{align*}

\subsection{Proof of Theorem {\color{red} 3}}

$\maltese$  \emph{Remark: This is a new proof we make for this paper.}

We consider $s(x,t) = s(t-\tau(x))$. Using Approximation 1, we write
\begin{align*}
s(x,t)
&= s(t-\tau(x)) = \calN(t \,|\, \tau(x), \sigma_t^2 )\\
&= \calN(t \,|\, \tau_n + c_n(x-x_n), \sigma_t^2 ).
\end{align*}
By Approximation 3, we can show that
\begin{align*}
\widetilde{s}(x,t)
&\bydef \phi(x) \circledast s(x,t) \\
&= \calN(x \,|\, 0, \sigma_x^2) \circledast \calN(t \,|\, \tau_n + c_n(x-x_n), \sigma_t^2)\\
&\overset{(a)}{=} \calN(x \,|\, 0, \sigma_x^2) \circledast \frac{1}{c_n} \calN\left(x \,\Big|\, \frac{t}{c_n}-\frac{\tau_n}{c_n}+x_n, \frac{\sigma_t^2}{c_n^2}\right)\\
&\overset{(b)}{=} \frac{1}{c_n} \calN\Big(x \,\Big|\, \frac{t}{c_n}-\frac{\tau_n}{c_n}+x_n, \sigma_n^2\Big)\\
&= \calN(t \,|\, \tau_n + c_n(x-x_n), \sigma_n^2),
\end{align*}
where $(a)$ is based on a simple switch between the roles of $x$ and $t$ in a Gaussian distribution, and $(b)$ is based on the fact that the convolution of two Gaussian probability density functions is equivalent to adding two Gaussian random variables.

If we restrict ourselves to $x = x_n$, then we will obtain $s_n(t) \bydef \widetilde{s}(x_n,t) = \calN(t \,|\, \tau_n, \sigma_n^2)$. Therefore, using the fact that $\phi(x) \circledast \lambda_b = \int_{-\infty}^{\infty} \phi(x) \lambda_b dx = \lambda_b$ since $\phi(x)$ integrates to 1, we can show that
\begin{align*}
\lambda_n(t)
&= \alpha \left\{\phi(x) \circledast s(x,t)\right\}|_{x = x_n} + \lambda_b\\
&= \alpha \calN(t \,|\, \tau_n, \sigma_n^2) + \lambda_b.
\end{align*}

\subsection{Proof of Theorem {\color{red} 4}}
In proving Theorem {\color{red} 4},
we need a few lemmas. The first lemma is the classical bias-variance trade-off.

\boxedthm{
	\begin{lemma}[MSE decomposition]
		\label{lemma: bias var decomp}
		The MSE can be decomposed as
		\begin{equation}
			\text{MSE}(\widehat{\tau}, \tau)  \bydef \text{bias} + \text{var},
		\end{equation}
		where
		\begin{align*}
			\text{bias}  &= \int_0^1 \left( \overline{\tau}(x) - \tau(x) \right)^2 \, dx\\
			\text{var} &= \E\left[ \int_0^1 \left( \widehat{\tau}(x) - \overline{\tau}(x) \right)^2 \, dx \right].
		\end{align*}
	\end{lemma}
}

\noindent $\maltese$  \emph{Remark: The decomposition of the MSE into bias and variance can be found in most of the machine learning textbooks. For completeness and in the context of our paper, we provide the derivations here.}

\noindent\textbf{Proof of \cref{lemma: bias var decomp}}. We start with the definition of the MSE:
\begin{align*}
	\text{MSE}(\widehat{\tau}, \tau)
	&\bydef \E\left[ \int_0^1 ( \widehat{\tau}(x) - \tau(x) )^2 \, dx\right]\\
	&= \E\left[ \int_0^1 \left( \widehat{\tau}(x) - \overline{\tau}(x) + \overline{\tau}(x) - \tau(x) \right)^2 \, dx\right]
\end{align*}
Expanding the terms, we can show that
\begin{align*}
	\text{MSE}(\widehat{\tau}, \tau)
	&= \underset{\text{variance}}{\underbrace{\E\left[ \int_0^1 \left( \widehat{\tau}(x) - \overline{\tau}(x) \right)^2 \, dx \right]}}\\
	&\quad + \underset{=0}{\underbrace{\E\left[ \int_0^1 \left( \widehat{\tau}(x) - \overline{\tau}(x) \right) \left( \overline{\tau}(x) - \tau(x) \right) \, dx \right]}}\\
	&\quad + \underset{\text{bias}}{\underbrace{\int_0^1 \left( \overline{\tau}(x) - \tau(x) \right)^2 \, dx}}.
\end{align*}

The second expectation in the above is zero because $\E[\widehat{\tau}(x)] = \overline{\tau}(x)$ for every $x$ according to Lemma~\ref{lemma: mu1, var1}.

\boxedthm{
	\begin{lemma}[Bias Term]
		\label{lemma: bias}
		Let $\overline{\tau}(x) = \sum_{n=0}^{N-1} \overline{\tau}_n \varphi(Nx-n)$, and let $N$ be the number of pixels in a unit space $0\le x \le 1$. The bias is approximately
		\begin{equation}
			\text{bias} = \frac{c^2}{12N^2},
		\end{equation}
		where $c_n = \tau'(x_n)$ is the slope of $\tau$ at $x_n$, and $c^2 = \frac{1}{N}\sum_{n=0}^{N-1}c_n^2$.
	\end{lemma}
}

\noindent$\maltese$  \emph{Remark: This is a new proof we make for this paper.}

\noindent\textbf{Proof of Lemma~\ref{lemma: bias}}

Integrating over the unit space $[0,1]$, we can show that
\begin{align*}
	&\int_0^1 \left( \overline{\tau}(x) - \tau(x) \right)^2  \, dx\\
	&= \int_0^1 \left( \sum_{n=0}^{N-1} \tau_n \varphi(Nx-n) - \tau(x) \right)^2  \, dx \\
	&= \sum_{n=0}^{N-1} \underset{=e_n^2}{\underbrace{\int_{n/N}^{(n+1)/N} \left(\tau_n - \tau(x) \right)^2 \, dx}}.
\end{align*}
Let's calculate the individual error $e_n^2$. Let $x_n$ be the mid-point of $n/N$ and $(n+1)/N$. That is,
\begin{align*}
	x_n = \frac{1}{2}\left(\frac{n}{N} + \frac{n+1}{N}\right) = \frac{2n+1}{2N}.
\end{align*}
Then, we can take the first order approximation of $\tau(x)$ around $x_n$:
\begin{align*}
	\tau(x) = \tau(x_n) + \tau'(x_n) (x-x_n).
\end{align*}

\noindent Since $\tau(x_n) = \tau_n$ by construction, it follows that
\begin{align*}
	e_n^2
	&= \int_{n/N}^{(n+1)/N} \left(\tau_n - \tau(x) \right)^2 \, dx\\
	&= \int_{n/N}^{(n+1)/N} \Big(\tau_n - (\tau(x_n) + \tau'(x_n) (x-x_n)) \Big)^2 \, dx\\
	&= \int_{n/N}^{(n+1)/N} \Big(\tau_n - (\tau_n + \tau'(x_n) (x-x_n)) \Big)^2 \, dx\\
	&= \int_{n/N}^{(n+1)/N} \Big(\tau'(x_n) (x-x_n) \Big)^2 \, dx\\
	&= [\tau'(x_n)]^2 \frac{(\frac{n+1}{N}-x_n)^3 - (\frac{n}{N}-x_n)^3}{3}, \qquad x_n = \frac{2n+1}{2N}\\
	&= \frac{[\tau'(x_n)]^2}{12N^3}.
\end{align*}

Summing over all $n$'s will give us
\begin{align*}
	\int_0^1 \left( \overline{\tau}(x) - \tau(x) \right)^2  \, dx  = \sum_{n=0}^{N-1} e_n^2 = \frac{c^2}{12N^2},
\end{align*}
where $c^2 = \frac{1}{N}\sum_{n=0}^{N-1}c_n^2$.

\boxedthm{
	\begin{lemma}[Variance Term]
		\label{lemma: var}
		Assume $s(t)$ is Gaussian and $\lambda_b = 0$. The variance is approximately
		\begin{equation}
			\text{var} = \frac{N}{\alpha_0}\left(c^2\sigma_x^2 + \sigma_t^2\right),
			\label{eq: var final}
		\end{equation}
		where $c_n = \tau'(x_n)$ is the slope of $\tau$ at $x_n$, and $c^2 = \frac{1}{N}\sum_{n=0}^{N-1}c_n^2$.
	\end{lemma}
}

\noindent$\maltese$  \emph{Remark: This is a new proof we make for this paper.}

\noindent\textbf{Proof of Lemma~\ref{lemma: var}}.

Firstly, we use Theorem {\color{red} 3}
and apply it to the single-pixel case in Theorem {\color{red} 2}.
Then, under the assumption that $\lambda_b = 0$, we can follow Example 3 to show that
\begin{align*}
\E[(\widehat{\tau}_n-\tau_n)^2]
&= \frac{\sigma_n^2}{\alpha} = \frac{c_n^2 \sigma_x^2 + \sigma_t^2}{\alpha}\\
&= \frac{N}{\alpha_0}\left(c_n^2 \sigma_x^2 + \sigma_t^2\right).
\end{align*}
Therefore, for a unit length, it follows that
\begin{align*}
	\text{var}
	&= \E\left[ \int_0^1 \left( \widehat{\tau}(x) - \overline{\tau}(x) \right)^2 \, dx \right]\\
	&= \E\left[ \sum_{n=0}^{N-1} \int_{n/N}^{(n+1)/N} \left( \widehat{\tau}(x) - \overline{\tau}(x) \right)^2 \, dx \right]\\
	&= \sum_{n=0}^{N-1} \int_{n/N}^{(n+1)/N}  \E\left[ \left( \widehat{\tau}(x) - \overline{\tau}(x) \right)^2 \right] \, dx\\
	&= \sum_{n=0}^{N-1} \int_{n/N}^{(n+1)/N}  \E\left[ \left( \widehat{\tau}_n - \overline{\tau}_n \right)^2 \, \right]  \, dx\\
	&= \sum_{n=0}^{N-1} \int_{n/N}^{(n+1)/N}  \frac{N}{\alpha_0}\left(c_n^2 \sigma_x^2 + \sigma_t^2\right) \, dx\\
	&= \frac{1}{N}\sum_{n=0}^{N-1} \frac{N}{\alpha_0}\left(c_n^2 \sigma_x^2 + \sigma_t^2\right) \\
	&= \frac{N}{\alpha_0}\left(c^2\sigma_x^2 + \sigma_t^2\right).
\end{align*}

\section{Proof of Additional Results}

\subsection{Proof of Corollary {\color{red} 1}}

$\maltese$  \emph{Remark: This proof is adopted from \cite{Bar-David_1969}.}

The probability of observing $M$ occurrence is
\begin{align*}
p(M) = \int_{-T}^T dt_1 \int_{t_1}^{T} dt_2 \ldots \int_{t_{M-1}}^T  dt_M p(\vt_M, M).
\end{align*}
The integration limit comes from the fact that the time stamps follow the order $-T \le t_1 < t_2 < \ldots < t_M \le T$. The integral is equivalent to $1/M!$ of the hypercube $(-T,T)$. Thus,
\begin{align*}
p(M)
&= \frac{1}{M!} \int_{-T}^T dt_1 \int_{-T}^{T} dt_2 \ldots \int_{-T}^T  dt_M  \quad e^{-Q} \prod_{j=1}^M \lambda(t_j)\\
&= \frac{e^{-Q}}{M!} \prod_{j=1}^M \int_{-T}^T \lambda(t_j) dt_j = \frac{e^{-Q}Q^M}{M!}.
\end{align*}

\subsection{Proof of Corollary {\color{red} 2}}

$\maltese$  \emph{Remark: Bar-David hinted the result in \cite{Bar-David_1969}, but the proof was missing. We provide the proof here.}

The sum of the integrals can be written as follows.
\begin{align*}
&\sum_{m=0}^{M} \int_{\Omega_M} p(\vt_M, M) \; d \vt_M\\
&= \sum_{m=0}^M \int_{-T}^{T} dt_1 \int_{t_1}^T dt_2 \ldots \int_{t_{M-1}}^T dt_M \; p(\vt_M,M).
\end{align*}
Since $\vt_M$ is an ordered sequence of time stamps, we can rewrite the integration limits by taking into consideration the $M!$ combinations of the orders. This will give us
\begin{align*}
&\sum_{m=0}^{M} \int_{\Omega_M} p(\vt_M, M) \; d \vt_M\\
&= \sum_{m=0}^M \int_{-T}^{T} dt_1 \int_{-T}^T dt_2 \ldots \int_{-T}^T dt_M \; \frac{1}{M!} p(\vt_M,M)\\
&= \sum_{m=0}^M \frac{e^{-Q}}{M!} \prod_{j=1}^M \left[\int_{-T}^T \lambda(t_j) d_{t_j}\right] = \sum_{m=0}^M \frac{e^Q}{M!} Q^M = 1.
\end{align*}

\subsection{Proof of Lemma {\color{red} 1}}

$\maltese$  \emph{Remark: This is a new proof we make for this paper}

\vspace{6pt}
\noindent The KL-divergence of the two distributions is
\begin{align*}
\text{KL}(\varphi \| \phi)
&= \int_{-\infty}^{\infty} \varphi(x) \log \frac{\varphi(x)}{\phi(x)} dx\\
&= \int_{-\frac{W}{2}}^{\frac{W}{2}} -\frac{1}{W} \left[
\log W + \log \left\{\frac{1}{\sqrt{2\pi\sigma^2}}e^{-\frac{x^2}{2\sigma^2}}\right\} \right] dx\\
&= \int_{-\frac{W}{2}}^{\frac{W}{2}} -\frac{\log W}{W} + \frac{1}{W}\log(\sqrt{2\pi\sigma^2}) + \frac{x^2}{2\sigma^2 W} dx\\
&= -\log W + \log(\sqrt{2\pi\sigma^2}) + \frac{W^2}{24\sigma^2}.
\end{align*}
Taking the derivative with respect to $\sigma^2$ will yield
\begin{align*}
\frac{d}{d\sigma^2} \text{KL}(\varphi \| \phi) = \frac{1}{2\sigma^2} - \frac{W^2}{24\sigma^4}.
\end{align*}
Equating this to zero and rearranging terms will give us
\begin{equation}
\sigma^2 = \frac{W^2}{12}.
\end{equation}

\subsection{Cramer Rao Lower Bound}

\boxedthm{
\begin{corollary}[Cramer Rao Lower Bound]
\label{cor: CRLB}
The Cramer-Rao lower bound for the ML estimate $\widehat{\tau}$ given $\calL(\tau)$ is
\begin{align*}
\E[(\widehat{\tau}-\tau_0)^2]
&\ge \left( -\E\left[\frac{\partial^2 \calL}{\partial \tau^2}\right]\right)^{-1} \\
&= \left[ \int_{-T}^T \frac{(\alpha \dot{s}(t))^2}{\alpha s(t) + \lambda_b} \; dt \right]^{-1}.
\end{align*}
\end{corollary}
}

\noindent$\maltese$  \emph{Remark: The Cramer-Rao lower bound for single-photon LiDAR has been previously mentioned in \cite{Shin_2013_ICIP}. We provide the proof here for completion.}

\vspace{60pt}

\noindent\textbf{Proof of \cref{cor: CRLB}}.

Let $\calL(\tau) = \log[\alpha s(t-\tau) + \lambda_b]$. Then
\begin{align*}
\frac{\partial \calL}{\partial \tau}
&= -\frac{\alpha \dot{s}(t-\tau)}{ \alpha s(t-\tau) + \lambda_b}\\
\frac{\partial^2 \calL}{\partial \tau^2}
&= -\frac{ -(\alpha s(t-\tau) + \lambda_b)\ddot{s}(t-\tau) + (\alpha\dot{s}(t-\tau))^2 }{(\alpha s(t-\tau) + \lambda_b)^2}\\
&= \frac{ \ddot{s}(t-\tau) }{\alpha s(t-\tau) + \lambda_b} - \frac{(\alpha\dot{s}(t-\tau))^2 }{(\alpha s(t-\tau) + \lambda_b)^2}
\end{align*}
Taking the expectation over $t$, we have
\begin{align*}
\E\left[-\frac{\partial^2 \calL}{\partial \tau^2}\right]
&= -\int_{-T}^{T} \frac{\partial^2 \calL}{\partial \tau^2} \cdot [\alpha s(t-\tau) + \lambda_b] dt\\
&= \underset{=0}{\underbrace{-\int_{-T}^{T} \ddot{s}(t-\tau)\, dt}} + \int_{-T}^T \frac{(\alpha\dot{s}(t-\tau))^2 }{\alpha s(t-\tau) + \lambda_b} \, dt\\
&= \int_{-T}^T \frac{(\alpha\dot{s}(t-\tau))^2 }{\alpha s(t-\tau) + \lambda_b} \, dt.
\end{align*}
Without loss of generality, we set $\tau = 0$. Taking the reciprocal completes the proof.

\section{Auxiliary Results and Proofs}
In this section, we present additional results and proofs that are essential to the development of our theories.

\subsection{Lemma about Product of Functions}

\boxedthm{
\begin{lemma}
\label{lemma: product of f}
Consider a function $f(t)$ and define its product:
\begin{equation}
f_{\pi}(\vt_M) = \prod_{j=1}^M f(t_j).
\end{equation}
Assume $\vt_M \sim p(\vt_M)$. The expectation of $f_{\pi}$ is
\begin{equation}
\E[f_\pi(\vt_M)] = e^{-Q+G(T)},
\end{equation}
where $G(T) = \int_{-T}^T f(t)\lambda(t) dt$.
\end{lemma}
}

\noindent$\maltese$  \emph{Remark: This proof is adopted from Bar-David.}

\noindent\textbf{Proof of Lemma~\ref{lemma: product of f}}.

The expectation can be shown as follows.
\begin{align*}
\E[f_\pi(\vt_M)]
&= \int_{\Omega} f_\pi(\vt_M) p(\vt_M) d\vt_M\\
&= \sum_{M=0}^{\infty} \underset{\bydef \E_M(f_{\pi})}{\underbrace{\int_{\Omega_M} f_\pi(\vt_M) p(\vt_M) d\vt_M}}.
\end{align*}

Substituting the definition of $f_{\pi}(\vt_M)$ into the expression above, we can show that
\begin{align*}
\E_M(f_{\pi})
&= \int_{\Omega_M} f_\pi(\vt_M) p(\vt_M, M) \; d \vt_M\\
&= e^{-Q}\int_{-T}^{T} dt_1 \int_{t_1}^T dt_2 \ldots \int_{t_{M-1}}^T dt_M \prod_{j=1}^M g(t_j),
\end{align*}
where $g(t)$ is defined as $g(t) = f(t)\lambda(t)$.

Now, consider a slightly different integration
\begin{align*}
&e^{-Q}\int_{-T}^{T} dt_1 \int_{-T}^T dt_2 \ldots \int_{-T}^T dt_M \prod_{j=1}^M g(t_j)\\
&= e^{-Q} \left[\int_{-T}^{T} g(t) dt \right]^M \\
&= e^{-Q}G^M(T),
\end{align*}
where $G(T) = \int_{-T}^T g(t) dt$. The difference between this integral and the previous integral is that the previous integral is based $-T \le t_1 \le \ldots \le t_M \le T$. To account for the shape from this triangle to the full hypercube, we add the $M!$ factors. This will give us
\begin{align*}
\E_M(f_{\pi}) = \frac{e^{Q} G^M(T)}{M!}.
\end{align*}
Therefore,
\begin{align*}
\E[f_{\pi}(\vt_M)] = e^{-Q}\sum_{j=1}^M \frac{G^M(T)}{M!} = e^{-Q+G(T)}.
\end{align*}

\subsection{Lemma about Characteristic Function}

\boxedthm{
\begin{lemma}{}
\label{lemma: f(t,iu)}
For the function $f(t;iu)$ defined as, where $i = \sqrt{-1}$,
\begin{equation}
f(t;iu) = \exp\left\{iu \cdot \frac{\alpha \dot{s}(t-\tau_n)}{\alpha  s(t-\tau_n)+\lambda_b}\right\}.
\end{equation}
The function $G(T) = \int_{-T}^T f(t;iu) \lambda(t) dt$ has derivatives
\begin{align*}
G(T)\Big|_{iu=0}                &= \int_{-T}^T \lambda(t) \, dt = Q\\
\frac{d}{d(iu)}G(T)\Big|_{iu=0} &= \int_{-T}^T \alpha \dot{s}(t-\tau_n) \, dt\\
\frac{d^2}{d(iu)^2}G(T)\Big|_{iu=0} &= \int_{-T}^T  \frac{(\alpha \dot{s}(t-\tau_n))^2}{\alpha s(t-\tau_n) + \lambda_b} \, dt
\end{align*}
\end{lemma}
}

\noindent$\maltese$  \emph{Remark: Bar-David mentioned the idea but we could not find the proof. Therefore, the proof here is new.}

\vspace{60pt}

\noindent\textbf{Proof of Lemma~\ref{lemma: f(t,iu)}}.

For the $f(t;iu)$ defined, we can show that
\begin{align*}
&\frac{d}{d(iu)} f(t;iu)\\
&= \frac{\alpha \dot{s}(t-\tau_n)}{\alpha s(t-\tau_n)+\lambda_b} \cdot \exp\left\{iu \cdot \frac{\alpha \dot{s}(t-\tau_n)}{\alpha s(t-\tau_n)+\lambda_b}\right\}\\
&= \frac{\alpha \dot{s}(t-\tau_n)}{\alpha s(t-\tau_n)+\lambda_b} \cdot f(t;iu)
\end{align*}
If we restrict $iu = 0$, then $f(t;iu)|_{iu=0} = 1$.Therefore,
\begin{align*}
\frac{dG(T)}{d(iu)}\Big|_{iu=0}
&= \int_{-T}^T \frac{d}{d(iu)} f(t;iu) \cdot \lambda(t) \, dt \Big|_{iu=0}\\
&= \int_{-T}^T f(t;iu) \cdot \frac{\alpha \dot{s}(t-\tau_n)}{\lambda(t)} \cdot \lambda(t) \, dt \Big|_{iu=0}\\
&= \int_{-T}^T \alpha  \dot{s}(t-\tau_n) dt = 0,
\end{align*}
where the last equality holds whenever $s(t)$ is a symmetric function.

\begin{align*}
&\frac{d^2G(T)}{d(iu)^2}\Big|_{iu=0}\\
&= \int_{-T}^T \frac{d}{d(iu)} f(t;iu) \cdot \frac{\alpha \dot{s}(t-\tau_n)}{\lambda(t)} \cdot \lambda(t) \, dt\Big|_{iu=0}\\
&= \int_{-T}^T  f(t;iu) \cdot \left(\frac{\alpha \dot{s}(t-\tau_n)}{\lambda(t)}\right)^2 \cdot \lambda(t) \, dt\Big|_{iu=0}\\
&= \int_{-T}^T  \frac{(\alpha \dot{s}(t-\tau_n))^2}{\alpha s(t-\tau_n) + \lambda_b} \, dt.
\end{align*}

\subsection{Lemma for $\mu_1$ and $F_1$}

\boxedthm{
\begin{lemma}{}
\label{lemma: mu1, var1}
Let $F_1 = \dot{F}(\tau_n)$, where the derivative is taken with respect to $\tau$. Write $F_1 = \mu_1 + F_1'$. It holds that
\begin{align*}
\mu_1 = 0, \qquad \E[F_1^{'2}] = \int_{-T}^T \frac{(\alpha \dot{s}(t-\tau_n))^2}{\alpha s(t-\tau_n)+\lambda_b} dt.
\end{align*}
\end{lemma}
}

\noindent$\maltese$  \emph{Remark: Similar to the previous Lemma, Bar-David mentioned the idea but we could not find the proof. Therefore, the proof here is new.}

\noindent\textbf{Proof of Lemma~\ref{lemma: mu1, var1}}.

Let the characteristic function of the random variable $F_1$ be
\begin{equation}
\Phi(iu) = \E\left[ \exp\left\{ix \cdot \sum_{j=1}^M \frac{\alpha \dot{s}(t_j-\tau_n)}{\alpha s(t_j-\tau_n)+\lambda_b}\right\} \right].
\end{equation}
Thus, by \cref{lemma: product of f}, we have $\Phi(iu) = \E[f_{\pi}(\vt_M)] = e^{-Q+G(T)}$ with $f_{\pi}$ defined by Lemma~\ref{lemma: f(t,iu)}.

The first moment of $F_1$ is therefore
\begin{align*}
\mu_1
&= \frac{d}{d(iu)} \Phi(iu) \Big|_{iu=0}
=  \frac{d}{d(iu)} e^{-Q+G(T)} \Big|_{iu=0} \\
&= e^{-Q+G(T)} \frac{d}{d(iu)} G(T)\Big|_{iu=0} = 0,
\end{align*}
where the last equality is due to Lemma~\ref{lemma: f(t,iu)}.

For $\E[F_1^{'2}]$, we can show that
\begin{align*}
& \left. \E[F_1^{'2}]
= \frac{d^2}{d(iu)^2} \Phi(iu) \right \vert_{iu=0}\\
&= \left. e^{-Q} \frac{d}{d(iu)} \left[e^{G(T)} \frac{d}{d(iu)}G(T)\right] \right \vert_{iu=0}\\
&= \left. e^{-Q} \left[e^{G(T)} \left(\frac{d}{d(iu)}G(T)\right)^2 + e^{G(T)}\frac{d^2}{d(iu)^2}G(T)\right]\right \vert_{iu=0}\\
&= e^{-Q} \left[e^{Q} \left(0\right)^2 + e^{Q} \int_{-T}^T \frac{(\alpha \dot{s}(t-\tau_n))^2}{\lambda(t)} \, dt  \right] \\
&= \int_{-T}^T \frac{(\alpha \dot{s}(t-\tau_n))^2}{\lambda(t)} \, dt = \int_{-T}^T \frac{(\alpha \dot{s}(t-\tau_n))^2}{\alpha s(t-\tau_n) + \lambda_b} \, dt.
\end{align*}
Since the width of the pulse is much smaller than the interval $(-T,T)$, we can make $\tau_n = 0$ without loss of generality.

\subsection{Lemma for $\mu_2$ and $F_2$}

\boxedthm{
\begin{lemma}{}
\label{lemma: mu2, var2}
Let $F_2 = \ddot{F}(\tau_n)$. Write $F_2 = \mu_2 + F_2'$. We can show that
\begin{align*}
\mu_2           &= \int_{-T}^T \frac{(\alpha \dot{s}(t-\tau_n))^2}{\alpha s(t-\tau_n)+\lambda_b} dt, \\
\E[F_2^{'2}]    &= \int_{-T}^T \left(\frac{d}{d\tau_n}\left(\frac{\alpha \dot{s}(t-\tau_n)}{\alpha s(t-\tau_n)+\lambda_b}\right)\right)^2 \lambda(t) \, dt.
\end{align*}
\end{lemma}
}

$\maltese$  \emph{Remark: Similar to the previous Lemma, Bar-David mentioned the idea but we could not find the proof. Therefore, the proof here is new.}

\noindent\textbf{Proof of Lemma~\ref{lemma: mu2, var2}}.

The characteristic function of $F_2$ is
\begin{equation}
\Phi(iu) = \E\left[ \exp\left\{iu \cdot \sum_{j=1}^M \frac{d}{d\tau_n} \frac{\alpha \dot{s}(t_j-\tau_n)}{\alpha s(t_j-\tau_n)+\lambda_b}  \right\} \right].
\end{equation}
Similar to Lemma~\ref{lemma: mu1, var1}, we define
\begin{equation}
h(t) = \exp\left\{iu \cdot \frac{d}{d\tau_n}\frac{\alpha \dot{s}(t-\tau_n)}{\alpha s(t-\tau_n)+\lambda_b}\right\}.
\end{equation}
Then, the function $H(T) = \int_{-T}^T h(t) \lambda(t) dt$, where $\lambda(t) = \alpha s(t-\tau_n) + \lambda_b$, and its derivatives are
\begin{align*}
H(T)\Big|_{iu=0}                    &= \int_{-T}^T [\alpha s(t-\tau_n) + \lambda_b] dt \\
\frac{d}{d(iu)}H(T)\Big|_{iu=0}     &= \int_{-T}^T \frac{d}{d\tau_n}\left(\frac{\alpha \dot{s}(t-\tau_n)}{\alpha s(t-\tau_n) + \lambda_b}\right) \lambda(t) \, dt\\
\frac{d^2}{d(iu)^2}H(T)\Big|_{iu=0} &= \int_{-T}^T \left(\frac{d}{d\tau_n}\left(\frac{\alpha \dot{s}(t-\tau_n)}{\alpha s(t-\tau_n) + \lambda_b}\right)\right)^2 \lambda(t) \, dt
\end{align*}
With some simplifications, we can show that
\begin{align*}
\mu_2
&= \frac{d}{d(iu)}H(T)\Big|_{iu=0}\\
&= \int_{-T}^T \frac{d}{d\tau_n}\left(\frac{\alpha \dot{s}(t-\tau_n)}{\alpha s(t-\tau_n)+\lambda_b}\right) \lambda(t) \, dt\\
&= \int_{-T}^T \left(\frac{-\lambda(t) \alpha \ddot{s}(t-\tau_n) - \alpha \dot{s}(t-\tau_n)\dot{\lambda}(t)}{\lambda(t)^2}\right) \lambda(t) \, dt\\
&= \underset{=0}{\underbrace{-\int_{-T}^T \alpha \ddot{s}(t) \, dt}} + \int_{-T}^T \frac{(\alpha \dot{s}(t-\tau_n))^2}{\alpha s(t-\tau_n)+\lambda_b} \,dt.
\end{align*}
Therefore,
\begin{align*}
\mu_2
&= \int_{-T}^T \frac{(\alpha \dot{s}(t-\tau_n))^2}{\alpha s(t-\tau_n)+\lambda_b} \,dt.
\end{align*}

\noindent Similarly, we can show that
\begin{align*}
&\left.\E[F_2^{'2}]
= \frac{d^2}{d(iu)^2} \Phi(iu) \right \vert_{iu=0}\\
&= \left.e^{-Q} \left[e^{H(T)} \left(\frac{d}{d(iu)}H(T)\right)^2 + e^{H(T)}\frac{d^2}{d(iu)^2}H(T)\right]\right \vert_{iu=0}\\
&= \left(\int_{-T}^T \frac{(\alpha \dot{s}(t-\tau_n))^2}{\alpha s(t-\tau_n)+\lambda_b} \, dt\right)^2\\
&\quad + \int_{-T}^T \left(\frac{d}{d\tau_n}\left(\frac{\alpha \dot{s}(t-\tau_n)}{\alpha s(t-\tau_n)+\lambda_b}\right)\right)^2 \Big(\alpha s(t-\tau_n)+\lambda_b\Big) \, dt.
\end{align*}
Subtracting the mean square will give us
\begin{align*}
&\E[F_2^{'2}]
= \E[F_2^2] - \mu_2^2 \\
&= \int_{-T}^T \left(\frac{d}{d\tau_n}\left(\frac{\alpha \dot{s}(t-\tau_n)}{\alpha s(t-\tau_n)+\lambda_b}\right)\right)^2 \Big(\alpha s(t-\tau_n)+\lambda_b\Big) dt.
\end{align*}

\section{Detailed Setups of Experiments}
\label{sec: Appendix Experiment Config}
For the experiments we presented in the main text, the parameters are configured as Table~\ref{table: parameters}. The parameters here are unit-free in the sense they are picked to elaborate the theory.

\begin{table}[h]
\centering
\caption{Parameters used in 1D experiments}
\begin{tabular}{lll}
\hline
\hline
$N$        & number of pixels      & \\
$x$        & spatial coordinate    & $0 \le x \le 1$\\
$t$        & temporal duration     & $0 \le t \le 10$\\
$\Delta N$ & width of each pixel   & $1/N$\\
$\Delta x$ & spatial grid spacing  & 1/2048 \\
$\Delta t$ & temporal grid spacing & 1/256 \\
$\sigma_t$ & pulse width           & 0.5\\
$\alpha_0$ & overall scene flux    & 10000\\
$\lambda_b$& noise floor           & 0\\
$\sigma_x$ & spatial Gaussian radius & $1/(\sqrt{12}N)$\\
$c_n$      & $n$th pixel of $\tau'(x)$ & use \texttt{gradient}\\
\hline
\end{tabular}
\label{table: parameters}
\end{table}

\subsection{Ground Truth Time-of-Arrival Function}
The true time of arrival function $\tau(x)$ can be any function defined on $0 \le x \le 1$. For example, in the main text, we consider a two-step scene with a smooth transition. Therefore, we can pick a sigmoid function of the following form.
\begin{equation}
\tau(x) = \frac{4}{1+e^{-20(x-0.5)}}+4,
\end{equation}
The shape for this particular model is shown in \cref{fig: true delay} below. We stress that this is just one of the many models we can pick.

\begin{figure}[h]
\centering
\includegraphics[width=\linewidth]{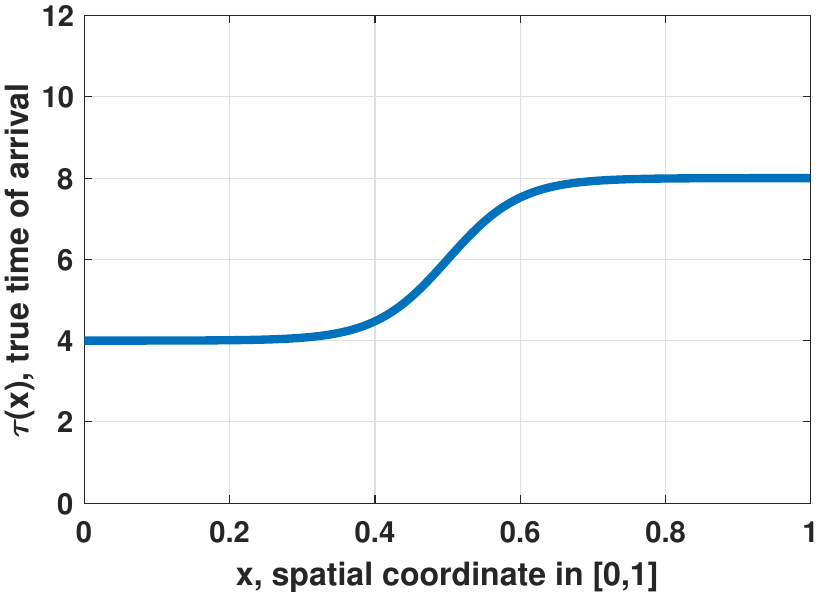}
\caption{The ground truth time of arrival function $\tau(x)$ as a function of the coordinate $0 \le x \le 1$.}
\label{fig: true delay}
\end{figure}

To compute the gradient $\tau'(x)$, the computationally efficient way is to perform finite difference instead of the analytic form. In MATLAB, this can be done using the command \texttt{gradient}.

\begin{lstlisting}[language=Matlab]
tau = 4./(1+exp(-20*(x-0.5)))+4;
tau_slope = gradient(tau)/dx;
\end{lstlisting}


The division by \texttt{dx} at the end is a reflection of the finite difference operation. For example, at $x_k$, the true gradient is
\begin{align*}
\tau'(x_k)
&= \lim_{\Delta x \rightarrow 0} \frac{\tau(x_k + \Delta x) - \tau(x_k)}{\Delta x} \\
&\approx \frac{\tau(x_{k+1}) - \tau(x_k)}{\Delta x}.
\end{align*}
Since finite difference \texttt{gradient} only computes $\tau(x_{k+1}) - \tau(x_k)$, we need to divide the result by $\Delta x$ to compensate for the spatial interval.

\subsection{What if we use a different $\tau(x)$?}
A critical observation of our theory is that the MSE can be decomposed into bias and variance. The variance has some subtle but important dependency on the shape of $\tau$, but the bias term will experience more impact.

\begin{figure}[h]
\centering
\begin{tabular}{cc}
\hspace{-2ex}\includegraphics[height=4cm]{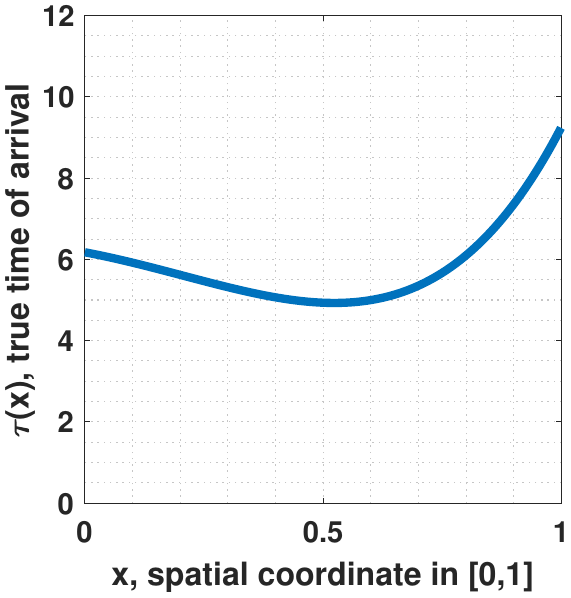}&
\hspace{-2ex}\includegraphics[height=4cm]{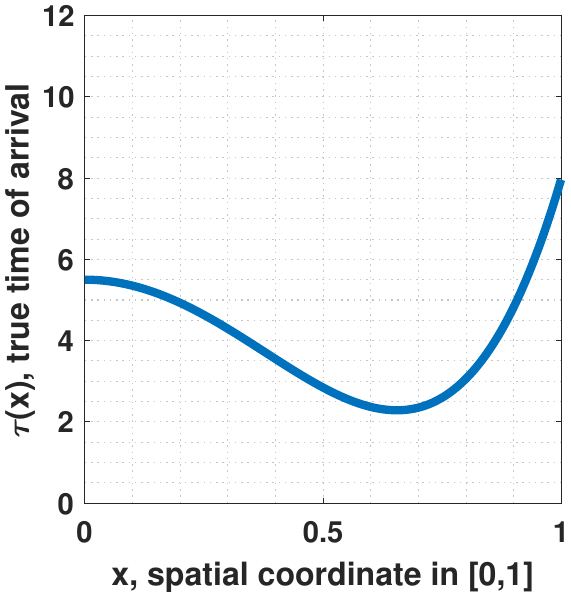}\\
\hspace{-2ex}\includegraphics[height=4cm]{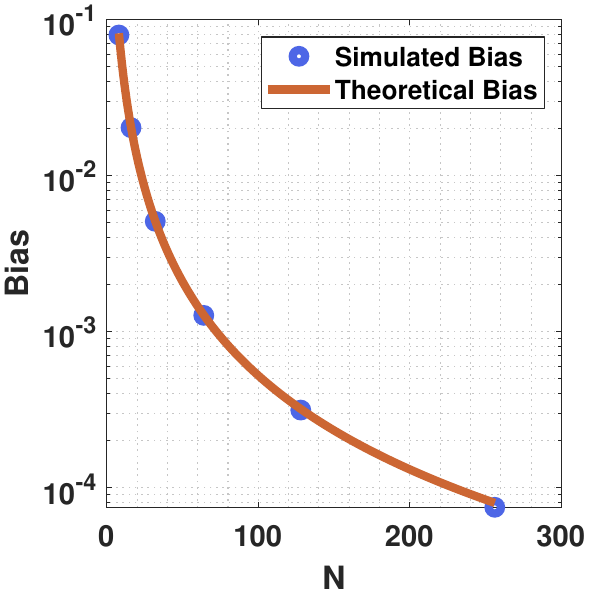}&
\hspace{-2ex}\includegraphics[height=4cm]{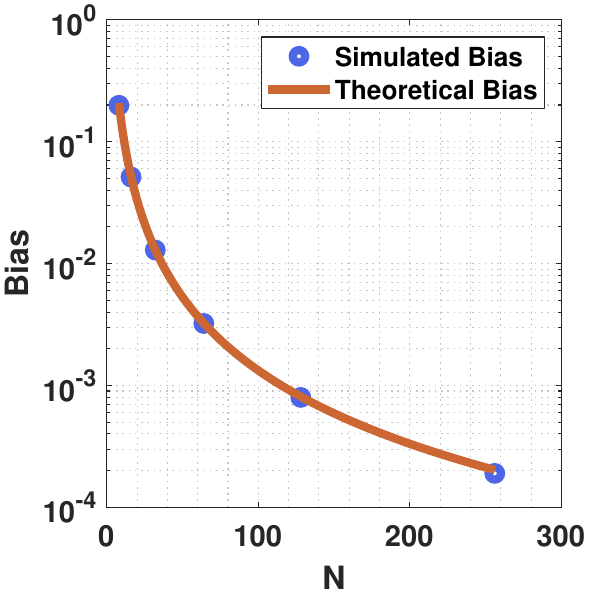}\\
\end{tabular}
\caption{$\tau(x)$ and the bias.}
\label{fig: tau examples}
\end{figure}

According to Lemma~\ref{lemma: bias}, the bias can be approximated by
\begin{equation}
\text{bias} = \int_{0}^{1} [\tau(x) - \overline{\tau}(x)]^2 dx \approx \frac{c^2}{12N^2},
\end{equation}
where $c^2 = \frac{1}{N}\sum_{n=0}^{N-1} c_n^2$, with $c_n = \tau'(x_n)$ being the gradient of $\tau$. Therefore, as $\tau$ changes, $c^2$ will change accordingly. \Cref{fig: tau examples} illustrates two examples. In the same figure, we also plot the theoretically predicted bias and the simulated result. When $\tau(x)$ is relatively smooth, the theory has an excellent match with the simulation.

\begin{figure}[h]
\centering
\begin{tabular}{cc}
\hspace{-2ex}\includegraphics[height=4cm]{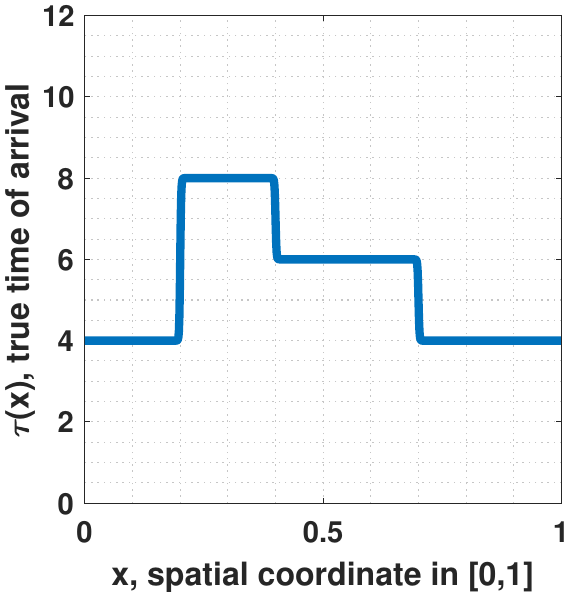}&
\hspace{-2ex}\includegraphics[height=4cm]{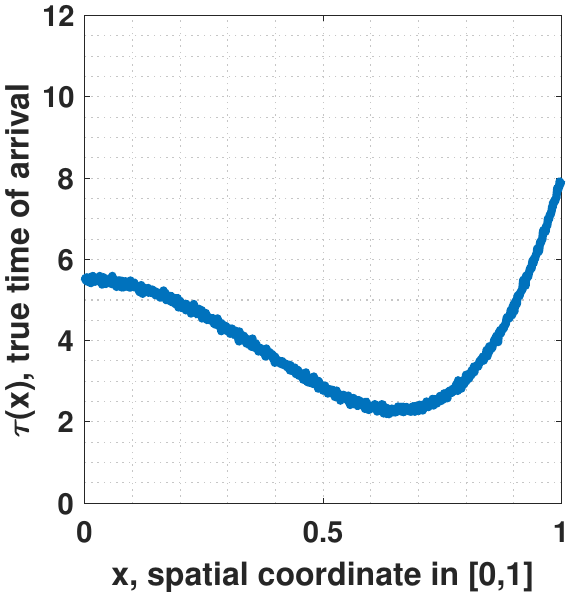}\\
\hspace{-2ex}\includegraphics[height=4cm]{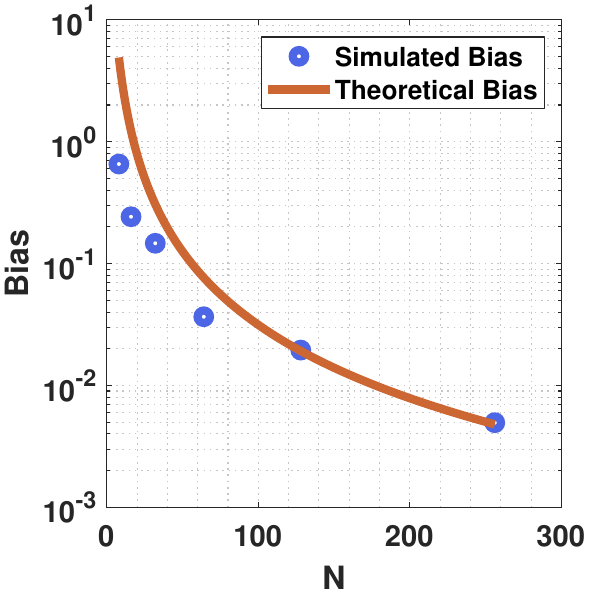}&
\hspace{-2ex}\includegraphics[height=4cm]{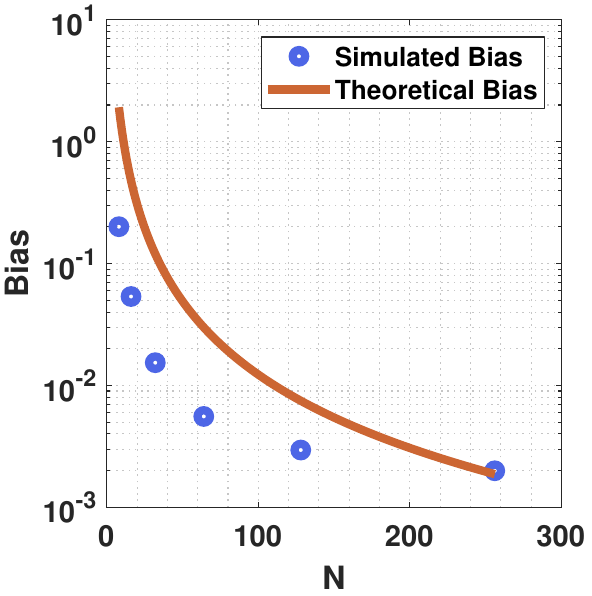}\\
\end{tabular}
\caption{Limitations of the approximations we presented in this paper. In the presence of discontinuous edges, and/or in the presence of noise, the performance of the theoretical prediction degrades.}
\label{fig: tau examples2}
\end{figure}

\textbf{Limitations}. With no surprise, the piecewise constant approximation has limitations. (1) When the function $\tau(x)$ is intrinsically discontinuous, the theory will not be able to keep track of the trend. (2) When the function $\tau(x)$ is noisy, then the gradient will be overestimated. Both will lead to degradation of the bias estimation, as illustrated in \cref{fig: tau examples2}.

Not all limitations can be analytically mitigated. For example, mitigating the discontinuity requires us to know the location of the discontinuous points. In practice, the likely solution is to numerically integrate the error. However, we will not be able to retain a clean, interpretable, and elegant formula.

\begin{figure}[h]
\centering
\begin{tabular}{cc}
\hspace{-2ex}\includegraphics[height=4cm]{./pix3/Figure13_Example_05a}&
\hspace{-2ex}\includegraphics[height=4cm]{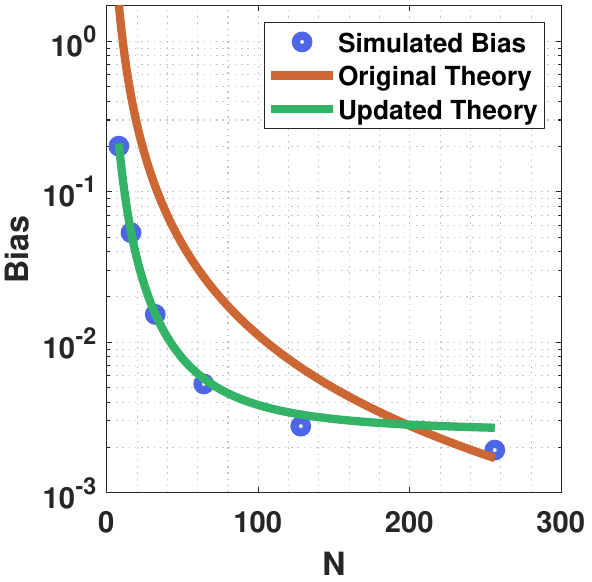}
\end{tabular}
\caption{When the true time of arrival is noise, it is possible to obtain a better bias estimate by compensating for the noise. However, this is rather an \emph{oracle} scenario because in practice, we never know the true noise-free $\tau(x)$ and we never know the probability distribution of noise.}
\label{fig: tau examples2 correction}
\end{figure}

For the noise problem, we can model the true time of arrival function as
\begin{align*}
\tau(x) = \tau_0(x) + \eta(x),
\end{align*}
where $\eta(x)$ is a random process denoting the noise, e.g., $\eta(x) \sim \calN(0,\sigma_e^2)$. In this case, the bias will become
\begin{align*}
\text{bias} = \sum_{n=0}^{N-1} \int_{\tfrac{n}{N}}^{\tfrac{n+1}{N}} [ \tau'(x_n)(x-x_n) + e(x) ]^2 dx.
\end{align*}
With some calculations, we can show that
\begin{align*}
\text{bias} = \frac{c^2}{12N^2} + \sigma_e^2,
\end{align*}
where $c^2 = \frac{1}{N}\sum_{n=0}^{N-1} c_n^2$, with $c_n = \tau_0'(x_n)$ being the slope of the \emph{clean} signal $\tau_0(x)$. Therefore, if we are given the noisy time of arrival $\tau(x)$, we first need to recover the clean $\tau_0(x)$ so that we can estimate the correct $c^2$. Then, the noise variance $\sigma_e^2$ is added to compensate for the noise term. The correction can be visualized in \cref{fig: tau examples2 correction}.

\section{Unit Conversion}
For our theoretical analysis to match a realistic sensor, a conversion between the size of a physical pixel and the size of a point in the numerical grid needs to be established. In what follows, we present a specific example with some specific numbers. These can be easily translated to other configurations.

Suppose that the overall size of the SPAD array is 10mm as shown in \cref{fig: grid analysis} below. This 10mm would correspond to 1 unit space. Suppose that we use 1024 points in the numerical grid to measure the array. Then, the width of each pixel is
\begin{equation*}
\Delta x \bydef 1 \text{ pixel} = \frac{10\text{mm}}{1024} = 9.76\text{um} = \frac{1}{1024} \text{unit space}.
\end{equation*}
For this paper's analysis, let's assume that we group 32 points (for example) as one \emph{super-pixel}. The width of each super-pixel is therefore
\begin{equation*}
1 \text{ super-pixel} = 32 \text{ pixels} = 312.5\text{um} = \frac{32}{1024} \text{unit space}.
\end{equation*}
Now, if we want to use a Gaussian to approximate the boxcar kernel, then the standard deviation of the Gaussian should be
\begin{align*}
\sigma_x &= \frac{1 \text{ super-pixel}}{\sqrt{12}} \\
&= 9.2372 \text{ pixels} = 90.21\text{um} = \frac{32}{1024\sqrt{12}} \text{unit space}.
\end{align*}

\begin{figure}[h]
\centering
\includegraphics[width=\linewidth]{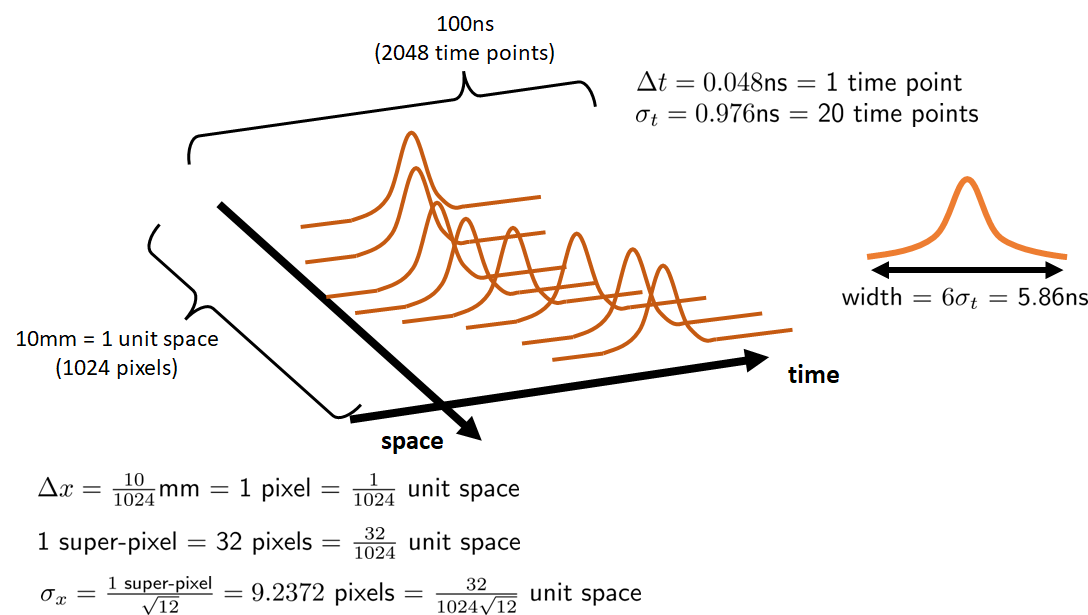}
\caption{Conversion between the physical units to the number of grid points used on a computer.}
\label{fig: grid analysis}
\end{figure}

For the temporal axis, we take a similar approach. Suppose that the total duration of the measurement is 100ns, and we use 2048 time points to measure the total duration. Then, each time point would correspond to
\begin{equation*}
\Delta t = 1 \text{ time point} = \frac{100 \text{ns}}{2048} = 0.048 \text{ns}.
\end{equation*}
A typical pulse has a width of around 2ns-6ns. So, suppose that we define the pulse standard deviation as $\sigma_t = 20$ time points, then
\begin{equation*}
\sigma_t = 0.976\text{ns} = 20 \text{ time points}.
\end{equation*}
Since $6\sigma_t$ of a Gaussian can capture 99\% of the energy, we can safely say that the width of the pulse is around
\begin{equation*}
\text{pulse width} = 6\sigma_t = 120 \text{ time points} = 5.86 \text{ns},
\end{equation*}
which is reasonably inside the 2ns-6ns range.

\section{Sampling Procedure}

In the main text, we outlined a procedure to generate time stamps according to a Gaussian pulse. This section discusses how to extend the idea to arbitrary pulse shapes.

\subsection{Inverse CDF Method}
When the pulse $\lambda(t)$ has a completely arbitrary shape and when the noise floor is not a constant, we will not be able to draw samples from distributions with known formulae. In this case, the sampling can be done using the inverse cumulative distribution function (inverse CDF) technique ~\cite[Ch.4.9]{Chan_2021Book}.

The concept of sampling from a known PDF is to leverage the cumulative distribution function (CDF) and a uniform random variable. A classical result shows the following. Suppose that there is a random variable $X$ generated from a distribution with a CDF $F_X$. If $U \sim \text{Uniform}[0,1]$, and if we send $U$ to the inverse CDF $F_X^{-1}$, then the transformed random variable $F_X^{-1}(U)$ will follow the distribution $F_X$.

The specific steps to perform the inverse CDF are as follows.
\begin{itemize}
\item Step 1: Compute the CDF $\Lambda(t) = \int_{-\infty}^t \lambda(r) dr$. Assuming that $\Lambda(t)$ is invertible, compute the inverse mapping $\Lambda^{-1}$. To obtain $\Lambda^{-1}(p)$ for a given $p$, we numerically build a lookup table.
\item Step 2: Compute the number of samples $M \sim \text{Poisson}(Q)$. Then, generate random samples by sending uniform random variables: $t_j = \Lambda^{-1}(U_j)$, where $U_j \sim \text{Uniform}[0,1]$, and $j = 1,\ldots,M$. Then $\{t_j\}_{j=1}^M$ will follow the probability distribution $\lambda(t)$.
\end{itemize}

\subsection{Implementation and Demonstration}
In terms of implementation, we can use the following MATLAB code. Suppose that we are given the received pulse $\lambda(t) = \alpha s(t-\tau) + \lambda_b$. Then, we can integrate $\lambda(t)$ to obtain the CDF $\Lambda(t) = \int_{-\infty}^t \lambda(r) dr$. On a computer, the command \texttt{cumsum} will serve the purpose of generating the CDF. Once the CDF is numerically determined, we send $M$ uniform random variables to $\Lambda^{-1}$. The inversion is performed numerically by finding the closest $t_j$ such that $t_j = \Lambda^{-1}(p)$ can give $\Lambda(t_j) = p$, where $p$ is the uniform random variable.

\begin{lstlisting}[language=Matlab]
function time_stamps = ...
    generate_time_stamps(lambda,t,M)

    c = cumsum(lambda)/sum(lambda);
    p = rand(1,M);
    c = repmat(c(:),[1,M]);
    [~,pos] = min(abs(c-p));
    time_stamps = t(pos);
end
\end{lstlisting}

\begin{figure}[h]
	\centering
	\begin{tabular}{c}
		\includegraphics[width=\linewidth]{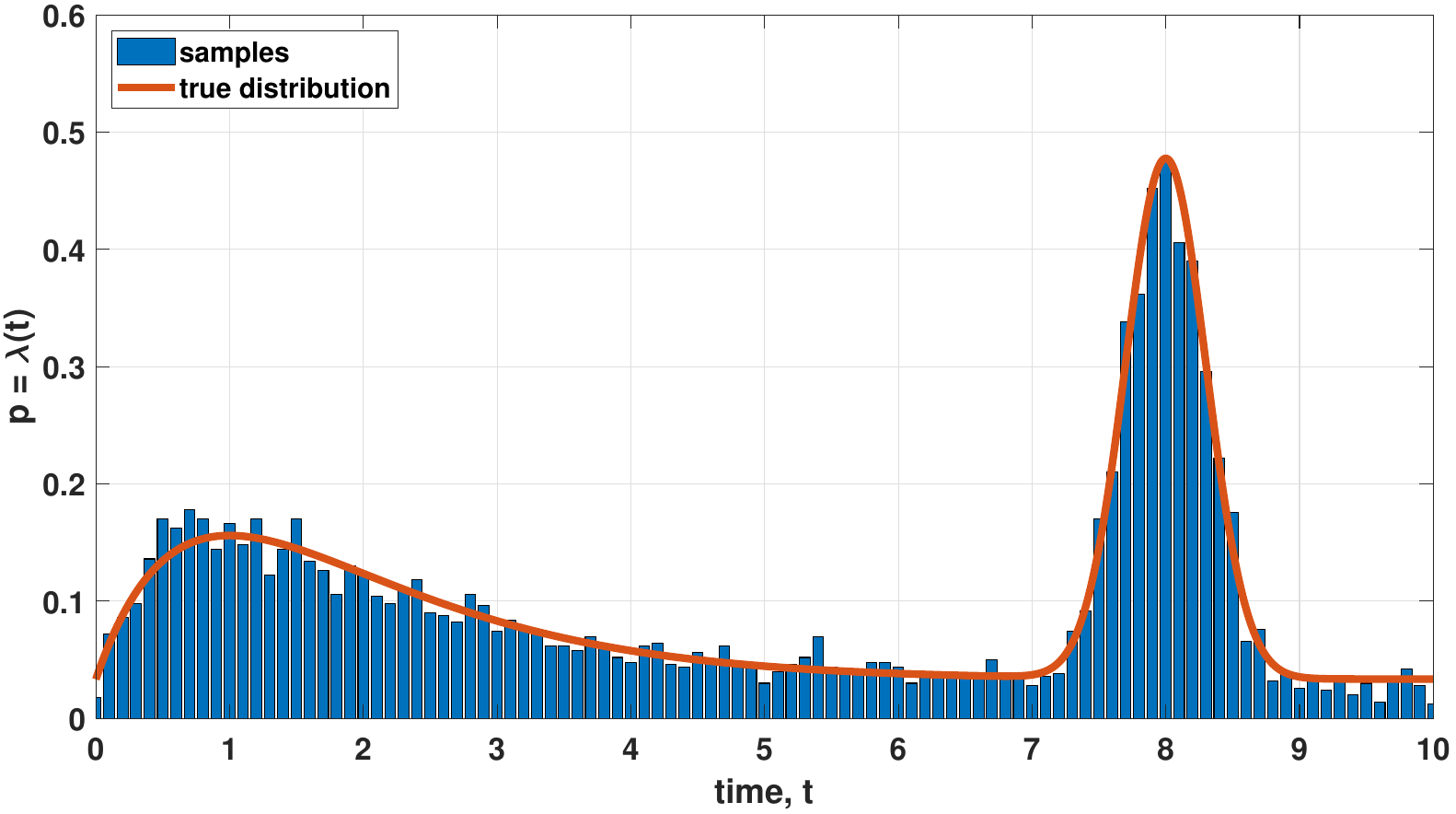}\\
		\includegraphics[width=\linewidth]{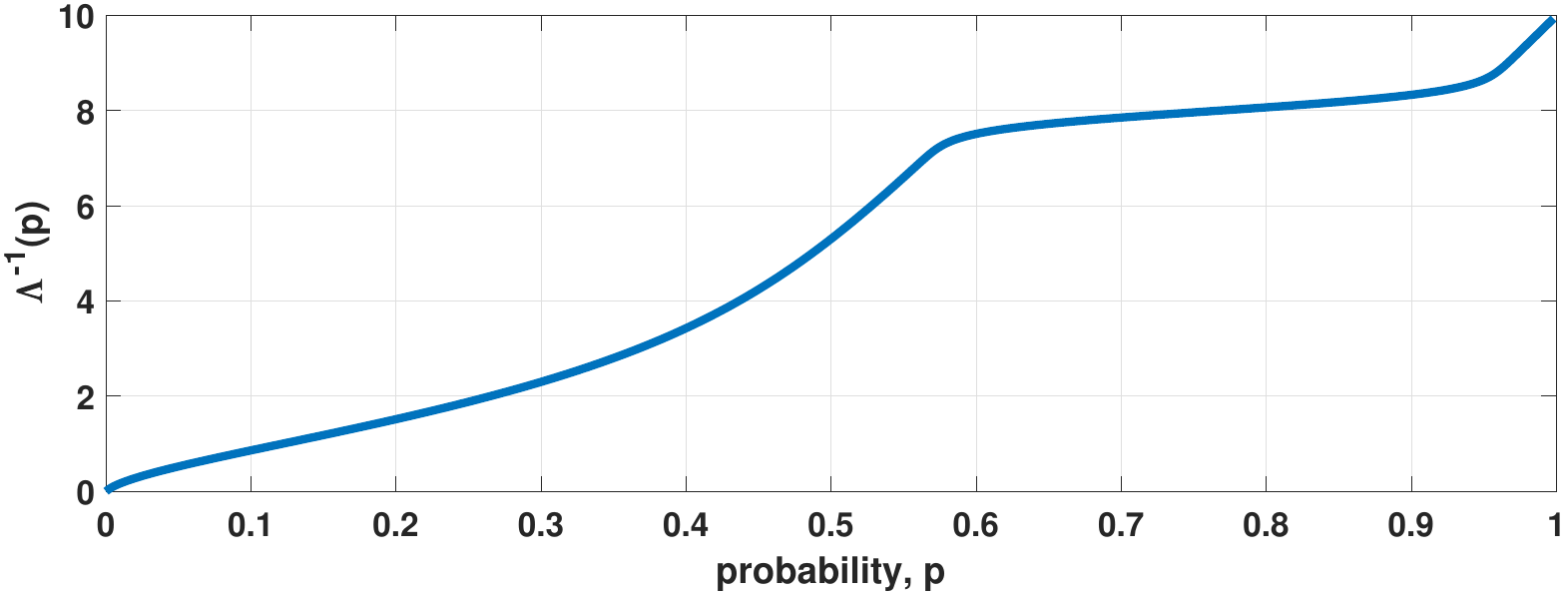}
	\end{tabular}
	\caption{[Top] Random samples $\vt_M$ drawn from a pulse $\lambda(t)$ which consists of a non-uniform noise floor and a Gaussian pulse. [Bottom] The inverse CDF $\Lambda^{-1}(p)$ is used to generate the samples from a uniform distribution. $\Lambda^{-1}(p)$ is generated numerically from the known equation $\lambda(t)$.}
	\label{fig: Figure 3 PDF and ICDF}
\end{figure}

\cref{fig: Figure 3 PDF and ICDF} shows a demonstration where we generate $M$ time stamps. The underlying pulse shape is Gaussian, but we also assume a nonlinear noise floor. Specifically, the noise floor mimics a scattering medium which is modeled by a Gamma distribution.

\begin{lstlisting}[language=Matlab]
s1 = 0.2*pdf('normal', t, tau, sigma_t);
s2 = 0.2*pdf('gamma', t, 2, 1);
lambda_b = 0.02*ones(1,length(t));
lambda   = s1+s2+lambda_b;

lambda_pdf = lambda/sum(lambda*dt);
lambda_cdf = cumsum(lambda/sum(lambda));
dp = 1/500;
p  = 0:dp:1-dp;
c = repmat(lambda_cdf(:),[1,length(p)]);
[~,pos] = min(abs(c-p));
lambda_icdf = t(pos);
\end{lstlisting}

\section{Non-zero Noise Floor}
The goal of the main text is to derive a simple and informative MSE equation. This is achieved by assuming that the pulse is Gaussian and there is no noise. In this section, we discuss how the theoretical results can be derived for arbitrary pulses, by means of a numerical scheme.

\subsection{Maximum Likelihood Estimation}
Assuming that the transmitted pulse is $s(t)$ and the received pulse is $\lambda(x,t) = \alpha s(t-\tau(x)) + \lambda_b$. Here, we assume that $s(t)$ can take an arbitrary shape and $\lambda_b > 0$. We also assume that there is a spatial kernel $\phi(x)$ which will be applied to $\lambda(x,t)$ to yield the resulting space-time function $\widetilde{\lambda}(x,t)$:
\begin{equation*}
\widetilde{\lambda}(x,t) = \phi(x) \circledast \lambda(x,t).
\end{equation*}
The convolution $\circledast$ can be implemented numerically. In MATLAB, we can use the command \texttt{imfilter}:

\begin{lstlisting}[language=Matlab]
lambda = alpha*s + lambda_b
h = ones(dN,1)/dN;
lambda_tilde = imfilter(lambda, h, ...
           replicate');
\end{lstlisting}

Following the same derivations as the main text, we let $x_n = (2n+1)/(2N)$ be the mid point of each spatial interval and define the effective pulse $\widetilde{\lambda}_n(t)$ as
\begin{equation*}
\widetilde{\lambda}_n(t; \tau_n) = \widetilde{\lambda}(x_n,t; \tau_n),
\end{equation*}
where we emphasize that $\widetilde{\lambda}_n$ has an underlying parameter $\tau_n$ which needs to be estimated. Then, the time of arrivals $\vt_M = [t_1,\ldots,t_M]$ will follow the distribution specified by the effective return pulse $\widetilde{\lambda}_n(t)$.

The questions we need to ask now are (1). how to estimate the time of arrival $\widehat{\tau}$, (2). what is the variance $\E[(\widehat{\tau}_n-\tau_n)^2]$?

For the purpose of deriving theoretical bounds, the estimator we use is the maximum likelihood estimator. The ML estimator is the one that maximizes the likelihood function. In our case, it is
\begin{equation}
\widehat{\tau}_n = \argmax{\tau_n} \; \underset{\calL(\tau_n)}{\underbrace{\sum_{j=1}^M \log \widetilde{\lambda}_n(t_j; \tau_n)}},
\end{equation}

\begin{figure}[b!]
	\centering
	\includegraphics[width=\linewidth]{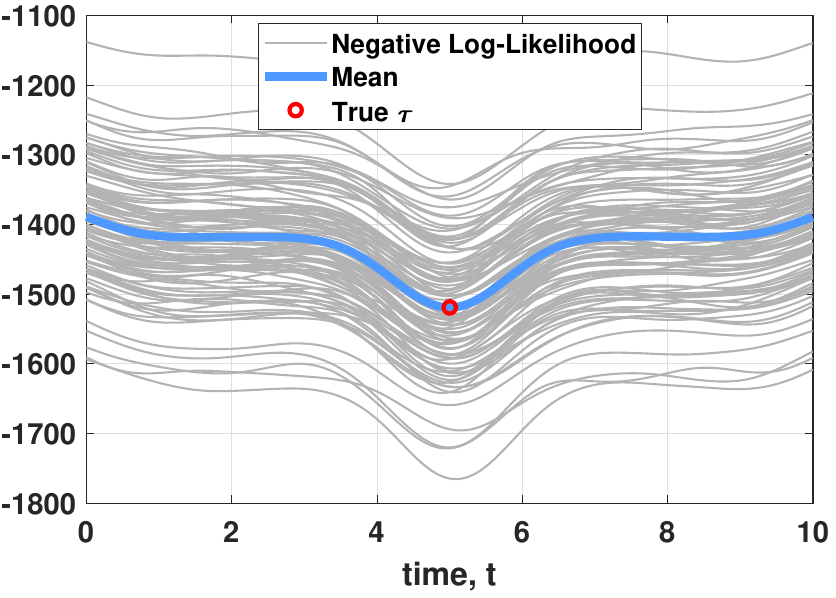}
	\caption{The shape of the negative log-likelihood function $\calL(\tau)$ as a function of the unknown parameter $\tau$.}
	\label{fig: likelihood 2}
\end{figure}

The shape of the likelihood function $\calL(\tau_n)$ as a function of the parameter $\tau_n$ is shown in \cref{fig: likelihood 2}, for a typical Gaussian pulse with a non-zero noise floor. The likelihood function has a clear maximum (minimum if we take the negative log likelihood) around the true parameter. For the specific example shown in \cref{fig: likelihood 2}, we use the following configurations: $\alpha = 100$, $\lambda_b = 30$, $s(t,\tau_0) = \calN(t \,|\, \tau_0, \sigma_t^2)$ where $\tau_0 = 5$ and $\sigma_t = 0.5$. We randomly generate $K = 100$ random vectors $\vt_M^{(k)}$ for $k = 1,\ldots,K$. We plot the negative log-likelihood $\calL(\tau_n \,|\, \vt_M^{(k)})$ as a function of the parameter $\tau_n$. The negative log-likelihood functions are random because $\vt_M^{(k)}$ are random. We take the expectation of these $K = 100$ random negative log-likelihood functions to visualize the mean function.

An equivalent alternative approach is to perform the estimation by solving a zero-finding problem. Taking the derivative of $\calL$ with respect to $\tau_n$, we can show that the solution $\widehat{\tau}_n$ must satisfy the equation
\begin{align}
\left. \frac{d\calL}{d\tau_n} \right \vert_{\tau_n = \widehat{\tau}_n}= \left. \sum_{j=1}^M \frac{\dot{\widetilde{\lambda}}_n(t_j; \tau_n) }{\widetilde{\lambda}_n(t_j; \tau_n)} \right \vert_{\tau_n = \widehat{\tau}_n}= 0.
\label{eq: derivative of L}
\end{align}
Then, the maximum likelihood estimate $\widehat{\tau}_n$ is determined by finding the zero-crossing of the derivative such that $\dot \calL(\widehat{\tau}_n) = 0$.

\boxedeg{
\begin{example}{}
Suppose that $\widetilde{\lambda}_n(t; \tau_n) = \alpha s(t - \tau_n) + \lambda_b$, then the likelihood function is
\begin{align*}
\calL(\tau_n) = \sum_{j=1}^M \log \Big\{ \alpha s(t_j - \tau_n) + \lambda_b \Big\}.
\end{align*}
The derivative of the likelihood is
\begin{align*}
\frac{d\calL}{d\tau_n} = -\sum_{j=1}^M \frac{\alpha s(t_j-\tau_n)}{\alpha s(t_j - \tau_n) + \lambda_b}.
\end{align*}
\end{example}
}

\begin{figure}[h]
\centering
\includegraphics[width=\linewidth]{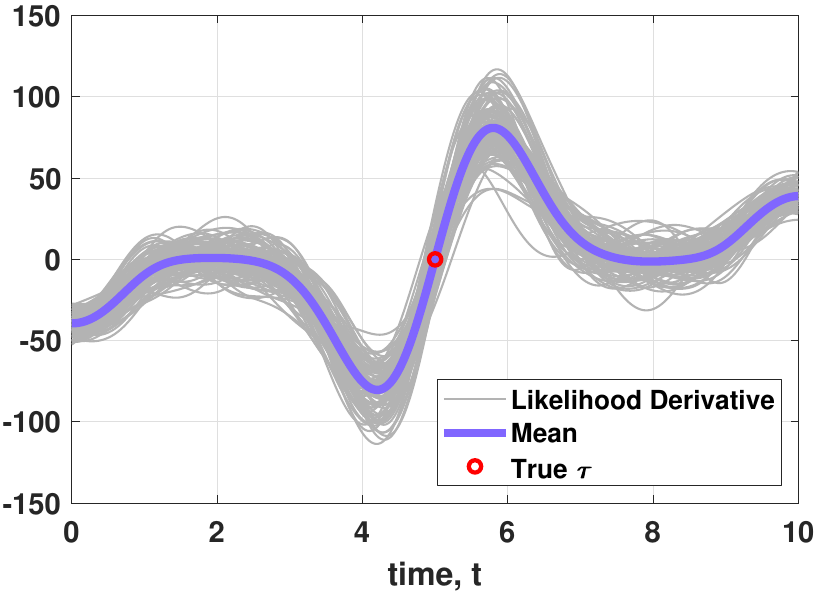}
\caption{The shape of the negative log-likelihood function's derivative $\frac{d}{d\tau}\calL(\tau)$ as a function of the unknown parameter $\tau$.}
\label{fig: likelihood 1}
\end{figure}

The shape of the likelihood derivative can be seen in \cref{fig: likelihood 1}, for a typical example outlined in the description of \cref{fig: likelihood 2}. The maximum likelihood estimate is the zero-crossing of the derivative. We emphasize that for arbitrary pulse shape and a non-zero floor noise, neither the zero-finding approach nor the matched filter approach would have an analytic solution. We will discuss the numerical implementation shortly.

\subsection{MSE Calculation}
Given the effective return pulse $\widetilde{\lambda}_n(t)$, the MSE follows from Theorem {\color{red} 2}. To make our notations consistent, we assume that $\widetilde{\lambda}_n(t)$ takes the form
\begin{equation}
\widetilde{\lambda}_n(t) = \alpha s_n(t-\tau_n) + \lambda_b,
\end{equation}
for some function $s_n(t)$. This assumption is valid whenever $\alpha$ and $\lambda_b$ are independent of $x$ so that any spatial convolution applied to construct $\widetilde{\lambda}(t)$ will not affect $\alpha$. Under this notation, we follow Theorem {\color{red} 2} and write the variance of the estimator
\begin{align}
\E[(\widehat{\tau}_n-\tau_n)^2] = \left[ \int_{-T}^T \frac{(\alpha \dot{s}_n(t))^2}{\alpha s_n(t) + \lambda_b} \; dt \right]^{-1},
\label{eq: Var arbitrary pulse}
\end{align}
where $\dot{s}_n(t)$ is the derivative of $s_n(t)$ with respect to $t$.

For arbitrary pulse shape, $s_n(t)$ does not have any analytic expression, so does $\dot{s}_n(t)$. Nevertheless, the integration in \cref{eq: MSE arbitrary pulse} can still be performed numerically. The MATLAB code below shows how we can do it numerically.

\begin{lstlisting}[language=Matlab]
tau = 4./(1+exp(-20*(x-0.5)))+4;
tau_slope = gradient(tau)/dx;
for n=1:N
 c(n) = mean( tau_slope(dN*(n-1)+[1:dN]) );
 s_dot(n,:) = gradient(s(n,:))/dt;
 var_n(n) = 1/sum(((alpha*s_dot(n,:)).^2) ...
       ./(alpha*s(n,:)+lambda_b)*dt);
end
var_theory  = mean( var_n );
bias_theory = mean((c.^2)/(12*N^2));
mse_theory  = bias_theory + var_theory;
\end{lstlisting}

With \cref{eq: Var arbitrary pulse} defined, the overall MSE as a function of the number of pixels is as follows.
\boxedthm{
\begin{theorem}{}
For arbitrary pulse shape and noise floor, the MSE takes the form
\begin{equation}
\text{MSE} = \frac{c^2}{12N^2} + \frac{N}{\alpha_0} \E[(\widehat{\tau}_n-\tau_n)^2].
\label{eq: MSE arbitrary pulse}
\end{equation}
\end{theorem}
}

Here, we use the fact that $\E[(\widehat{\tau}_n-\tau_n)^2]$ has the same value for all $n$ so it does not matter which $n$ we use to calculate the variance. Although \cref{eq: MSE arbitrary pulse} is no longer a closed-form expression, it still preserves the shape of the resolution limit where the MSE decays quadratically with respect to $1/N^2$ and increases linearly with respect to $N$.

\subsection{Solving the ML Estimation}
\label{sec: Appendix solving ML estimation}
In this subsection, we explain how to implement the maximum likelihood estimation for an arbitrary pulse and noise floor. There are three ways to do this.

\noindent\textbf{Approach 1: Gradient-based Likelihood Maximization}. For pulses with a known analytic expression, the most straightforward approach is to perform gradient descent with the gradient $\frac{d}{d\tau}\calL$:
\begin{align*}
\tau^{(k+1)} = \tau^{(k)} - \gamma^{(k)} \frac{d}{d\tau}\calL(\tau^{(k)}),
\end{align*}
for some step size $\gamma^{(k)}$. Most numerical software such as MATLAB has built-in optimization packages to accomplish this step. For example, using \texttt{fminunc}, we will be able to find the maximum of the likelihood function.

The example below assumes a Gaussian pulse so that we have a closed-form expression for the likelihood function.

\begin{lstlisting}[language=Matlab]
function L = myL(tau, tj, ...
        sigma_t, alpha, lambda_b)
L  = -sum(log(alpha * ...
        1/sqrt(2*pi*sigma_t^2) * ...
   * exp(-(tj-tau).^2/(2*sigma_t^2)) + ...
   + lambda_b));
\end{lstlisting}

To solve the actual maximization we do

\begin{lstlisting}[language=Matlab]
fun = @(tau) myL(tau, tj, ...
    sigma_t, alpha, lambda_b);
tau_hat = fminunc(myL, tau0, option);
\end{lstlisting}

where \texttt{tau0} is the initial guess. For the purpose of deriving the ``oracle'' maximum likelihood estimate (so that we are not penalized by having a bad algorithm), we use the ground truth $\tau_0$ as the initial guess. Readers may worry that this would return us the ground truth $\tau_0$ as the maximum likelihood estimate, i.e., $\widehat{\tau} = \tau_0$. We note that this will never happen because the likelihood $\calL$ needs to fit the measured time stamps $\vt_M$ and so $\calL$ is a random function. Since $\calL$ is a random function, the maximum location is a random variable too. Therefore, while the expected value of the estimate $\widehat{\tau}$ is the true $\tau_0$, for a particular realization $\widehat{\tau}$ will never be the same as $\tau_0$.

\noindent\textbf{Approach 2: Search-based Likelihood Maximization}. For arbitrary pulse shape, the gradient-based approach is difficult to implement because the pulse $s(t-\tau)$ is numerically a different function when we use a different $\tau$. The viable approach, as we mentioned in the previous subsection, is to run a matched filter. A matched filter requires us to shift the known pulse to the left or to the right until we see the best fit to the data. On computers, we need to perform two steps:
\begin{itemize}
\item Given a current estimate $\tau$ and a perturbation $\Delta \tau$, translate it to the time index of the numerical array $\widetilde{\lambda}_n(t;\tau)$.
\item Shift the index based on the amount corresponding to $\Delta \tau$, and calculate $\widetilde{\lambda}_n(t;\tau+\Delta \tau)$.
\end{itemize}

In MATLAB, the shifting operation can be implemented using the commands below.

\begin{lstlisting}[language=Matlab]
function L = myL(tau, tj, lambda_bar_n, ...
            t, tau0)
lengthT     = length(t);
[~,pos_new] = min(abs(t-tau));
[~,pos_ini] = min(abs(t-tau0));
lambda1_pad = [lambda_bar_n(1)...
                    *ones(1,lengthT), ...
               lambda_bar_n, ...
               lambda_bar_n(end) ...
                    *ones(1,lengthT)];
lambda2_pad = circshift(...
                lambda1_pad, ...
                pos_new-pos_ini);
lambda2     = lambda2_pad(...
                lengthT+1:2*lengthT);
L = -sum(log(interp1(t, ...
            lambda2, tj, 'spline')));
\end{lstlisting}

The padding is a band-aid for a MATLAB specific command \texttt{circshift} which circularly shifts the indices. If we do not pad the array appropriately, a pulse located near the end of the time axis will cause erroneous values to the likelihood function after they get circularly flipped to the beginning of time.

Because of the shifting operations defined above, the function call is significantly harder for automatic differentiation unless we customize the step size. The reason is that if the default step size is smaller than the temporal grid we used to define the pulse, two adjacent indices will remain the same. Thus the gradient will be zero. A workaround solution is to go with a non-gradient optimization. For MATLAB, we can use the package \texttt{fminsearch}.

\begin{lstlisting}[language=Matlab]
fun = @(tau) myL(tau, tj, lambda_n, t, tau0);
tau_hat = fminsearch(fun, t0, option);
\end{lstlisting}

\noindent\textbf{Approach 3: Zero-finding of the Likelihood Gradient}. The third approach we can use is to directly solve for the maximum likelihood solution. Recall from \cref{eq: derivative of L} that the maximum likelihood estimate must be the zero-crossing point of \cref{eq: derivative of L}, we can directly implement the derivative as a function call such as the example below.

\begin{lstlisting}[language=Matlab]
function L = myL_dt(tau, tj, sigma_t, ...
                        alpha, lambda_b)
my_s    =  alpha ...
    * (1/sqrt(2*pi*sigma_t^2)) ...
    * exp( -(tj-tau).^2/(2*sigma_t^2) ) ...
    + lambda_b;
my_s_dot= -alpha*(1/sqrt(2*pi*sigma_t^2)) ...
    * exp( -(tj-tau).^2/(2*sigma_t^2) ) ...
    .* ((tj-tau)./(sigma_t^2));
L   = sum( my_s_dot./my_s );
\end{lstlisting}

Then, we can use \texttt{fzero} to find the zero-crossing point.

\begin{lstlisting}[language=Matlab]
fun = @(tau) myL_dt(tau, tj, sigma_t, ...
                       alpha, lambda_b);
tau_hat= fzero(fun,tau0);
\end{lstlisting}

\textbf{Which approach is faster?} Since the goal of this paper is not to provide an algorithm, we skip a formal complexity analysis of the methods. To give readers a rough idea of the comparison between the three approaches, on a typical experiment using the same machine, the runtime is as follows.

\begin{table}[h]
\centering
\begin{tabular}{llll}
\hline
   & Method            & Command             & Runtime \\
\hline
1 & Gradient descent  & \texttt{fminunc}    & 0.064 sec\\
2 & Matched filter    & \texttt{fminsearch} & 0.034 sec\\
3 & Zero crossing     & \texttt{fzero}      & 0.016 sec\\
\hline
\end{tabular}
\end{table}
We remark that all three methods give nearly identical solutions up to the precision of the numerical grid we set.

\subsection{Experimental Results}
\begin{figure}[h]
\centering
\includegraphics[width=\linewidth]{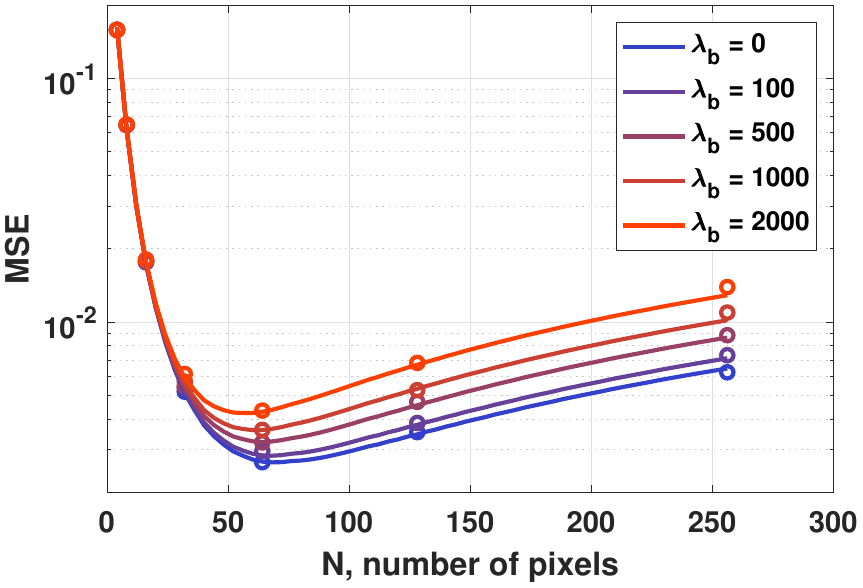}
\caption{Excellent match between the theoretical bound and the simulation for various levels of ambient noise. }
\label{fig: MSE 2}	
\end{figure}

To convince readers that \cref{eq: MSE arbitrary pulse} can be numerically implemented and matches with simulation, we show in \cref{fig: MSE 2} the comparison between the simulation and the theoretical prediction for a range of $\lambda_b$ values. As is evident from the plot, the theoretical prediction matches very well with the simulation.

As we have demonstrated in this example, it is sometimes possible to numerically compute the bias and variance so that the theory will match with the simulation. However, by doing so, we will no longer be able to write down the resolution limit in a simple and interpretable closed-form expression. This is not a deficiency of our theory, it is just a sacrifice of clarity and interpretability in exchange for better theoretical precision.

\section{Pile-up Effects}
Pile up effects refer to the situation where the photo detector responds earlier than the actual arrival of the pulse signal, typically due to the presence of a strong background.

\subsection{Distribution}
\begin{figure}[h]
\centering
\includegraphics[width=\linewidth]{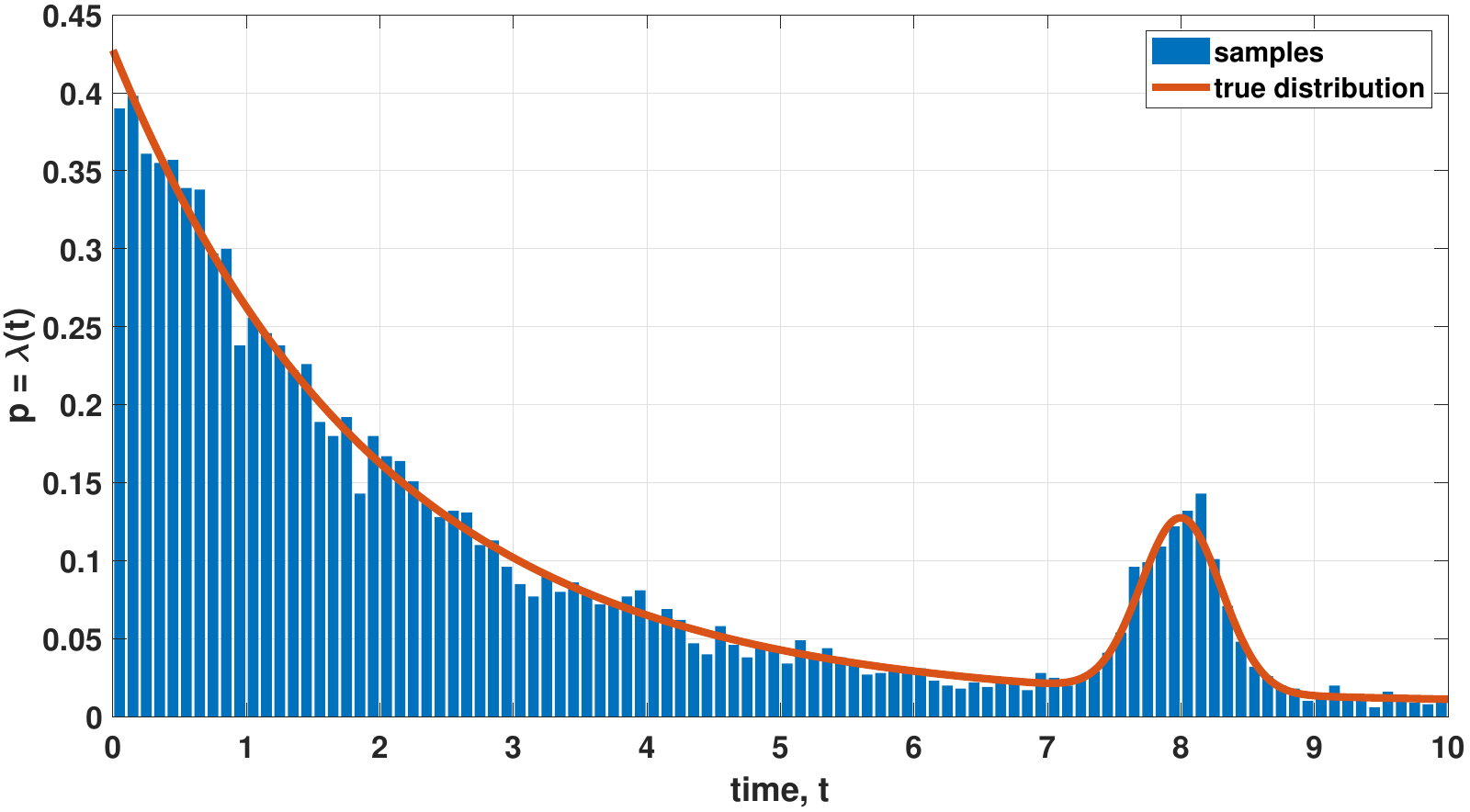}
\caption{Pile up effect can be modeled as an exponential decaying distribution. Added to the signal pulse $s(t-\tau)$, pile up effect makes the ML estimation of the true delay significantly harder.}
\label{fig: pile up histogram}
\end{figure}

For simplicity, we model the pile up effect as an exponentially decaying distribution in the background, that is,
\begin{equation}
\lambda(x,t) = \alpha s(t-\tau(x)) + \beta \lambda_p(t) + \lambda_b,
\end{equation}
where $\lambda_p(t) = \gamma e^{-\gamma t}$ is an exponential function parameterized by the reflected laser event rate $\gamma$, and the background event rate $\lambda_b$ \cite{Tontini_2023_histogramless}. In this case, the sampling of the time stamps will follow from three steps:
\begin{itemize}
\item Draw $M_s = \text{Poisson}(\alpha)$ samples. The distribution of these $M_s$ samples is the shape of the pulse $s(x,t)$. The parameter $\alpha$ specifies the average number of incident signal photons.
\item Draw $M_p = \text{Poisson}(\beta \gamma)$ samples. These $M_p$ samples follow the distribution $\text{Exponential}(\gamma)$ (whose mean is $1/\gamma$). The parameter $\beta\gamma$ specifies the average number of pile-up photons (coming from the background).
\item Draw $M_b = \text{Poisson}(\lambda_b T)$ samples, assuming that $0 \le t \le T$. These $M_b$ samples follow the distribution $\text{Uniform}(\lambda_b T)$.
\end{itemize}
An example of the time stamp histogram is shown in \cref{fig: pile up histogram}.

\subsection{MLE}
The maximum likelihood estimation in the presence of the pile up effect needs a separation of background and signal. Because of the simplified model we choose (for the purpose of theoretical illustration), we can exploit the oracle knowledge we know about the pile up and background. Given $\lambda(x,t)$, we know that the binning is such that
\begin{align*}
\lambda_n(t)
&= \int_{\tfrac{n}{N}}^{\tfrac{n+1}{N}} \lambda(x,t) dx \\
&= \alpha \underset{=s_n(t-\tau_n)}{\underbrace{\int_{\tfrac{n}{N}}^{\tfrac{n+1}{N}}s(t-\tau(x)) dx}}  + \int_{\tfrac{n}{N}}^{\tfrac{n+1}{N}} [\beta \lambda_p(t) + \lambda_b] dx  \\
&= \alpha s_n(t-\tau_n)  + \frac{\beta \lambda_p(t) + \lambda_b}{N},
\end{align*}
Where we approximate the integral by a simplified function $s_n(t)$ delayed by a time constant $\tau_n$.

Given this decomposition of $\lambda_n(t)$ as the signal $s_n(t)$ and the noise part $\beta \lambda_p(t) + \lambda_b$, we use Theorem {\color{red} 2}
to estimate the variance of the $n$th pixel:
\begin{align*}
\E[(\widehat{\tau}_n-\tau_n)^2] = \int_{-T}^{T} \frac{[\alpha \dot{s}_n(t)]^2}{\lambda_n(t)} dt.
\end{align*}
This integration needs to be evaluated numerically.

In terms of simulation, we need to run the ML estimation. The ML estimation requires us to search for an optimal $\tau_n$ such that the candidate distribution matches with the measurement. To this end, we follow the procedure outlined in \cref{sec: Appendix solving ML estimation} to numerically run a search scheme to pick up the ML estimate.

\subsection{Theoretical MSE} 
\begin{figure}[h]
\centering
\includegraphics[width=\linewidth]{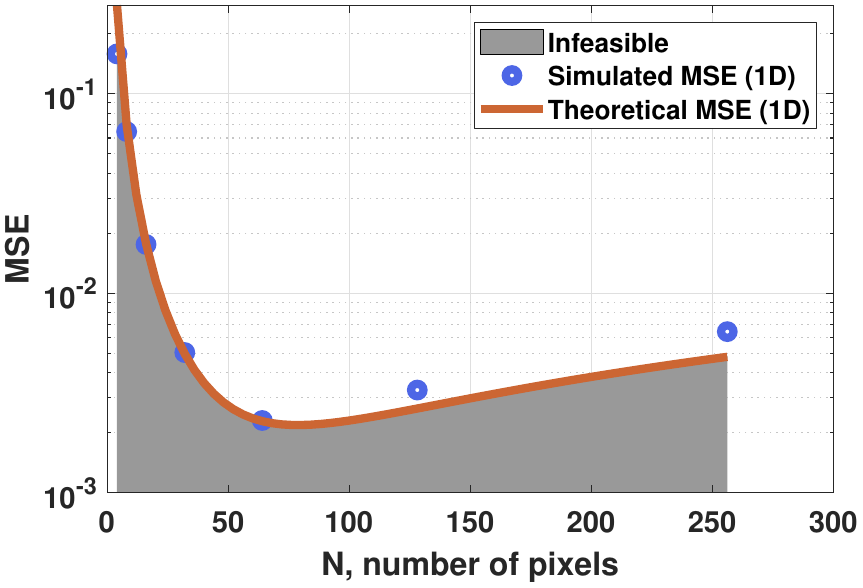}
\caption{MSE comparison in the presence of pile up. The theoretically predicted MSE has a very good match with the simulated MSE. }
\label{fig: pile up MSE}
\end{figure}

A comparison between the simulated MSE and the theoretically predicted MSE is shown in \cref{fig: pile up MSE}. For this particular example, we set $\alpha = 1\times10^4$, $\beta = 1\times10^{5}$, and $\lambda_b = 100$. The pile up distribution has a constant $\gamma = 4$. Again, we emphasize that these units are unit-free, in the sense they are chosen to illustrate the validity of the theorem rather than matching any particular real sensor.

As we can see in \cref{fig: pile up MSE}, the theoretical prediction matches extremely well with the simulated MSE. The small deviation towards a larger $N$ is due to the fact that the ML estimation faces difficulty when the total number of samples is few. As an example, when $N = 256$, the number of samples per pixel is around 400 whereas when $N = 8$, the number of pixels is around 28,000. When there is no pile up noise, the variance of ML estimate is basically limited by the number of samples. But in the presence of complicated noise (such as pile up), the ML estimation procedure has limitations where it cannot differentiate signal from noise. Therefore, when the number of samples is few, the simulation performance will be worse than what the theory predicts. Nevertheless, \cref{fig: pile up MSE} still shows a high degree of match between the theory and the simulation.

\section{Dark Count}
The model we presented in the main text does not consider dark count. In this section, we briefly discuss how it can be included.

Dark count is a type of sensor noise caused by the random generation of electrons in the depletion region. These random electrons are present even when there is no photoelectric event. Therefore, one way to model dark count is to treat it as a uniform noise floor that does not change over time and does not depend on the scene flux. However, dark counts scale with the sensing area. So, if there are $N$ pixels and if we assume that the sensing area scales linearly with $N$, then the dark counts will be reduced by $N$ times on average. Therefore, a simple model in the context of this paper is to write
\begin{equation}
\lambda(t) = \frac{\alpha}{N} s(t-\tau) + \frac{\lambda_b}{N} + \frac{\lambda_{\text{dark}}}{N}.
\end{equation}

Because of this simple addition to the background, we can treat $\lambda_{\text{dark}}$ as part of the background noise. In a typical scenario when the dark count rate is in the order 10 per second while a SPAD can count up to 1 million photons per second (e.g., ThorLab's single-photon detectors), the impact of dark current is not visible unless we operate it in extremely low-light.

\section{Extension to 2D}

\label{sec: theory 2D}
In this section, we briefly summarize how the results can be extended to 2D. We will derive the bias and the variance.

\subsection{Bias}
For the bias term, in 2D, we need to consider a 2D coordinate $\vx \in [0,1]\times[0,1]$. The time of arrival function $\tau$ is denoted as $\tau(\vx)$. The average function $\overline{\tau}(\vx)$ is
\begin{align*}
\overline{\tau}(\vx) = \sum_{m=0}^{M-1} \sum_{n=0}^{N-1} \tau_{m,n} \varphi(Mx-m)\varphi(Ny-n),
\end{align*}
Where $\tau_{m,n}$ is the $(m,n)$th value defined as
\begin{align*}
\tau_{m,n} = \int_{\tfrac{m}{M}}^{\tfrac{m+1}{M}} \int_{\tfrac{n}{N}}^{\tfrac{n+1}{N}} \tau(x,y) \, dx\,dy.
\end{align*}

The derivative of the time of arrival function, in the 2D case, will become the gradient. This means that
\begin{align*}
\nabla_{\vx} \tau(\vx) = \left[ \frac{\partial \tau}{\partial x}, \; \frac{\partial \tau}{\partial y} \right]^T.
\end{align*}
For notational simplicity, we write $\nabla_{\vx}\tau(\vx) = \vc = [c^x, \, c^y]^T$. If we are interested in the gradient of the $(m,n)$th pixel, we write
\begin{align*}
\vc_{m,n} = [c^x_{m,n}, \, c^y_{m,n}]^T = \nabla_{\vx} \tau(\vx_{m,n}).
\end{align*}
The magnitude square of the gradient is denoted as $\|\vc_{m,n}\|^2 = (c^x_{m,n})^2 + (c^y_{m,n})^2$, and the overall gradient summed over all the pixels is
\begin{align*}
\|\vc\|^2
&= \frac{1}{MN}\sum_{m=0}^{M-1}\sum_{n=0}^{N-1} \|\vc_{m,n}\|^2 \approx \int_{0}^{1} \int_{0}^{1} \|\nabla_{\vx} \tau(\vx)\|^2 \, d\vx,
\end{align*}
where the last approximation holds by assuming that $(m,n)$ covers the unit squares.

The main result for the bias term in the 2D case is summarized in the lemma below.
\boxedthm{
\begin{lemma}
\label{lemma: 2D bias}
Let $\tau(\vx)$ be a 2D time of arrival function over the unit square $\vx \in [0,1]\times[0,1]$. Suppose that we use $M \times N$ pixels to approximate the unit square, and for simplicity assume that $M = N$. If we approximate $\tau(\vx)$ using a piecewise constant function $\overline{\tau}(\vx)$, the bias is given by
\begin{equation}
\text{Bias} = \frac{\|\vc\|^2}{12N^2}.
\end{equation}
\end{lemma}
}

\noindent\textbf{Proof of Lemma~\ref{lemma: 2D bias}}.

We start with the definition of the bias:
\begin{align*}
&\text{Bias}
=\int_0^1\int_{0}^{1} \left( \overline{\tau}(\vx) - \tau(\vx) \right)^2  \, d\vx\\
&= \sum_{m=0}^{M-1} \sum_{n=0}^{N-1} \underset{=e_{m,n}^2}{\underbrace{\int_{\tfrac{m}{M}}^{\tfrac{m+1}{M}} \int_{\tfrac{n}{N}}^{\tfrac{n+1}{N}} \left(\tau_{m,n} - \tau(x,y) \right)^2 \, dx \, dy}}.
\end{align*}
Let $\vx_{m,n}$ be the mid-point such that
\begin{align*}
\vx_{m,n} = \left[\frac{2m+1}{2M}, \frac{2n+1}{2N}\right].
\end{align*}
We can approximate the first order approximation to $\tau(\vx)$:
\begin{align*}
\tau(\vx) = \tau(\vx_{m,n}) + \vc_{m,n}^T (\vx-\vx_{m,n}),
\end{align*}
where $\vc_{m,n} = \nabla \tau(\vx_{m,n})$ is the gradient. Then, the error is
\begin{align*}
e_{m,n}^2
&= \int_{\tfrac{m}{M}}^{\tfrac{m+1}{M}} \int_{\tfrac{n}{N}}^{\tfrac{n+1}{N}}  \left(\tau_{m,n} - \tau(\vx) \right)^2 \, d\vx\\
&= \int_{\tfrac{m}{M}}^{\tfrac{m+1}{M}} \int_{\tfrac{n}{N}}^{\tfrac{n+1}{N}}  \Big(\nabla \tau(\vx_{m,n})^T (\vx-\vx_{m,n}) \Big)^2 \, d\vx\\
&= \left[c_{m,n}^x \right]^2 \frac{1}{12M^3N} + \left[c_{m,n}^y\right]^2 \frac{1}{12N^3M}.
\end{align*}
Assuming $M = N$, the above is simplified to
\begin{align*}
e_{m,n}^2 =  \frac{\left[c_{m,n}^x \right]^2 + \left[c_{m,n}^y\right]^2}{12N^4} = \frac{\|\vc_{m,n}\|^2}{12N^4}.
\end{align*}
Summing over all $(m,n)$'s will give us
\begin{align*}
\text{Bias}
&= \int_0^1 \int_0^1 \left( \overline{\tau}(\vx) - \tau(\vx) \right)^2  \, d\vx\\
&= \sum_{n=0}^{N-1}\sum_{m=0}^{N-1} \frac{ \|\vc_{m,n}\|^2 }{12N^4} = \frac{\|\vc\|^2}{12N^2}.
\end{align*}

\begin{figure}[h]
\centering
\begin{tabular}{cccc}
\hspace{-2ex}\includegraphics[height=2cm]{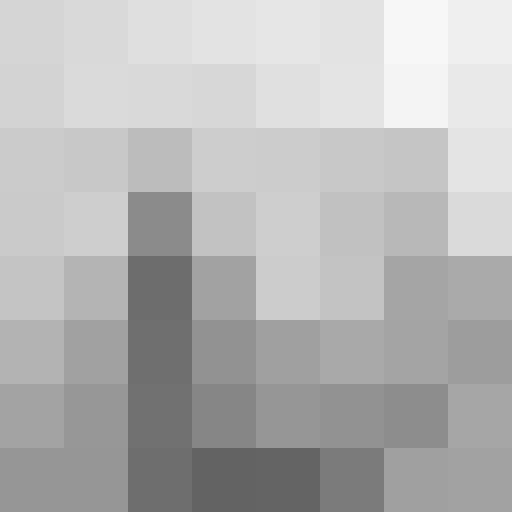}&
\hspace{-2ex}\includegraphics[height=2cm]{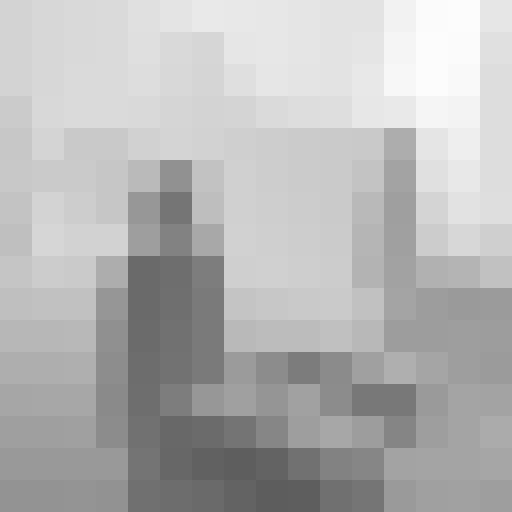}&
\hspace{-2ex}\includegraphics[height=2cm]{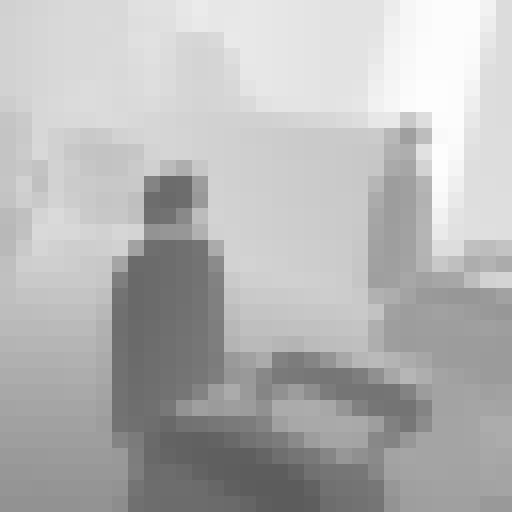}&
\hspace{-2ex}\includegraphics[height=2cm]{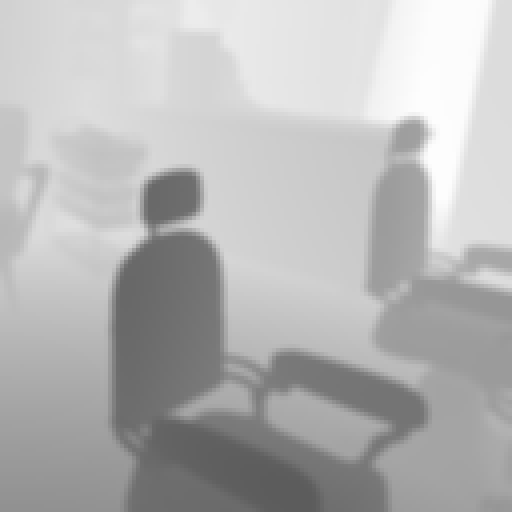}\\
\end{tabular}
\includegraphics[width=\linewidth]{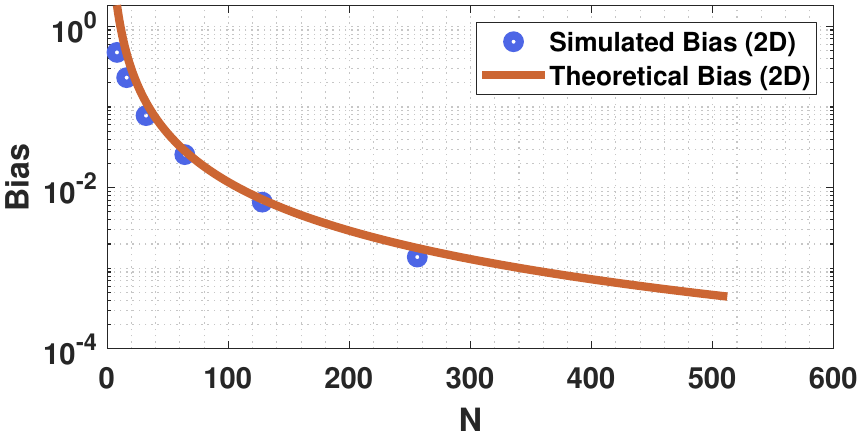}
\caption{Visualization of the bias in 2D. The full resolution of the depth map is $512\times512$. We consider multiple resolutions to study the effect of the bias. This plot highlights the excellent match between the theoretical prediction and the simulation results.}
\label{fig: 2D bias}
\end{figure}

\noindent\textbf{Visualization of the Bias}. A visualization of the bias can be seen in \cref{fig: 2D bias}. In this example, we use a real depth map as the ground truth $\tau(\vx)$. We scale the depth map so that the maximum value is 20, and the minimum value is 10. The full resolution is $512 \times 512$. To reduce the discontinuity of the depth map at object boundaries, we apply a spatial lowpass filter to smooth out the edges. The estimated gradient magnitude is $\|\vc\|^2 = \int_{0}^{1} \nabla_{\vx}\tau(\vx) d\vx = 1.42 \times 10^{3}$. We consider four different resolutions $8 \times 8$, $16\times 16$, $32\times32$, and $128\times128$. For each resolution, we use the equation outlined in Lemma~\ref{lemma: 2D bias} to calculate the theoretically predicted bias. We compare this predicted bias with the simulation. As we can see from \cref{fig: 2D bias}, the theoretical prediction offers an excellent match with the simulation.

\subsection{Variance}
For the variance term, we note that the derivation follows similarly to the 1D case:
\begin{align*}
\text{Var}
&= \E\left[ \int_0^1\int_0^1 \left( \widehat{\tau}(\vx) - \overline{\tau}(\vx) \right)^2 \, d\vx \right]\\
&= \E\left[ \sum_{m=0}^{M-1}\sum_{n=0}^{N-1} \int_{\frac{m}{M}}^{\frac{m+1}{M}} \int_{\frac{n}{N}}^{\frac{n+1}{N}} \left( \widehat{\tau}(\vx) - \overline{\tau}(\vx) \right)^2 \, d\vx \right]\\
&= \sum_{m=0}^{M-1} \sum_{n=0}^{N-1} \int_{\frac{m}{M}}^{\frac{m+1}{M}} \int_{\frac{n}{N}}^{\frac{n+1}{N}}  \E\left[ \left( \widehat{\tau}_{m,n} - \overline{\tau}_{m,n} \right)^2\right] \, d\vx.
\end{align*}
Therefore, as long as we can calculate $\E\left[ \left( \widehat{\tau}_{m,n} - \overline{\tau}_{m,n} \right)^2\right]$, we will be able to determine the variance.

Based on the arguments we described before stating the lemma, we shall focus on deriving the variance for each individual pixel. To this end, we need to derive the effective return pulse $\lambda_{m,n}(\vx,t)$. Following the steps of the proof of  Theorem {\color{red} 3}, the critical step is the convolution of a 2D spatial Gaussian and a 1D temporal Gaussian
\begin{align*}
\text{2D spatial Gaussian}  &= \calN(\vx \,|\, 0, \sigma_s^2\mI)\\
\text{1D temporal Gaussian} &= \calN(t   \,|\, \tau_0 + \vc_0^T(\vx - \vx_0), \sigma_t^2).
\end{align*}
Here, $\sigma_s = 1/(\sqrt{12}N)$ is the spatial radius of the 2D Gaussian, which serves the same role as $\sigma_x$ in the 1D case. The linear approximation $\tau_0 + \vc_0^T(\vx - \vx_0)$ is the 2D version of the 1D linear approximation $\tau_0 + \tau'(x_0)(x-x_0)$, where $\vx_0$ is the center pixel in each interval of the grid, and $\vc_0$ is the gradient at $\vx_0$.

Before we proceed to the proof, we state the main result.
\boxedthm{
\begin{lemma}
\label{lemma: 2D var}
Let $\tau(\vx)$ be a 2D time of arrival function over the unit square $\vx \in [0,1]\times[0,1]$. Suppose that we use $M \times N$ pixels to approximate the unit square, and for simplicity assume that $M = N$. Let $\alpha_0$ be the total flux over the unit square, then The variance is given by
\begin{equation}
\text{Var} = \frac{N^2}{\alpha_0}\left(\|\vc\|^2 \sigma_s^2 + \sigma_t^2\right),
\end{equation}
where $\sigma_s = 1/(\sqrt{12}N)$ is the spatial radius of the Gaussian approximation, and $\sigma_t$ is the temporal spread of the pulse.
\end{lemma}
}

\noindent\textbf{Proof of Lemma~\ref{lemma: 2D var}}. The core derivation is the convolution between the 2D Gaussian in space and the 1D Gaussian in time. By separability of a 2D Gaussian, we can write the 2D Gaussian as the convolution of two orthogonal 1D Gaussian functions:
\begin{align*}
\calN(\vx \,|\, 0, \sigma_s^2\mI) = \calN(x \,|\, 0, \sigma_s^2) \circledast \calN(y \,|\, 0, \sigma_s^2).
\end{align*}
The order of $x$ versus $y$ does not matter. Then, substituting it into the main convolution, we can perform two consecutive convolutions:
\begin{align*}
\calN(t \,|\, \tau_0 + \vc_0^T(\vx - \vx_0), \; \sigma_t^2) \circledast \calN(\vx \,|\, 0, \sigma_s^2\mI)\\
= \calN(t \,|\, \tau_0 + \vc_0^T(\vx - \vx_0), \; \sigma_t^2)
&\circledast \calN(x \,|\, 0, \sigma_s^2) \\
&\circledast \calN(y \,|\, 0, \sigma_s^2)
\end{align*}

The first convolution, following the derivation of  Theorem {\color{red} 3}, will give us
\begin{align*}
&\calN(t \,|\, \tau_0 + \vc_0^T(\vx - \vx_0), \; \sigma_t^2)  \circledast \calN(x \,|\, 0, \sigma_s^2) \\
&= \calN(t \,|\, \tau_0 + \vc_0^T(\vx - \vx_0), \; \sigma_t^2 + [c_0^x]^2 \sigma_s^2),
\end{align*}
where $c_0^x$ is the horizontal component of the gradient $\vc_0 = [c_0^x, c_0^y]^T$.

The second convolution, which is applied after the first convolution, will give us
\begin{align*}
&\calN(t \,|\, \tau_0 + \vc_0^T(\vx - \vx_0), \; \sigma_t^2 + [c_0^x]^2 \sigma_s^2) \circledast \calN(y \,|\, 0, \sigma_s^2) \\
&= \calN(t \,|\, \tau_0 + \vc_0^T(\vx - \vx_0), \; \sigma_t^2 + [c_0^x]^2 \sigma_s^2 + [c_0^y]^2 \sigma_s^2)\\
&= \calN(t \,|\, \tau_0 + \vc_0^T(\vx - \vx_0), \; \sigma_t^2 + \|\vc_0\|^2 \sigma_s^2).
\end{align*}

Restricting ourselves to $\vx = \vx_0$, we can show that the effective return pulse is
\begin{align*}
\lambda_{m,n}(t) =
\alpha \calN(t \,|\, \tau_{m,n} , \|\vc_{m,n}\|^2\sigma_s^2+\sigma_t^2) + \lambda_b.
\end{align*}

Finally, we recognize that since there are $MN$ pixels in the unit space, (and assuming that $M=N$ for simplicity), then each pixel will see a total flux of $\alpha_0/N^2$. Therefore,
\begin{align*}
\E\left[ \left( \widehat{\tau}_{m,n} - \overline{\tau}_{m,n} \right)^2\right]
&= \frac{\text{variance of $\lambda_{m,n}(t)$}}{\alpha}\\
&= \frac{N^2}{\alpha_0}\left(\|\vc_{m,n}\|^2\sigma_s^2+\sigma_t^2\right).
\end{align*}
Summing over $m$ and $n$ will give us
\begin{align*}
\text{Var}
&=
\sum_{m=0}^{M-1} \sum_{n=0}^{N-1} \int_{\frac{m}{M}}^{\frac{m+1}{M}} \int_{\frac{n}{N}}^{\frac{n+1}{N}}  \E\left[ \left( \widehat{\tau}_{m,n} - \overline{\tau}_{m,n} \right)^2\right] \, d\vx\\
&= \frac{N^2}{\alpha_0} \left(\|\vc\|^2\sigma_s^2+\sigma_t^2\right).
\end{align*}

\begin{figure}[h]
\centering
\begin{tabular}{cccc}
\hspace{-2ex}\includegraphics[height=2.1cm]{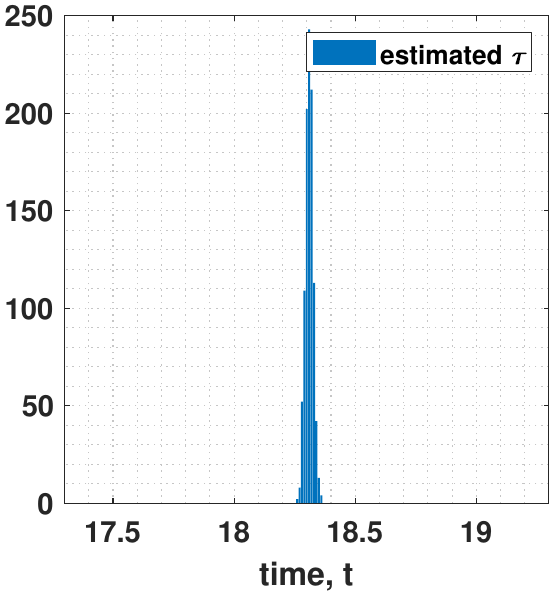}&
\hspace{-2ex}\includegraphics[height=2.1cm]{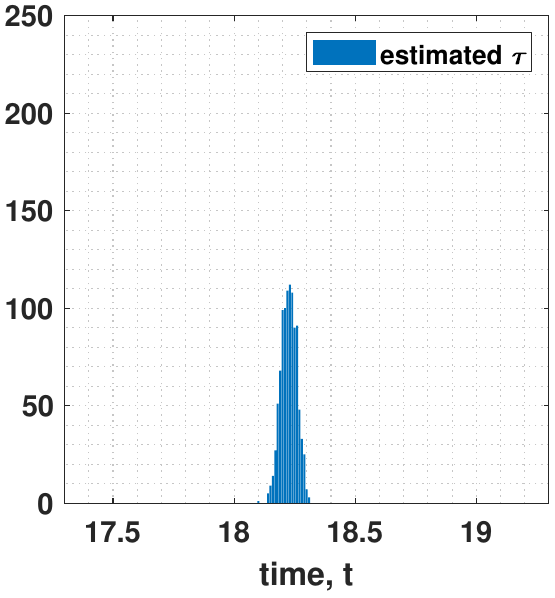}&
\hspace{-2ex}\includegraphics[height=2.1cm]{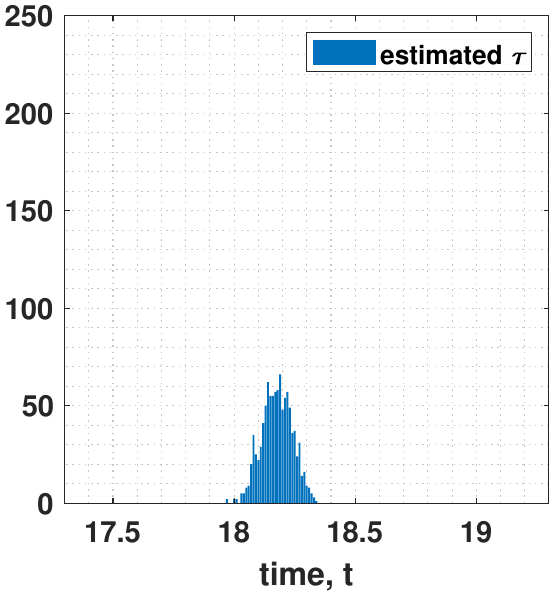}&
\hspace{-2ex}\includegraphics[height=2.1cm]{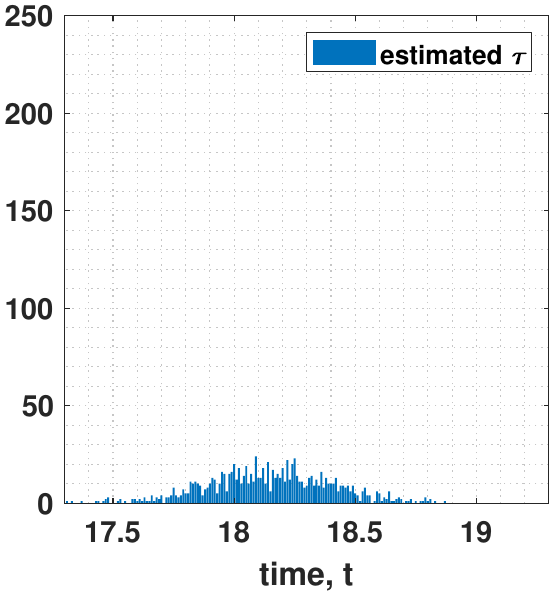}\\
\end{tabular}
\includegraphics[width=\linewidth]{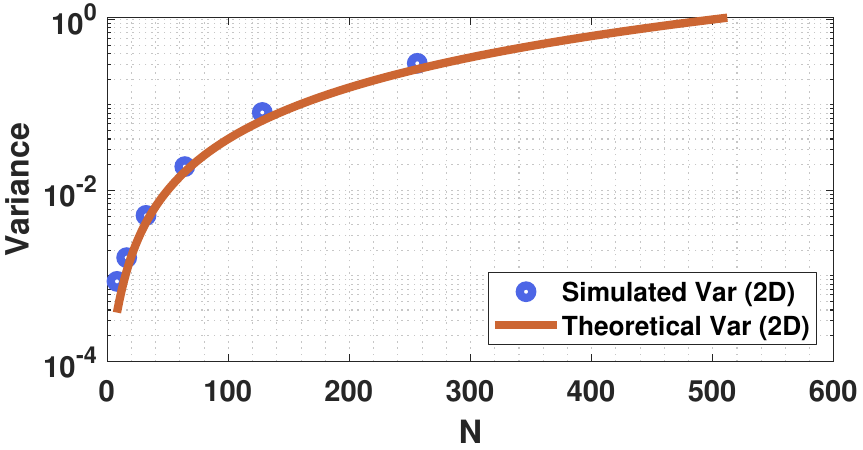}
\caption{Visualization of the variance for a 2D example. For every pixel in the scene, we run a Monte Carlo simulation to compute the variance of the estimated depth values. The histograms shown in this figure correspond to the four resolutions we showed in the bias case. As we increase the spatial resolution (so that each pixel becomes smaller), the histograms become flat and so the variance increases. The plot at the bottom highlights the excellent match between the theoretically predicted variance and the simulated variance.}
\label{fig: 2D var}
\end{figure}

\noindent\textbf{Visualizing the Variance}. \cref{fig: 2D var} shows the visualization of the variance as a function of $N$. The experimental configuration is identical to the one we used in the 2D bias analysis. For the purpose of analyzing the variance, we need to further set up a Monte Carlo simulation. To this end, we first set the spatial spacing of the grid as $\Delta x = 1/512$ along each direction. The total amount of flux is set as $\alpha_0 = 1e^{6}$. Suppose that we have $N$ pixels on each side, the unit square will have $N^2$ pixels. Each pixel will therefore observe $\alpha_0/N^2$ amount of flux. When $N = 8$, the per pixel flux is 15625 photons per pixel. When $N = 256$, the per pixel flux is 15.26 photons per pixel.

For simplicity, we assume that the pulse $s(t)$ is Gaussian. The width of the pulse is 2 arbitrary units, which is roughly 10 percent of the true depth value. We also assume that the noise floor is $\lambda_b = 0$. These two assumptions allow us to use a simple ensemble average as the maximum likelihood estimator. Otherwise, we need to numerically implement the matched filter.

In this particular experiment, we use $\sigma_s = 1/(\sqrt{12}N)$ as the spatial radius, and $\|\vc\|^2 = 1.42 \times 10^{3}$. The variance is calculated for each pixel and then summed over the entire unit square. A total number of 1000 random trials are conducted to obtain the variance per pixel (and for the whole image).

\section{Real 2D Experiment}
In this section, we provide additional detail on how real 2D experiments are conducted.

\subsection{Snapshot of Real Data}
\begin{figure}[h]
\centering
\begin{tabular}{cc}
\includegraphics[width=0.45\linewidth]{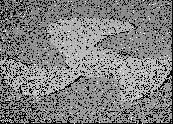}&
\includegraphics[width=0.45\linewidth]{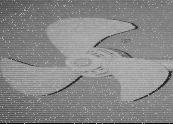}\\
1 frame & average of 10,000 frames
\end{tabular}
\caption{Raw data produced by the SPAD array.}
\label{fig: Figure 25 real data}
\end{figure}

As depicted in the main text, the real data is collected by a $192\times128$ SPAD array previously reported by Henderson et al. \cite{Henderson_2019_192x128}. The system has a time-to-digital converter (TDC) timing resolution of 35 ps. For convenience, we crop the center $128\times128$ region for analysis. \cref{fig: Figure 25 real data} shows a one-frame snapshot of the real data, and the 10,000-frame average. We observe that the 10,000-frame average is noisy, although the shape of the object is visible.

\begin{figure}[h]
\centering
\includegraphics[width=\linewidth]{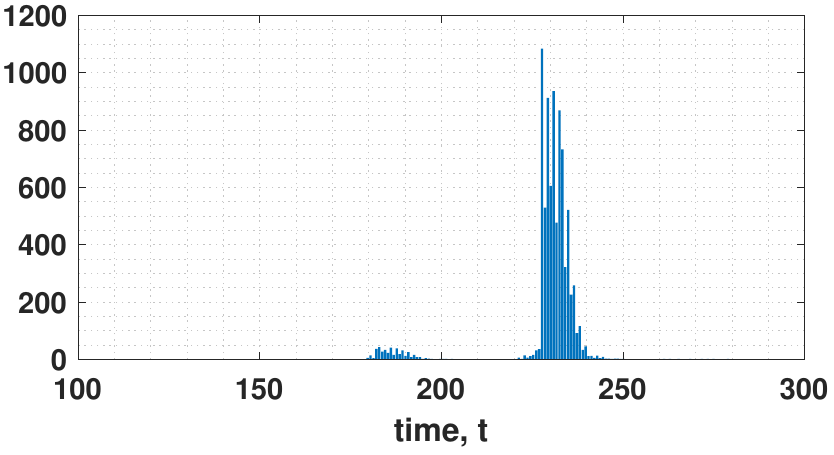}
\caption{Histogram of one of the pixels in the SPAD data array.}
\label{fig: Figure 24 pulse}
\end{figure}

Inspecting the source of the noise, we plot the histogram of one of the pixels of the data volume in \cref{fig: Figure 24 pulse}. We recognize two issues in the data: (i) There is a secondary pulse, likely caused by secondary bounces from the background; (ii) there is a strong peak happening at the beginning (and sometimes at the end) of the sensing period, likely caused by failed detections of the SPAD. We did not show these samples in the histogram because it is just a strong spike in the histogram. The presence of these two issues makes the average of the time stamps problematic, hence generating noise.

\subsection{Pseudo Ground Truth}
For the purpose of MSE analysis, we need to construct a pseudo ground truth --- a ground truth signal generated from all the available samples, hoping as noise free as possible. To achieve this goal, we need to reject the outliers so that we can retain only the primary pulse.

Our pre-processing step is done by identifying the center of the primary pulse. Once the center is identified, we crop around the center histogram with $\pm 3\sigma_t$ where $\sigma_t$ is the standard deviation of the pulse.

Two remarks are worth mentioning: (i) The identification of the center of the primary pulse involves a few steps. The first step is to find a coarse estimate of the center. For this purpose, we compute the mean of the entire 10,000 data points and crop a large temporal window around the mean. We smooth out the histogram so that we can pick a more reliable peak. The smoothing step is done by adding Gaussian random variables. (ii) Once the peak is identified, we move to the second stage by putting a small temporal window around the peak. We reject all samples that fall outside this small temporal window, thus effectively removing the secondary pulse and the spike in the beginning and at the end.

\begin{figure}[h]
\centering
\begin{tabular}{cc}
\hspace{-2ex}\includegraphics[height=3.5cm]{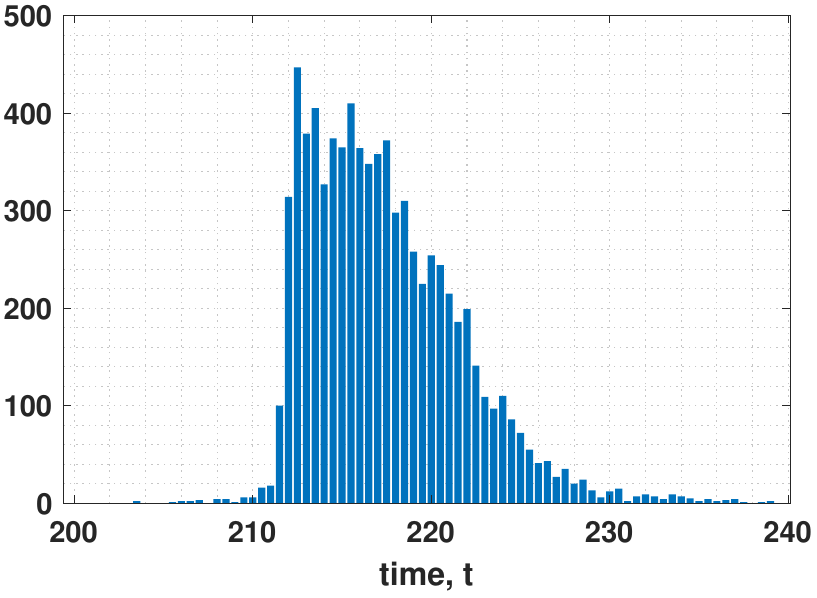}&
\hspace{-2ex}\includegraphics[height=3.5cm]{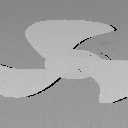}
\end{tabular}
\caption{[Left] The processed histogram of one of the pixels and [Right] the resulting depth map.}
\label{fig: Figure 24 processed pulse}
\end{figure}

After we have cleaned up the data, we can plot the resulting histogram as shown in \cref{fig: Figure 24 processed pulse}. The pulse is significantly more concentrated around the true peak. The resulting depth map is slightly over-smoothed. However, since all our subsequent analysis is done based on the processed histogram (as we will be re-sampling with replacement from the processed histogram), computing the MSE with respect to this smoothed depth map will not cause issues to the analysis.

\subsection{Estimating $\sigma_t$ and $\alpha_0$}
In the theoretical analysis of the variance, we need to know the pulse width $\sigma_t$ and the total amount of flux $\alpha_0$.

\begin{figure}[h]
\centering
\begin{tabular}{cc}
\hspace{-2ex}\includegraphics[width=0.475\linewidth]{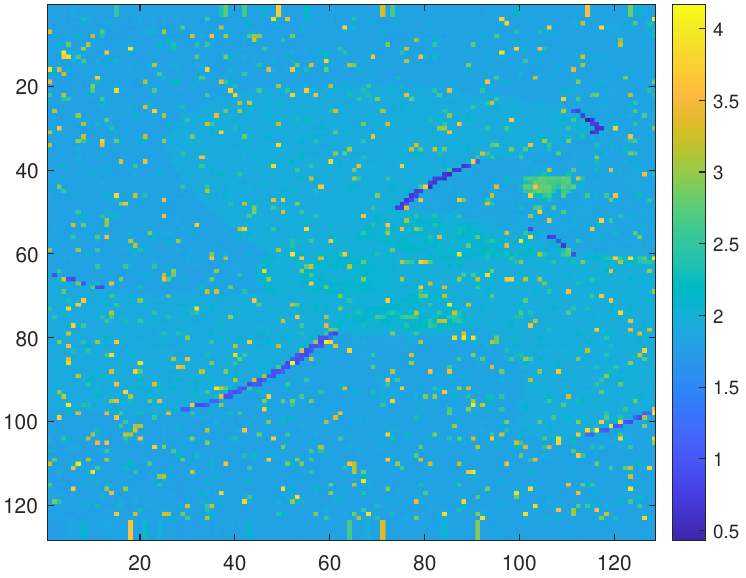}&
\hspace{-2ex}\includegraphics[width=0.475\linewidth]{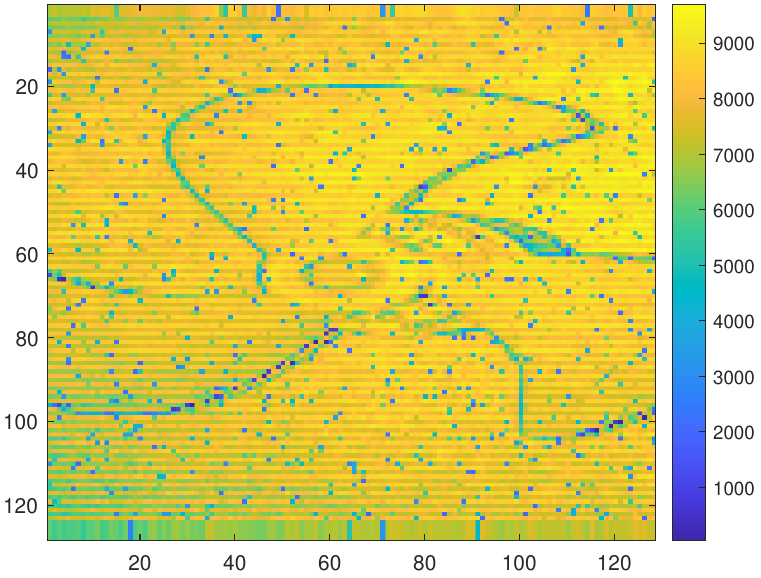}\\
$\sigma_t$ & $\alpha_0$
\end{tabular}
\caption{The values of $\sigma_t$ and $\alpha_0$ over the unit square.}
\label{fig: Figure 26 sigma alpha}
\end{figure}

For $\sigma_t$, we use the histogram shown in \cref{fig: Figure 24 processed pulse} to compute the variance at every pixel location. Although the pulse is asymmetric, we treat it as a symmetric pulse and compute the standard deviation regardless. We repeat the process for every pixel, and we generate a map of $\sigma_t$ as shown in \cref{fig: Figure 26 sigma alpha}. For simplicity, we use the average of the $\sigma_t$ in this map as the true value for our theoretical analysis. Our estimated $\sigma_t$ is $\sigma_t = 3.93$.

For $\alpha_0$, we again use the histogram shown in \cref{fig: Figure 24 processed pulse} to count the number of elements in the histogram. Recall that during the pre-processing stage, we have already eliminated all the outliers from the raw data. This will leave us with a smaller set of samples compared to the original data array. The number of active samples will inform us about the number of photons arriving at the sensor. During our analysis, we will sample with replacement from the histogram. For each trial, we draw $K$ samples from the available data points per pixel. Therefore, the total flux we have in the data is the sum of the values in \cref{fig: Figure 26 sigma alpha}, divided by 10,000 because we have 10,000 frames, and multiplied by $K$. Approximately our $\alpha_0$ is $3.86\times10^{4}$.

\subsection{Variance Estimation via Bootstrap}
The main tool we use to evaluate the variance is bootstrap. The idea is that given a dataset of $M$ data points, we sample with replacement $K$ samples and calculate the estimate of interest. The variance of the estimate is called the bootstrapped variance. When we repeat the random trials for long enough, the bootstrapped variance will converge to the true variance in probability.

Following this idea, for every pixel, we sample with replacement $K = 3$ times to generate the samples. From these samples, we construct the ML estimate (for every pixel) where the ML estimation is done by taking the simple average of the samples. Here, we assume that the pulse is symmetric even though it is not. We find that this approximation does not cause too many issues. Once the ML estimate is obtained, we repeat the process $20$ times to evaluate the variance.

For lower resolutions, we bin the samples to form a bigger pool. If we use a $2\times 2$ bin, then the total number of samples to be generated for the bootstrap purpose is $K \times 2^2 = 12$ samples; If we use a $3 \times 3$ bin, then the total number of samples is $K \times 3^2 = 27$ samples. These samples will give us an estimate of the real variance of the data.

The theoretical formula of the variance is based on a simpler form $\sigma_t^2 N^2/\alpha_0$. We omitted the $\|\vc\|^2\sigma_s^2$ because $\|\vc\|^2\sigma_s^2$ is small compared to $\sigma_t^2$. (Recall that $\sigma_s = 1/(\sqrt{12}N)$).

\subsection{Bias Estimation via Numerical Integration}
The bias is computed numerically via integration. The reason is that the pseudo ground truth depth map is overall a piecewise constant function. As we explained in the limitation subsection of the Bias section, there is no simple analytic formula for calculating the bias for piecewise constant functions unless we perform the numerical integration. The numerical integration is done by downsampling and upsampling the cleaned raw data volume. For example, to calculate the bias caused by a $2 \times 2$ bin, we take the average over all the available samples in the $2 \times 2 \times T$ (where $T$ is the number of available samples). After binning, we scale it back to the full resolution and compute the sum squared error with respect to the pseudo ground truth.

\section{Q\&A}
In this section, we list a few questions and answers which might be of interest to readers.

\emph{Q1. Can you just use as many pixels as possible during acquisition, and then perform binning (of the time stamps) as post-processing?}

Answer: Yes, this is completely possible. In fact, in our real data analysis, we see that when the pulse is short enough, the MSE decays monotonically with the $N$. Theoretically, the optimal $N$ still exists but this $N$ could be larger than what the physical resolution of the sensor can support. In this case, we should just maximize the resolution.

From a practical point of view, we also agree that post-processing of noisy time stamps can be cost-effective. Analog processing on the sensor front could also be a solution.

\vspace{2ex}
\emph{Q2. What if you denoise the estimated time of arrival map? Will it beat your MSE bound?}

Answer: Yes, it will. The theory we presented here uses the maximum likelihood (ML) estimation. ML estimation allows us to say things concretely so that we can offer a simple and interpretable MSE estimate. If we do denoising, then we will be doing maximum a posteriori or minimum mean square estimation. In those cases, we need to specify the prior distribution for which no one has a formula. Even if we do pick a prior (e.g., total variation), the derivation of the MSE will be substantially harder if not intractable. So, we lose the capability of writing a simple and interpretable formula. In short, while we are almost certain that a well-defined post-processing will ``beat'' our MSE bound, this ``victory'' offers little to no theoretical benefit to the scope of this paper.

\vspace{2ex}
\emph{Q3. What is the utility of this paper?}

Answer: The MSE we show in this paper is, in our honest opinion, simple, interpretable, and elegant. For the first time in the LiDAR literature, we provide the exact characterization of the resolution limit.

\vspace{2ex}
\emph{Q4. You need to show more comparisons by sending the raw time stamps through a state-of-the-art depth reconstruction neural network.}

Answer: Beating a SOTA depth reconstruction neural network is not the purpose of this paper.

\vspace{2ex}
\emph{Q5. Is MSE the right metric? }

Answer: Yes, if you want to derive equations, especially a simple equation to give you intuitions about the problem. No, if you are more interested in practical scenarios. There is no better or worse.

\vspace{2ex}
\emph{Q6. Your model is inaccurate. It made too many assumptions such as Gaussian pulse, single-bounce, no dark current, constant reflectivity, etc.}

Answer: While we also want to be as accurate as possible, accuracy and simplicity are mutually exclusive in this paper. As we have explained in the supplementary material, all these situations can be handled \emph{numerically} by integrations. But this will defeat the purpose of deriving a closed-form expression.

\vspace{2ex}
\emph{Q7. Some papers in the literature use the Markov chain/self-excitation process to model the photon arrivals. Why are you skipping all these?}

Answer: We are just assuming that there is no dead time. If there is dead time, then you are correct that self-excitation processes are needed to provide a better model. However, this would be substantially more complicated than what we present here. By assuming no dead time, we can go back to the standard inhomogeneous Poisson process by exploiting independence. Even in this significantly simplified case, we see that the derivation of the MSE is non-trivial.

\vspace{2ex}
\emph{Q8. Can your model handle fog?}

Answer: Yes, but it will be complicated. Scattering medium such as fog affects the reflectivity $\alpha_0$ and causes additional background as shown in Figure~\ref{fig: Figure 3 PDF and ICDF}. You will need numerical integration to evaluate the bound.

\end{document}